\documentclass{aa}
\usepackage[varg]{txfonts}
\usepackage{natbib}
\usepackage{graphicx}
\usepackage{braket}
\usepackage{hyperref}
\begin{document}

\title{Mass accretion rates of clusters of galaxies: CIRS and HeCS}

\author{M. Pizzardo \inst{\ref{1},\ref{2}}
\and S. Di Gioia \inst{\ref{3},\ref{4},\ref{5}}
\and A. Diaferio \inst{\ref{1},\ref{2}}
\and C. De Boni \inst{\ref{6}}
\and A. L. Serra \inst{\ref{1},\ref{2}}
\and M. J. Geller \inst{\ref{8}}
\and J. Sohn \inst{\ref{8}}
\and \\K. Rines \inst{\ref{9}}
\and M. Baldi \inst{\ref{10},\ref{11},\ref{12}}
}

\institute{\label{1}Dipartimento di Fisica, Universit\`a di Torino, via P. Giuria 1,  I-10125 Torino, Italy \\
e-mail: {\tt michele.pizzardo@unito.it}
\and \label{2}Istituto Nazionale di Fisica Nucleare (INFN), Sezione di Torino, via P. Giuria 1,  I-10125 Torino, Italy
\and \label{3}Dipartimento di Fisica, Universit\`a di Trieste, via A. Valerio, 2, I-34127 Trieste, Italy
\and \label{4}Istituto Nazionale di Astrofisica (INAF), Sezione di Trieste, via G.B. Tiepolo 11, I-34143 Trieste, Italy
\and \label{5}Istituto Nazionale di Fisica Nucleare (INFN), Sezione di Trieste, via A. Valerio 2, I-34127 Trieste, Italy
\and \label{6}OmegaLambdaTec GmbH, Lichtenbergstra{\ss}e 8, D-85748 Garching, Germany
\and \label{8}Smithsonian Astrophysical Observatory, 60 Garden Street, Cambridge, MA-02138, USA
\and \label{9}Department of Physics and Astronomy, Western Washington University, Bellingham, WA-98225, USA
\and \label{10}Dipartimento di Fisica e Astronomia, Alma Mater Studiorum Universit\`a di Bologna, via Gobetti 93/1, I-40129 Bologna, Italy
\and \label{11}Astrophysics and Space Science Observatory Bologna, via Gobetti 93/2, I-40129, Bologna, Italy
\and \label{12}Istituto Nazionale di Fisica Nucleare (INFN), Sezione di Bologna, viale Berti Pichat 6/2, I-40127, Bologna, Italy
} 

\date{Received date / Accepted date}

\abstract
{We use a new spherical accretion recipe tested on N-body simulations to measure the observed mass accretion rate (MAR) of 129 clusters in the Cluster Infall Regions in the Sloan Digital Sky Survey (CIRS) and in the Hectospec Cluster Survey (HeCS). The observed clusters cover the redshift range of $0.01<z<0.30$ and the mass range of $\sim 10^{14}-10^{15}h^{-1}$~M$_\odot$. Based on three-dimensional mass profiles of simulated clusters reaching beyond the virial radius, our recipe returns MARs that agree with MARs based on merger trees. We adopt this recipe to estimate the MAR of real clusters based on measurements of the mass profile out to $\sim 3R_{200}$. We use the caustic method to measure the mass profiles to these large radii. We demonstrate the validity of our estimates by applying the same approach to a set of mock redshift surveys of a sample of 2000 simulated clusters with a median mass of $M_{200}= 10^{14} {h^{-1}~\rm{M_{\odot}}}$ as well as a sample of 50 simulated clusters with a median mass of $M_{200}= 10^{15} {h^{-1}~\rm{M_{\odot}}}$: the median MARs based on the caustic mass profiles of the simulated clusters are unbiased and agree within 19\% with the median MARs based on the real mass profile of the clusters. The MAR of the CIRS and HeCS clusters increases with the mass and the redshift of the accreting cluster, which is in excellent agreement with the growth of clusters in the $\Lambda$CDM model. }


\maketitle

\section{Introduction}

In the current cold dark matter (CDM) model of the formation and evolution of cosmic structures, smaller dark matter halos hierarchically aggregate into larger and more massive halos. The accretion occurs  through mergers with halos of comparable (major mergers) or lower (minor mergers) mass and with the capture of diffuse dark matter particles \citep[e.g.][]{press1974formation,10.1093/mnras/183.3.341,laceyCole93,10.1093/mnras/248.2.332,10.1046/j.1365-8711.2002.04950.x,10.1111/j.1745-3933.2008.00472.x,10.1111/j.1365-2966.2011.19638.x,PhysRevLett.106.241302,2014JCAP...10..077A,2018MNRAS.476.4877M}.

In principle, the mass accretion of dark matter halos is a valuable tool for testing different models of structure formation. Specifically, the mass evolution $M(z)$ of a dark matter halo, which describes its mass assembling history, or its time derivative, the mass accretion rate $\dot M(z)$ can be used to determine the halo formation redshift \citep{laceyCole93,10.1046/j.1365-8711.2002.05171.x,ragone2010relation,10.1111/j.1365-2966.2012.20594.x}. The rates are correlated with halo properties including concentration \citep{Wechsler_2002,Tasitsiomi_2004,Zhao_2009,10.1111/j.1365-2966.2012.20594.x,2013MNRAS.432.1103L}, shape \citep{kasun2005shapes,allgood2006shape,bett2007spin,ragone2010relation}, spin \citep{Vitvitska_2002,bett2007spin}, degree of internal relaxation \citep{10.1111/j.1365-2966.2011.19820.x}, and fraction of substructures \citep{10.1111/j.1365-2966.2004.08360.x,10.1111/j.1365-2966.2005.08964.x,2013MNRAS.432.1103L}.  
The assembly of dark matter halos can trace the accretion rate of baryons from the cosmic web onto the dark matter halo \citep{10.1046/j.1365-8711.2002.05171.x,2009MNRAS.398.1858M,2010MNRAS.406.2267F,2020MNRAS.498.1668W}. 

In theoretical studies, the estimates of the mass accretion history (MAH) $M(z)$ and the mass accretion rate (MAR) $\dot M(z)$ of a dark matter halo evolved at the present time, $z=0$, are usually tackled by tracing the merger tree of the halo, either with numerical simulations \citep[e.g.][]{genel2008mergers,KUHLEN201250,2012MNRAS.422.1028B} or with Monte Carlo methods \citep{laceyCole93,10.1093/mnras/264.1.201,somerville1999plant,parkinson2007generating,jiang2014generating}.
In hierarchical clustering scenarios, when we follow the growth history of the dark matter halo backwards in time, we see that the halo separates into two or more halos; at each epoch, the growing halo, known as the descendant, has a main progenitor in the form of the most massive halo, which merges with smaller halos and thereby generates the descendant. By identifying the main progenitor of the halo at each epoch, we can trace the formation history, or MAH, of a simulated halo by tracing the main branch of its merger tree.  

Within this framework, numerous studies investigate the MAH in $\Lambda$CDM cosmologies. Analytical approximations can describe the MAH as a function of the final mass of the halo and other additional parameters.
For example, \citet{2009MNRAS.398.1858M}, \citet{2010MNRAS.406.2267F}, and \citet{2015MNRAS.450.1514C} adopt the relation $M(z)=M_0(1+z)^\beta e^{-\gamma z}$ for the growth in mass with redshift $z$, whereas \citet{10.1046/j.1365-8711.2002.05171.x} and \citet{2014MNRAS.445.1713V} adopt $M(z)=M_0 \exp \left \{ \ln (1/2) \left[\frac{\ln(1+z)}{\ln(1+z_f)}\right]^\nu \right \}$. In the latter case, the formation redshift, $z_f$, and $\nu$ are left as free parameters, whereas in the former expression, $\beta$ and $\gamma$ are left either free or fixed by the linear power spectrum of matter \citep{2015MNRAS.450.1514C}. In both formulae, $M_0$ is the halo mass at redshift $z=0$. 

These studies point out that the MAH can be separated into two regimes. In the first regime, at early times, the mass accretion is relatively large and the growth is nearly exponential in redshift; here, major mergers are frequent and keep the system unrelaxed. In the second regime, at later times, the accretion slows down and the growth is governed by a power-law in redshift, thus enabling the halo to reach virial equilibrium. These studies also show that more massive halos, which form at relatively low redshifts, have a greater MAR than less massive halos. This correlation is supported by (1) the correlation between the age and the concentration and (2) the anti-correlation between the mass and the concentration of dark matter halos \citep{Zhao_2009}. In other words, old, low-mass, and highly concentrated dark matter halos are expected to have lower MAR than young, high-mass and less concentrated halos. 

The MAH can be a probe of the cosmological parameters. \cite{hurier2019cosmological} uses the thermal Sunyaev-Zel'dovich (SZ) effect as a proxy for the mass of the clusters from the second Planck SZ catalogue \citep{ade2016planck} and the fit by \cite{2015MNRAS.450.1514C} to the MAH of dark matter halos in simulations to derive values for the power spectrum normalisation $\sigma_8 $, the cosmic mass density $\Omega_m$, and the Hubble parameter $H_0$, which yield $\sigma_8 (\Omega_m/0.3)^{-0.3} (H_0/70)^{-0.2} = 0.75 \pm 0.06$. This value is in rough agreement with other analyses of galaxy cluster samples and of the power spectrum of the cosmic microwave background \citep{2016ApJ...832...95D,2018arXiv180706209P,2019MNRAS.489..401Z}. 

The investigation of the MAH and the MAR is also connected to the splashback radius, which roughly corresponds to the first apocentre of the orbits of recently accreted material. This radius is larger than the radii $R_{200}$ or $R_{\mathrm {vir}}$\footnote{ $R_{\Delta}$ is commonly defined as the radius within which the average mass density is $\Delta$ times the critical density of the Universe at the appropriate redshift. The value $\Delta=200$ thus defines $R_{200}$. $R_{\mathrm {vir}}$ is the radius within which the cluster dynamics satisfies the virial theorem. This radius is usually estimated by assuming the spherical top-hat collapse of the density perturbation. For a $\Lambda$CDM model with cosmic mass density $\Omega_m=0.3$ and cosmological constant $\Omega_\Lambda=0.7$, $R_{\mathrm {vir}}\sim R_{100}$ \citep{1998ApJ...495...80B}. } that are usually adopted to quantify the halo size and is close to $\sim 2R_{200}$, on average \citep{2015ApJ...810...36M}. In their simulations, \cite{2014ApJ...789....1D} find that the steepness of the slope of the halo mass profile at large radii increases with increasing MAR. Moreover, the cluster-centric radius of this change of slope decreases with increasing MAR. \cite{2014JCAP...11..019A} associate the location of this feature with the splashback radius.

A change of slope that is consistent with the expectations from the simulations is indeed present in the profile of the surface number density of galaxies  from the Dark Energy Survey (DES) cross-correlated with the SZ clusters from  the South Pole Telescope (SPT) and the Atacama Cosmology Telescope (ACT) \citep{2019MNRAS.487.2900S}, as well as in the deprojected cross-correlation of the SZ clusters from the Planck Survey with  galaxies detected photometrically in  the PanSTARRS survey \citep{2019ApJ...874..184Z}. Similarly, the splashback radius is detected in the inferred dark matter density profiles of the redMaPPer clusters \citep{2016ApJ...825...39M} and in clusters from either the Sloan Digital Sky Survey (SDSS) \citep{2017ApJ...841...18B} or DES \citep{Chang_2018}. 

Although interlopers along the line of sight affect the inference of this feature both from optically selected clusters \citep{2017MNRAS.470.4767B,2019MNRAS.487.2900S,2019MNRAS.490.4945S} and from weak lensing analyses of X-ray selected clusters \citep{umetsu2017lensing,2019MNRAS.485..408C}, dense galaxy redshift surveys \citep{2011AJ....142..133G,2013ApJ...768..116S,2018ApJ...856..172S,2018ApJ...862..172R}, and upcoming lensing surveys might potentially overcome the effects of this contamination and constrain the relation between the splashback radius and the accretion rate \citep{2020MNRAS.499.3534X}.

All of these studies highlight the relevance of an observational estimate of the MAR of galaxy clusters. Unfortunately, only a handful of direct measurements have been attempted thus far.
In fact, we can observe a real cluster only at a specific time and we cannot clearly identify its merger tree to quantify the MAR, as we would usually do when $N$-body simulations are available. 
A viable approach that has been pursued with real clusters is to identify galaxy groups that surround the cluster and appear to fall onto it \citep[e.g.][]{Rines_2001,Rines2002MassA2199}; unfortunately, the estimate of their masses does not provide an estimate of the MAR, offering only a lower limit instead \citep[e.g.][]{Lemze_2013,2018MNRAS.477.4931H}. 

More importantly, the cluster outer region needs to be  chosen properly. For example, \citet{Lemze_2013} investigate the region slightly beyond $R_{200}$ in the X-ray and optical bands, whereas \citet{tchernin2016xmm} detect infalling gas clumps of A2142 in X-ray and SZ out to $\sim 1.3R_{200}$.  Similarly, \citet{2018MNRAS.477.4931H} identify the infalling groups in the range ${(0.28-1.35)R_{200}}$. According to studies of the splashback radius, these radii may be too small to return a full estimate of the MAR: in fact, we expect that these regions contain matter with rather different dynamical histories: matter that is falling onto the cluster for the first time, matter that is moving outwards, and matter that is falling back again. 

To avoid this complex dynamical structure and to return a proper estimate of the MAR, we must consider a region that is further out, beyond, at least, the splash-back radius of $\sim 2R_{200}$. We expect that this region mostly contains matter that will fall onto the cluster in the near future \citep{2016ApJ...818..188D, 2018MNRAS.477.4931H} and that the fraction of matter that is actually moving out from the central region  is limited \citep{2009PhDT........13L, 2017ApJ...843..140D, 2020MNRAS.499.3534X, 2020arXiv200805475B, Aung2020}. 
 
In this paper, we pursue this idea that was originally suggested by \cite{2016ApJ...818..188D}. We estimate the MAR from the amount of mass in the cluster's outer region beyond $\sim 2R_{200}$.
Unfortunately, compared to the cluster central region, the cluster outskirts are a large and low-density region, where the system is not dynamically relaxed. The methods used to estimate the mass cannot rely on the hypothesis of virial equilibrium. The caustic method and weak gravitational lensing are two complementary methods that do not rely on this hypothesis \citep{Geller_2013} and are thus appropriate to estimate the amount of mass in these outer regions.  

Here, we present the first estimation of the MAR of real clusters based on the spherical accretion model developed by \cite{2016ApJ...818..188D}. To estimate the cluster mass profiles in their outer region, we use the caustic technique \citep{1997ApJ...481..633D,1999MNRAS.309..610D}. The caustic technique is known to return an unbiased mass estimate with a relative uncertainty of 50\%, or better, in the regions where accretion is taking place \citep{2011MNRAS.412..800S}. Unlike methods based on weak gravitational lensing, where the signal is stronger at intermediate redshift and rapidly drops at low and high redshift \citep{2003MNRAS.339.1155H,2011MNRAS.412.2095H}, the caustic technique can be applied to clusters at any redshift, provided that the number of cluster galaxies with spectroscopic redshift is large enough to sample the velocity field properly. 
 
In Sect. \ref{model}, we briefly summarise the spherical infall method introduced in \citet{2016ApJ...818..188D}. In Sect. \ref{simulation}, we use $N$-body simulations to test our method of estimating the MAR.  
Section \ref{data} describes the CIRS and HeCS catalogues of real clusters that span the redshift range $0-0.3$ and the mass range $\sim 10^{14}-10^{15}h^{-1}$~M$_\odot$. In Sect. \ref{mar}, we illustrate and discuss the estimates of the MAR of individual clusters and the mean MAR of the cluster samples as a function of mass and redshift. We discuss our results in Sect. \ref{discussion} and we present our conclusions in Sect. \ref{conclusion}.  
We adopt $H_0=100 h\, {\mathrm {km\, s^{-1}Mpc^{-1}}}$ throughout.

\section{Spherical accretion recipe} 
\label{model}

In this section, we briefly review the model proposed in \citet{2016ApJ...818..188D} to evaluate the MAR of clusters from spectroscopic redshift surveys. 

\citet{2016ApJ...818..188D} estimate the MAR from the merger trees of dark matter halos, of a mass of $\sim 10^{14}h^{-1}~$M$_\odot$, extracted from $N$-body simulations. They find that in the redshift range of $z = [0, 2]$, a spherical accretion recipe returns an unbiased MAR within $\sim 20\%$ of the average MAR from the merger trees.  For
the more massive halos of a mass of $\sim 10^{15}h^{-1}$~M$_\odot$, even though the statistics of \citet{2016ApJ...818..188D} are rather poor and the MAR estimated with the spherical accretion recipe is $\sim 10-40\%$ biased towards the low end, the recipe still returns a MAR within the 1$-\sigma$ spread of the MAR derived from the merger trees. 

In the spherical accretion recipe, the MAR is estimated by assuming that the mass within a spherical shell of a proper thickness, centred on the cluster, will fully accrete onto the cluster within a given time interval, $t_{\rm inf}$. 
We assume that the massive shell falls onto the cluster with constant acceleration and a given initial proper velocity $v_i$. By solving the equation of motion, we end up with an equation whose unknown is the thickness $\delta_sR_i$ of the shell, where $R_i$ is the physical inner radius of the shell: 
\begin{equation}\label{thickness}
t_{\rm inf}^2 GM(<R_i) - t_{\rm inf} 2 R_i^2 (1+  \delta_s/2)^2 v_i  - R_i^3  \delta_s (1+  \delta_s/2)^2 = 0 \ ,
\end{equation}
where $M(<R_i)$ is the mass of the cluster within the radius $R_i$, and $G$ is the gravitational constant.

We assume that the inner radius of the shell is $R_i = 2 R_{200}$. As anticipated in the introduction, this radius, $R_i$, is close to the average splashback radius of massive dark matter halos of cluster size at redshift $z< 2$ found in the $N$-body simulations \citep{2015ApJ...810...36M}. This radius also approximates the inner radius of the region containing the mass that will fall onto clusters in the near future. In addition, it is close to $R_i$ that the absolute value of the infall radial velocity generally reaches its maximum (see Sect. \ref{radialvel}) and the recipe thus includes the largest contribution to the MAR.
The solution of Eq.~(\ref{thickness}) yields the shell thickness, $\delta_sR_i$, and the mass of the shell, $M_{\rm shell}$, if the mass profile of the cluster at radii larger than $2 R_{200}$ is known. 

The MAR is thus simply estimated as: 
\begin{equation}\label{MAR_recipe}
{\rm MAR}=\frac{dM}{dt}\equiv \frac{M_{\rm shell}}{t_{\rm inf}} \, .
\end{equation}

Our analysis below shows that Eq.~(\ref{thickness}) typically yields $\delta_sR_i\sim 0.5R_{200}$ and, thus, $M_{\rm shell}$ is the mass of the shell of inner and outer radii $2R_{200}$ and $\sim 2.5R_{200}$, respectively. The $N$-body simulations suggest that a fraction of this $M_{\rm shell}$ has already been within $R_{200}$ and is thus not actually falling onto the central region of the cluster for the first time \citep[e.g.][]{Aung2020}. According to \citet{2009PhDT........13L} and \citet{2020arXiv200805475B}, no more than $30\%$ of the subhalos in the radial range $2-3R_{200}$ have already been within $R_{200}$. Similarly, \citet{2017ApJ...843..140D} and \citet{2020MNRAS.499.3534X} find that $75\%$ to $87\%$ of the apocenters of the dark matter particles on their first orbit agree with the splashback radius identified by the maximum slope of the logarithmic derivative of the halo density profile \citep{2014ApJ...789....1D, 2015ApJ...810...36M}. This feature generally is within $2 R_{200}$ \citep{2014ApJ...789....1D, 2020MNRAS.499.3534X}. We can thus conclude that less than $\sim 20-25\%$ of the dark matter particles in the radial range $\sim 2-2.5R_{200}$ are not falling onto the inner region of radius $2R_{200}$ for the first time.

In principle, we could introduce a correction factor $\alpha_M<1$ for $M_{\rm shell}$ in Eq.~(\ref{thickness}), to take this effect into account. However, avoiding the introduction of this parameter has the advantage of keeping the recipe as simple as possible without affecting the effectiveness of our method: indeed, applying exactly the same recipe to real and simulated clusters yields a sensible comparison between the observed and the expected MAR. Alternatively, we could make the introduction of $\alpha_M$ irrelevant by adopting the inner radius of the shell larger than $2 R_{200}$ and thus decrease the fraction of $M_{\rm shell}$ that is not actually falling in for the first time; however, this choice would return a spherical shell that extends to radii larger than $\sim 2.5-3 R_{200}$. As we discuss in Sect. \ref{sec:selection}, in these regions, real clusters currently suffer from spectroscopic incompleteness and, consequently, the estimate of the MAR would be biased.

Our recipe for the MAR estimate still requires the values of $t_{\rm inf}$ and $v_i$, that are not currently measurable in real clusters.
For $t_{\rm inf}$ and $v_i$, we resort to $N$-body simulations of dark matter halos of cluster size, assuming that these systems resemble real clusters. Clearly, these values can differ widely from halo to halo. To apply Eq.~(\ref{MAR_recipe}) properly, we adopt an average value for both $t_{\rm inf}$ and $v_i$ for samples of
dark matter halos within proper mass and redshift bins. If the ranges covered by these bins are sufficiently small, as we detail below, 
this approach can provide the MAR of a real cluster if its mass profile is known.
 
We measure the mass profile of real clusters to radii larger than $R_{200}$ with the caustic technique. This technique 
uses redshift data alone and does not require the assumption of dynamical equilibrium; however, it assumes spherical symmetry and deviations from this symmetry are responsible for most of the uncertainty in the mass profile and, consequently, in the MAR. 

\section{Testing the mass accretion recipe on mock redshift surveys of clusters} 
\label{simulation}

Before applying the MAR recipe to real clusters, we need to evaluate the reliability of our MAR estimate and its possible systematic errors.
Here, we test the recipe on mock redshift surveys of clusters extracted from an $N$-body simulation of a $\Lambda$CDM model. We also use this simulation to provide the proper values for $t_{\rm inf}$ and $v_i$ for clusters in different bins of mass and redshift.

In Sects. \ref{CoDECS} and \ref{MarCodecs}, we describe the $N$-body simulation and the construction of the mock redshift surveys, respectively. In Sect. \ref{radialvel} we discuss the radial velocity profiles of the clusters in the simulation and in Sect. \ref{MarSimulated}, we illustrate how our estimate of the MAR from the redshift surveys compares with the MAR estimated with the full three-dimensional information.

\subsection{CoDECS simulations} 
\label{CoDECS}

For our tests, we relied on the L-CoDECS simulations \citep{2012MNRAS.422.1028B}, which is a set of $N$-body numerical simulations of a $\Lambda$CDM cosmology and other quintessence models. For our purposes, we used only the $\Lambda$CDM run.

The simulations are normalised at the cosmic microwave background epoch with cosmological parameters according to the {\it{WMAP7}} analysis \citep{2011ApJS..192...18K}: cosmological dark matter density $\Omega_{m0}=0.226$, cosmological constant $\Omega_{\Lambda 0}=0.729$, baryonic mass density $\Omega_{b0}=0.0451$, Hubble constant $H_0=70.3$~km~s$^{-1}$~Mpc$^{-1}$, power spectrum normalisation $\sigma_8=0.809$, and power spectrum index $n_s=0.966$. The box size is $1 h^{-1}~{\rm{Gpc}} $ on a side in comoving coordinates. The simulation contains $1024^3$ dark matter particles with mass $m_{\rm DM} = 5.84 \times 10^{10} {h^{-1}~\rm{M_{\odot}}} $ and the same number of baryonic particles with mass $m_b = 1.17 \times 10^{10} {h^{-1}~\rm{M_{\odot}}} $. Baryons are included only to check fifth-force effects in the quintessence cosmologies, but no hydrodynamics is included in the simulation. The $\Lambda$CDM run is a standard collisionless $N$-body simulation.\footnote{In the $\Lambda$CDM simulation we use here, both dark matter and baryonic particles are present: the baryonic particles are collisionless and the fifth-force effects are absent. Particles of unequal mass can generate collisional effects on small scales \citep[e.g.][]{Binney2002TwoBodyRel,Ludlow2019EnEquip}. However, these effects have a negligible impact on our analysis. We prove this statement with an $N$-body simulation with a single dark matter fluid and similar cosmological parameters to the L-CoDECS simulation. In this test simulation, the estimated MARs are within a few percent of the MARs that we present here. These differences are well within the dispersions of the MAR distributions of the clusters in the same mass and redshift bins that we show in Fig. \ref{marsimulated}. }

Groups and clusters in the simulations were identified with a friends-of-friends algorithm with linking length $\lambda=0.2\bar{d}$, where $\bar{d}$ is the mean Lagrangian inter-particle separation. The algorithm is only run over the dark matter particles. Each baryonic particle is associated with the closest dark matter particle at the end of the procedure. 

We considered only the clusters in two different mass bins with median mass, within $R_{200}$ at $z=0$, $M_{200} \simeq 10^{14} h^{-1}~\rm{M_{\odot}}$, and $M_{200} \simeq 10^{15} h^{-1}~\rm{M_{\odot}}$. The two samples contain $N=2000$ and $N=50$ clusters, respectively. We identify the main progenitors of these clusters and consider the samples of these progenitors at each redshift. We consider the outputs of the simulation at six different redshifts: $z = 0.0, 0.12, 0.19, 0.26, 0.35, 0.44$. Table \ref{binStat} shows the medians and the $68$th percentile ranges of the two mass bins at different redshifts.

\begin{table*}[htbp]
\begin{center}
\caption{ \label{binStat} Samples of simulated clusters.}
\begin{tabular}{ccc}
\hline
\hline
 $z$ & median $M_{200}$ & $68$th percentile range \\
 & $\left[10^{14}h^{-1}~\text{M}_{\odot }\right]$  & $\left[10^{14}h^{-1}~\text{M}_{\odot }\right]$ \\
\hline
 & & \\
0.00 & 1.00 & 0.97-1.04 \\
0.12 & 0.95 & 0.75-1.07 \\
0.19 & 0.90 & 0.64-1.06 \\
0.26 & 0.85 & 0.56-1.02 \\
0.35 & 0.78 & 0.49-0.96 \\
0.44 & 0.69 & 0.43-0.89 \\
 
\hline
 & & \\
0.00 & 10.0 & 9.5-11.2 \\ 
0.12 & 8.2 & 6.8-10.9 \\
0.19 & 7.3 & 5.2-9.6 \\
0.26 & 6.1 & 4.4-8.4 \\
0.35 & 5.4 & 4.0-7.5 \\
0.44 & 4.7 & 3.2-6.1 \\

\hline
 \end{tabular}
 \end{center}
 \footnotesize{{\bf Notes.} The upper (lower) part of the table is for the clusters in the low-mass (high-mass) bin. The low- and high-mass bins contain 2000 and 50 clusters, respectively.}
 \end{table*}

\subsection{Mock catalogues and mass profiles}
\label{MarCodecs}

The first ingredient of the MAR recipe is the three-dimensional mass profile of the cluster extending to large radii. We use the caustic method to estimate this profile in real clusters. With a sufficiently dense redshift survey of the outer regions of an individual cluster, the caustic mass profile deviates from the real three-dimensional mass profile by less than 10\%, on average, in the radial range $[0.6,4]R_{200}$, with a $1\sigma$ relative uncertainty of $\sim 50$\% \citep{2011MNRAS.412..800S}. 

The uncertainty in the caustic mass profile is due mostly to the assumption of spherical symmetry and clearly propagates into our estimate of the MAR. To quantify how this uncertainty propagates, we apply our MAR recipe to synthetic observations of the simulated clusters.
We assume that the dark matter halos we model in our $N$-body simulation are a realistic representation of galaxy clusters, and that their dark matter particles trace the same velocity field of the real galaxies. 
This latter assumption is supported by $N$-body/hydrodynamical simulations that suggest that the velocity bias, $b_v$, between the velocity dispersions of galaxies and dark matter particles is negligible in the outskirts of galaxy clusters \citep{10.1111/j.1365-2966.2004.07940.x,2016MNRAS.461L..11H,2018MNRAS.474.3746A}. Specifically, \citet{2018MNRAS.474.3746A} show that in their 30 dark matter halos with mass $M_{200}$ in the range $(7\times 10^{13}-1.7\times 10^{15})h^{-1}$~M$_\sun$, $b_v$ steadily decreases from $\sim 1.4$ in the halo centre to $\sim 1$ beyond $2 R_{200}$; this is mainly because dynamical friction, which dominates in the central region \citep[e.g.][]{Tormen1998Substruct,Taylor2004Substruct}, is less relevant in the outer regions.

\begin{figure}
\centering
\includegraphics[trim=0cm 1.8cm 0cm 1.8cm, clip, width=0.47\textwidth]{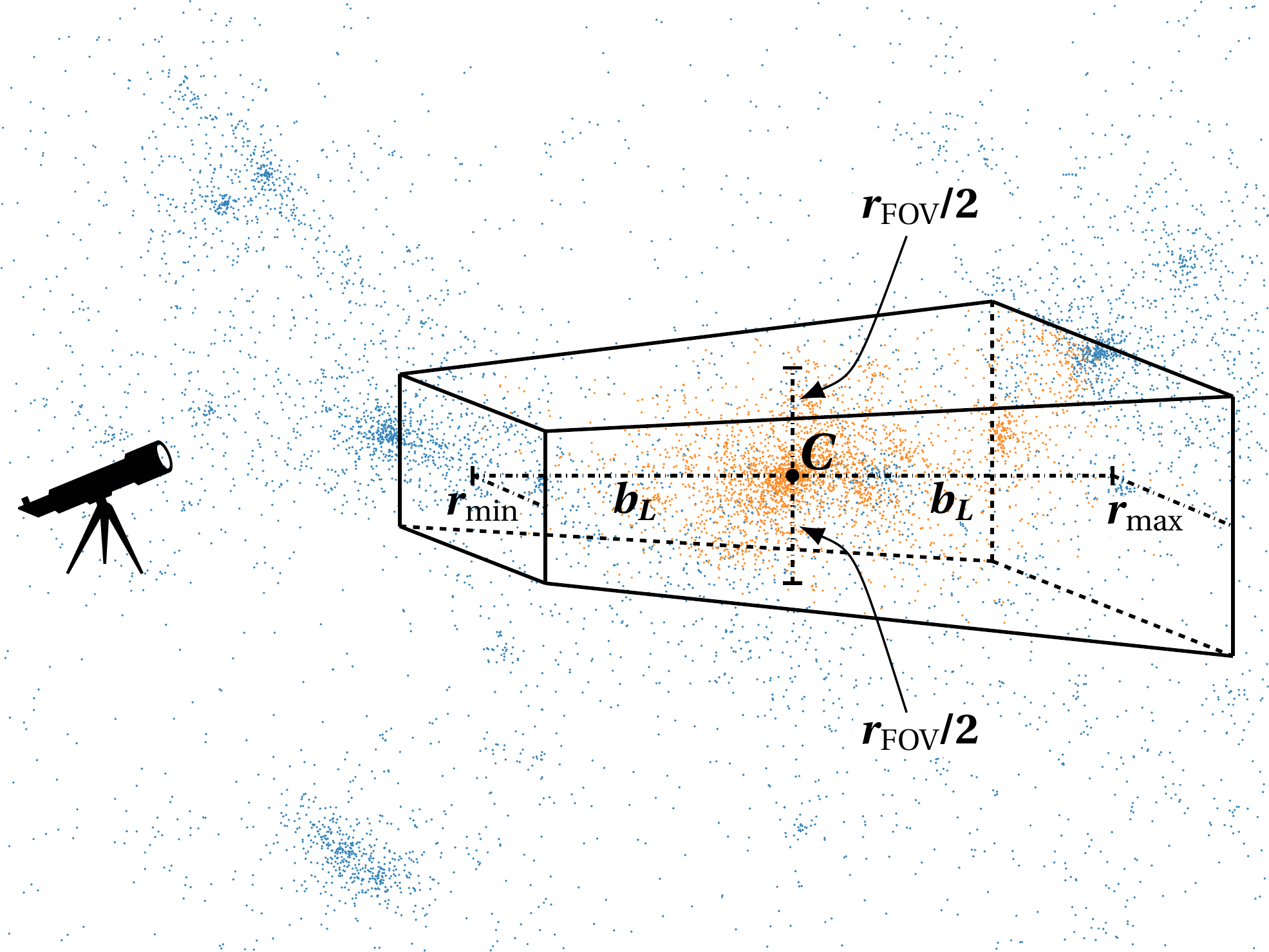}
\caption{Schematic figure of the truncated-pyramidal volume of the mock catalogue of a cluster extracted from the simulation. The solid dots show the positions of the dark matter particles in a slice of the simulation box centred on the cluster $C$. The orange (blue) dots are within (outside) the volume. This figure has only an illustrative purpose: the actual mock catalogues have $b_L\gg r_{\mathrm {FOV}}$ and include a substantially larger number of particles than shown here.}\label{mock_skt} 
\end{figure}

To create the mock redshift survey of a simulated cluster, we extract a squared-basis truncated pyramid centred on the cluster, with the smaller basis closer to the observer (see Fig. \ref{mock_skt}); the pyramid axis is aligned along one of the three cartesian coordinates chosen as the line of sight and has a height of $2b_L=140 h^{-1}$~Mpc. The sizes, $r_{\mathrm {min}}$ and $r_{\mathrm {max}}$, of the two bases are defined by the intercept theorem $r_s:r_{\mathrm {FOV}}/2=\left( r_s\mp b_L \right):r_{\mathrm {min, max}}$, where $r_s$ is the comoving distance of the cluster centre from the observer and $r_{\mathrm {FOV}}$ is the size of the field of view (FOV). We can use the previous relation that is appropriate for Euclidean geometry because we are considering the flat $\Lambda$CDM model.
Similarly to \citet{2011MNRAS.412..800S},  we take $r_{\mathrm {FOV}}=12 h^{-1}$~Mpc. 
This value easily covers even the most massive clusters out to a few virial radii.

For each cluster, we built three mock catalogues, one for each cartesian coordinate chosen as the line of sight. In general, the clusters are not spherically symmetric. Thus, to improve the statistics, we can consider these three mock catalogues as independent clusters even though they are not completely independent systems. 

Each selected volume contains the number $N_{14}=35_{-10}^{+16}\times 10^3$ and $N_{15}= 62_{-14}^{+22}\times 10^3$ of dark matter particles for the low- and high-mass bins, respectively; the ranges shown indicate the $10^{\rm th}-90^{\rm th}$ percentile ranges. In the same field of view, the densest survey of a real cluster contains a few thousand galaxies  \citep[e.g.][]{2014ApJ...797..106H,2019ApJ...871..129S}). Therefore, to identify each dark matter particle with an individual galaxy, we randomly sample a limited number of dark matter particles within the volume. According to \citet{2011MNRAS.412..800S}, the caustic method performs better when the velocity field of the cluster is sampled by $\sim 200$ galaxies within $3R_{200}$ from the cluster centre. We follow this approach and sample the  dark matter particles until we reach  $N_{\mathrm {sample}}=200$ particles within $3R_{200}$. This procedure yields the number $N^{\rm red}_{14}= 3000_{-900}^{+1700}$ and $N^{\rm red}_{15}= 610_{-130}^{+290}$ of particles within each mock redshift survey for the low- and the high-mass bin, respectively.

The observed redshift $z$ of each particle is set by its cosmological redshift and its peculiar velocity ${\mathbf v}_p={\mathbf u}_p/(1+z_s)$ in the comoving frame of the simulation box:  $z_s$ is the redshift of the simulation snapshot and ${\mathbf u}_p$ is the comoving peculiar velocity provided by the $N$-body simulation.\footnote{As velocity variable, the code Gadget-II used for the CoDECS simulations actually uses and returns the quantity ${\mathbf w}=a^{1/2}{\mathbf u}_p$ as a remedy to the divergence of the comoving peculiar velocity field at small scale factors $a$; this strategy maximises the computational efficiency of the integration of the equations of motion at early times \citep{2001NewA....6...79S}.}

The comoving distance from the observer to the centre of the simulation box, which coincides with the cluster centre, is  $r_s=c/H_0\int_0^{z_s}dz'/E(z')$, where $E(z)=H(z)/H_0=[(\Omega_{m 0}+\Omega_{b 0})(1+z)^3+\Omega_\Lambda]^{1/2}$ in the flat $\Lambda$CDM model.
The particle position vector in the observer reference frame is thus ${\mathbf r_i}={\mathbf r}_s+ {\mathbf r}_{c,i}$, where ${\mathbf r}_{c,i}$ is the comoving position vector of the particle in the reference frame of the simulation box. This sum of vectors is derived in the Euclidean geometry of the $\Lambda$CDM model of the simulation.

The cosmological redshift, $z_i$, of the particle satisfies the relation $r_i=c/H_0\int_0^{z_i}dz'/E(z')$ and the observed component of the peculiar velocity is $v_{\mathrm {los,}i}={\mathbf v}_p \cdot {\mathbf r_i}/r_i$. 
The redshift due to the peculiar velocity is $z_{p,i}=v_{\mathrm {los,}i}/c$ because $v_{\mathrm {los,}i}\ll c$. Combining $z_{p,i}$ with $z_i$ yields the observed redshift $(1+z_{\mathrm {obs,}i})=(1+z_i)(1+z_{p,i})$, namely: 
\begin{equation}
        cz_{\mathrm {obs}}=cz_i+v_{\mathrm {los},i}(1+z_i).
\end{equation} 

Standard geometrical transformations finally yield the celestial coordinates $(\alpha, \delta)$ from the cartesian components of ${\mathbf r}_i$. For all the clusters, we chose the celestial coordinates of the cluster centre $(\alpha, \delta ) = (6^h, 0)$. For the snapshot corresponding to $z=0$, we located the cluster centre at $cz=32,000$~km~s$^{-1}$, similarly to the $z=0$ mock catalogues described in \citet{2011MNRAS.412..800S}. 

We have two samples of simulated clusters: 2000 clusters with $M_{200}\sim 10^{14} h^{-1}~ {\rm{M_{\odot}}}$ and 50 clusters with $M_{200}\sim 10^{15} h^{-1}~ {\rm{M_{\odot}}} $ at $z=0$. By projecting each clusters along three lines of sight, we obtain two samples of 6000 and 150 cluster redshift surveys for the low- and high-mass bins, respectively.  

\begin{figure*}[!h]
  \centering
  \includegraphics[scale=1.1]{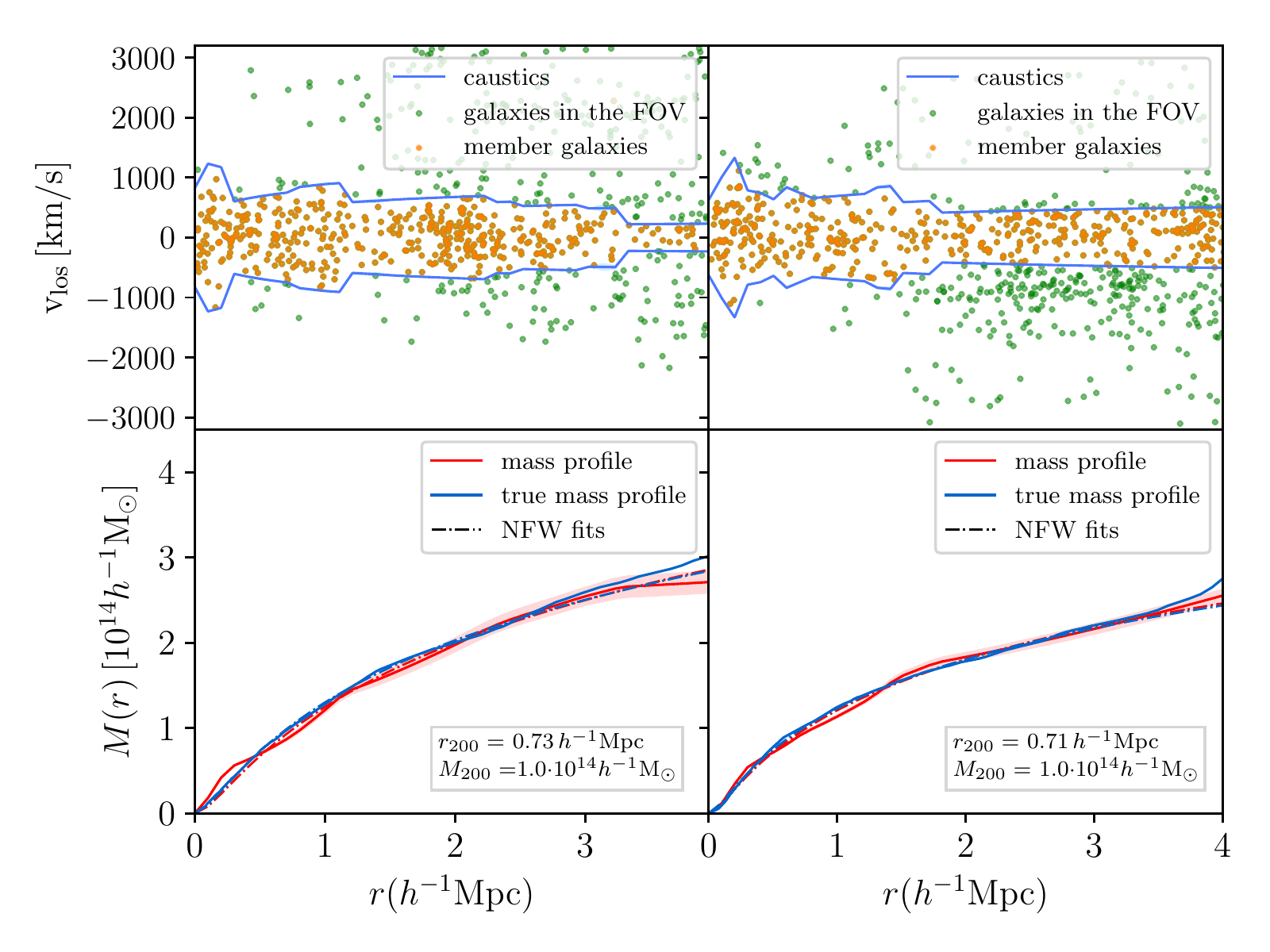}
  \caption{$R-v_{\rm los}$ diagram ({\it top panels}) and mass profile ({\it bottom panels}) of two simulated clusters in the low-mass bin. {\it Left (right) column}: cluster at $z=0.12$ ($z=0.19$). In the {\it bottom panels}, the red (blue) solid curves show the caustic (real) mass profile, whereas the red (blue) dot-dashed curves show the NFW fits to the caustic (real) mass profile. In these two examples, the two NFW fits are indistinguishable. The shaded areas show the 50\% confidence level of the caustic location and of the caustic mass profile according to the caustic technique recipe. In the ${R-v_{\rm los}}$ diagrams the shaded areas are present, but very thin.}\label{sim14}
\end{figure*}

\begin{figure*}[!h]
  \centering
  \includegraphics[scale=1.1]{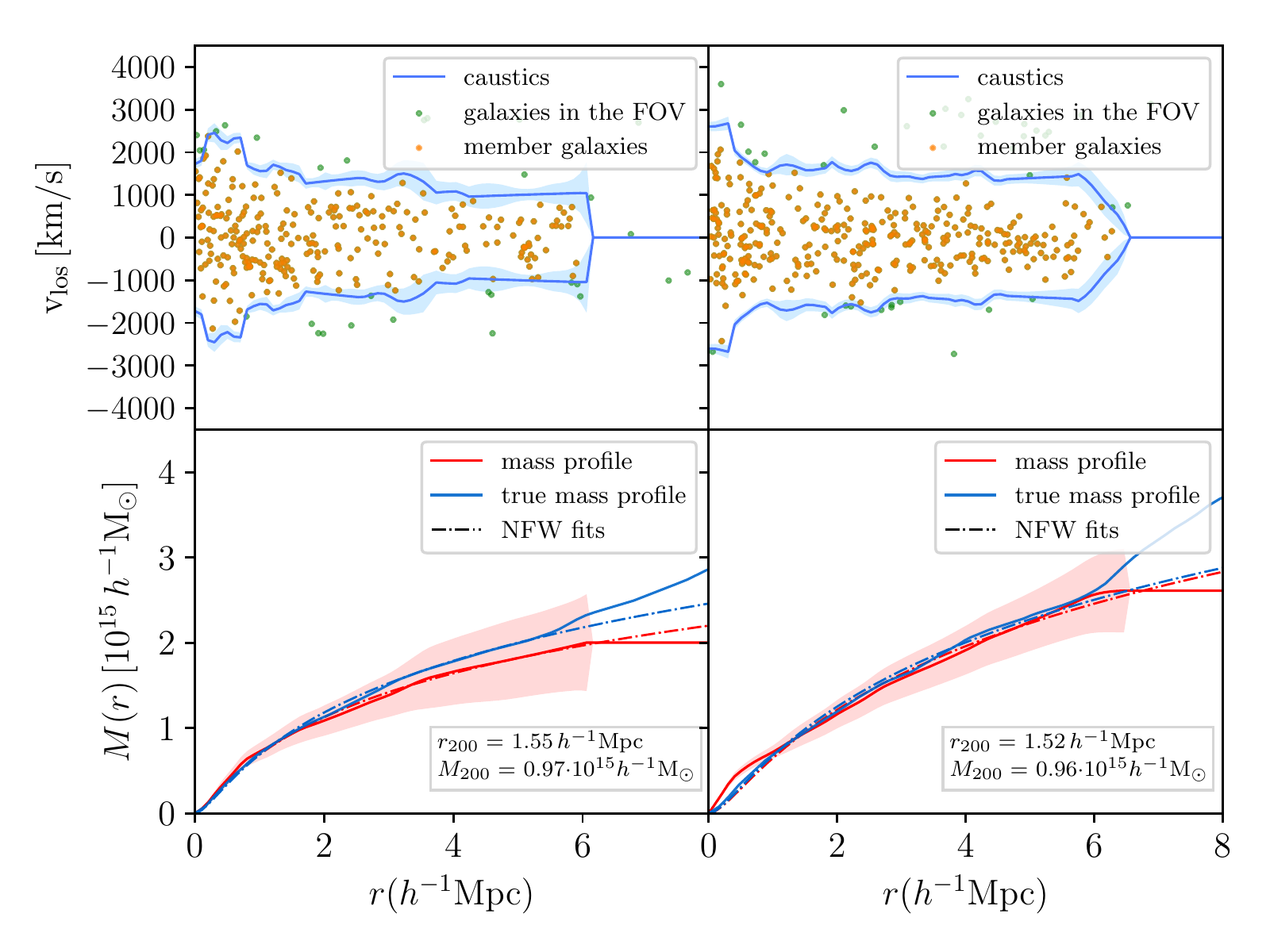}
  \caption{Same as Fig. \ref{sim14} but for two simulated clusters in the high-mass bin.}\label{sim15}
\end{figure*}

For each mock catalogue, we construct the ${R-v_{\rm los}}$ diagram, the line-of-sight velocity relative to the cluster mean as a function of the projected distance from the cluster centre. The caustic method returns the mass profile estimated from the amplitude of the caustics \citep[see][for further details on the caustic technique]{1999MNRAS.309..610D, 2011MNRAS.412..800S}. Figures \ref{sim14} and \ref{sim15} show some examples of the procedure for the low- and high-mass bins. In both figures, the left and right columns show simulated clusters at $z=0.12$ and $z=0.19$, respectively; the upper and lower panels show the ${R-v_{\rm los}}$ diagrams and the mass profiles. 
 
The mass profiles estimated from the caustic amplitude are within the expected uncertainty from the real mass profile within $\sim 4 R_{200}$. At larger radii, the caustic mass profiles generally flattens, unlike the real mass profile. These regions can include large nearby groups and clusters that would increase the caustic amplitude with increasing radius; however, the caustic technique is designed to favour decreasing amplitudes with increasing radius, to avoid the inclusion of background and foreground systems, as detailed in \citet{1999MNRAS.309..610D} and \citet{2011MNRAS.412..800S}. Thus, at radii where the amplitude would increase by an anomalous amount, the algorithm sets the caustic amplitude to zero and the cumulative mass profile gets flattened. 

We fit both the real and caustic mass profiles with the \citet{1997ApJ...490..493N} (NFW) model. We fit the caustic mass profiles up to $4 R_{200}$, unless the caustic amplitude shrinks to zero at a smaller radius. Similarly, we fit the real mass profiles up to $4R_{200}$. In the bottom panels of Figs. \ref{sim14} and \ref{sim15}, the red and blue dot-dashed lines show the NFW fits to the caustic and the real mass profiles, respectively.
When averaged over the entire radial range, the NFW fits overestimate, by $\sim 5\%,$ on average, the caustic and the real mass profile. This analysis shows that the NFW profile properly models the mass profiles of the clusters in our simulation to radii substantially larger than $R_{200}$. This result is also observed in real galaxy clusters in the CAIRNS \citep{2003AJ....126.2152R}, CIRS \citep{2006AJ....132.1275R}, and HeCS catalogues \citep{2013ApJ...767...15R}. Indeed, by applying the caustic technique to a dense redshift survey of the Coma cluster, \citet{Geller_1999} were the first to prove that the NFW model describes the mass profile of real clusters to these large radii; a few years later, this result was confirmed by weak gravitational lensing observations of other clusters \citep{Clowe2001WLA1689,Kneib2003,Lemze2009A1689}. 

The concentration parameters derived from  the NFW fits to the real mass profiles are in the range between $2-6$ for both mass bins, which is consistent with the concentration parameters of dark matter halos  with similar mass and redshift in other $N$-body simulations \citep[e.g.][]{2014MNRAS.441..378L,10.1093/mnras/stw1046}. The concentrations derived from the fits to the caustic mass profiles are, on average, $15\%$ larger than those derived from the real mass profiles. This overestimation confirms the results of \citet{2011MNRAS.412..800S} and is derived from the fact that the caustic method generally overestimates  the mass  of the cluster within $\sim 0.6R_{200}$.

\subsection{Radial velocity profiles and the accretion time, $t_{\mathrm {inf}}$}
\label{radialvel}

The second ingredient of our MAR recipe is the radial velocity profile of the cluster extending to large radii. This piece of information cannot be derived from observations. Therefore, we need to rely on the modelling of clusters in the $N$-body simulations.

For each cluster, we construct the radial velocity profile by computing the mean radial velocity $v_{\mathrm {rad}}$ of the particles within individual radial bins. The radial velocity of each particle is $v_i=[{\mathbf v}_p +H(z_s)a(z_s){\mathbf r}_{c,i}]\cdot {\mathbf r}_{c,i}/r_{c,i}$, where, as in the previous section, ${\mathbf v}_p={\mathbf u}_p/(1+z_s)$ is the proper peculiar velocity and ${\mathbf r}_{c,i}$ is the comoving position vector of the particle from the cluster centre; $H(z_s)$ and $a(z_s)$ are the Hubble parameter and the cosmic scale factor at  the snapshot redshift $z_s$, respectively. 

\begin{figure*}[!h]
  \centering
  \includegraphics[scale=0.55]{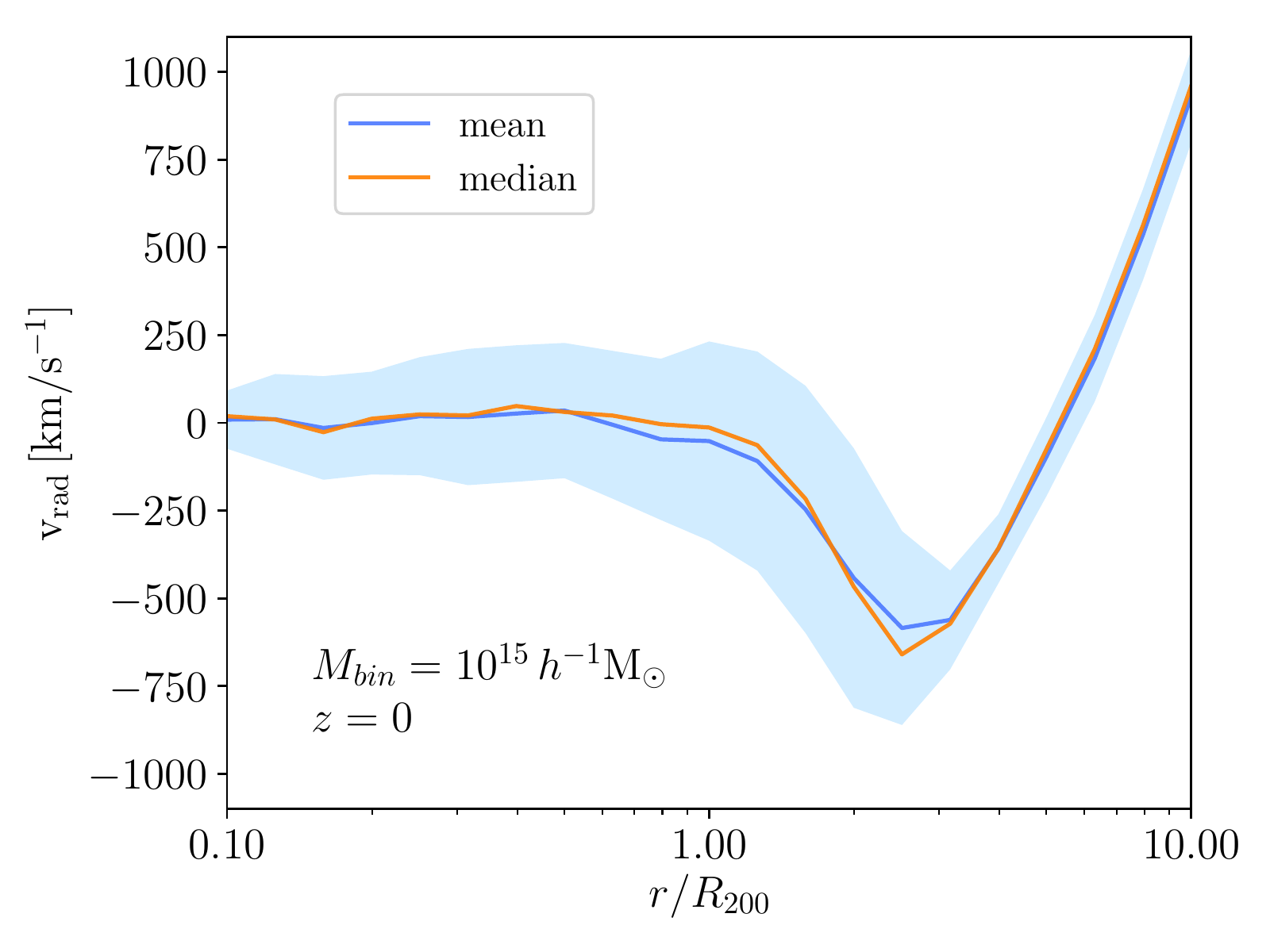}%
  \includegraphics[scale=0.55]{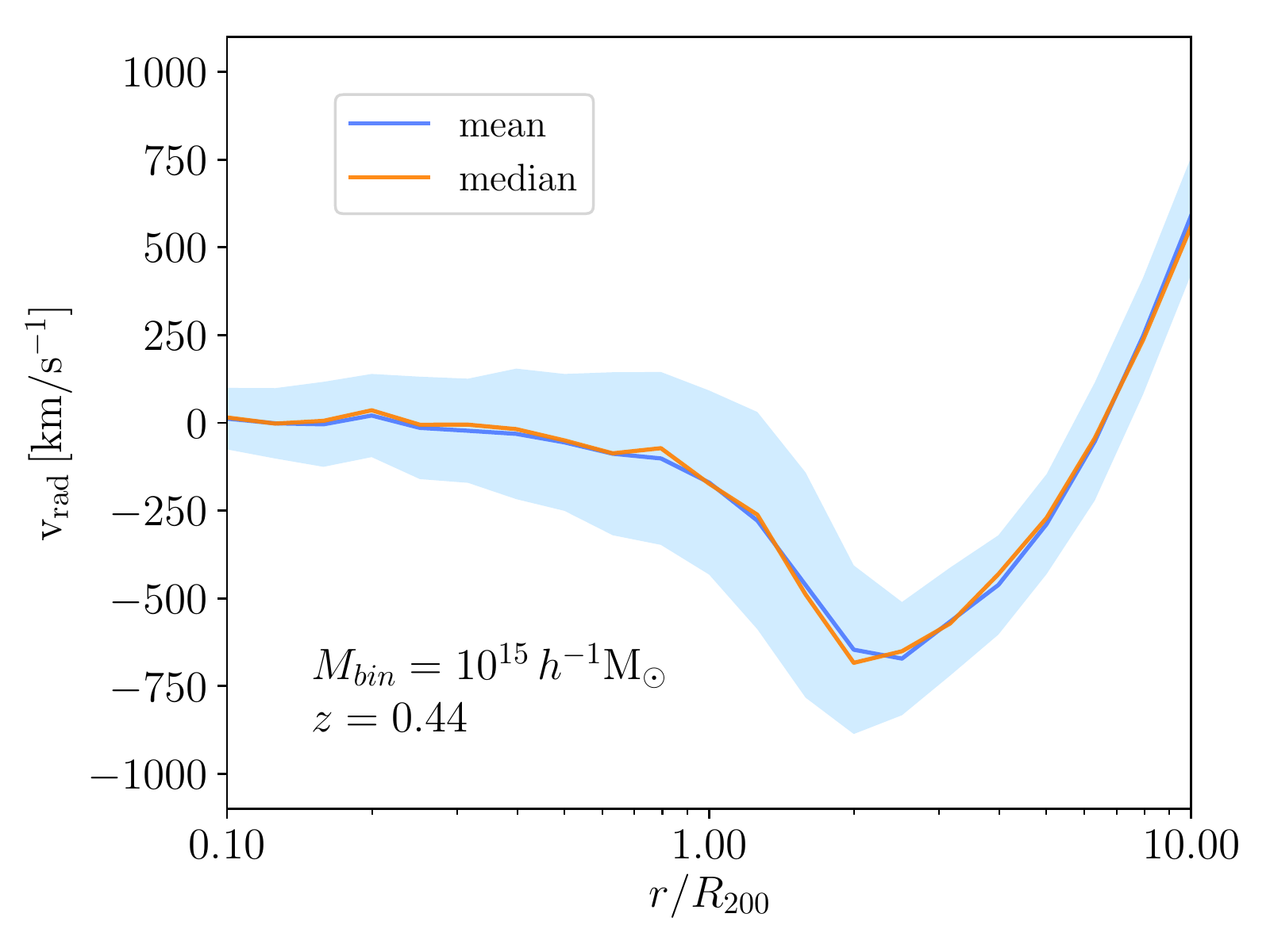}%

  \includegraphics[scale=0.55]{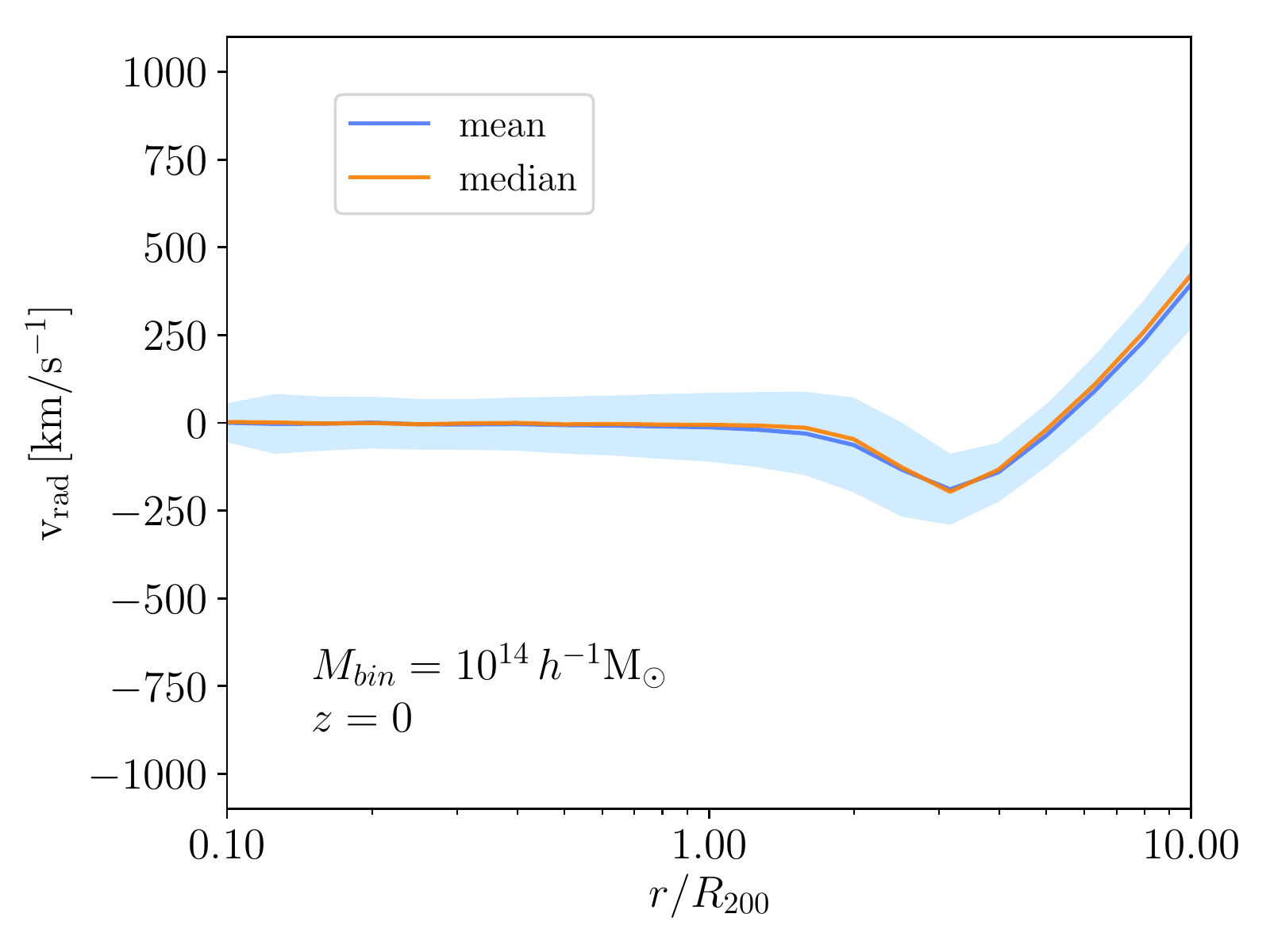}%
  \includegraphics[scale=0.55]{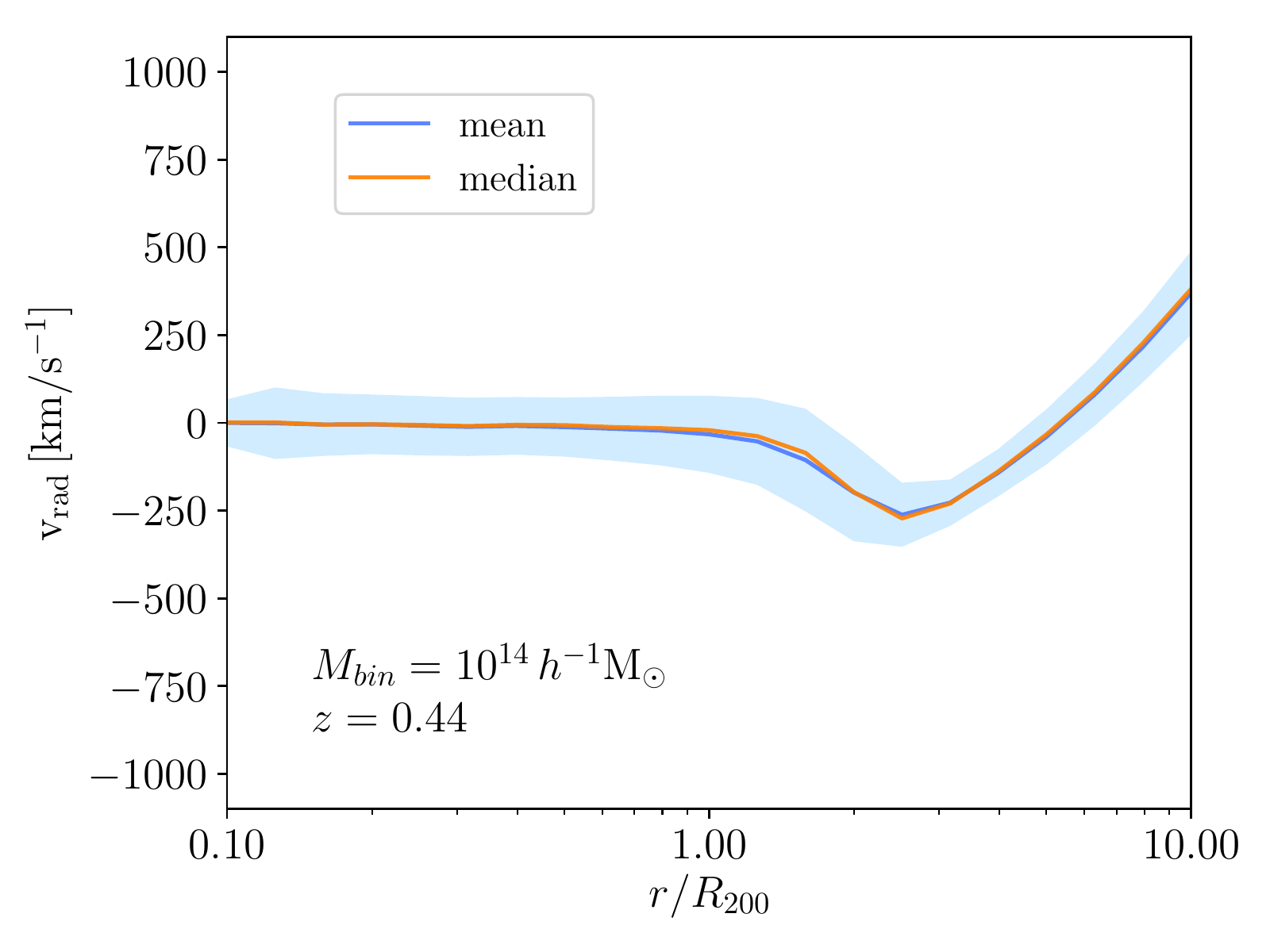}
  \caption{Mean (blue) and median (orange) profiles of the radial velocity of the particles within the dark matter halos extracted from the simulation for the two mass bins at $z=0$ and $z=0.44$, as indicated in the panels. $68\%$ of the profiles of the individual halos lie within the light blue areas.}
  \label{fig:radvel}
\end{figure*}

Figure \ref{fig:radvel} shows the mean and median profiles of the radial velocity of the simulated clusters in our low- and high-mass samples at $z=0$ and $z=0.44$. We also show the $68$th percentile range of the profile distribution. In the estimate of the velocity profiles, we include both the dark matter and the baryonic particles. The simulated clusters at other redshifts show qualitatively similar results.

By inspecting the radial velocity profile, we can identify three regions: an internal region with radial velocity $v_{\mathrm {rad}} \simeq 0$, where the matter is moving close to the centre within an isotropic velocity field; an infall region, where
$v_{\mathrm {rad}}$ becomes negative and indicates an actual infall of matter towards the centre of the cluster; and a Hubble region at radii $r\gtrsim 4 R_{200}$, where $v_{\mathrm {rad}}$ becomes positive and the Hubble flow dominates.
Broadly speaking, the infall radius $R_{\mathrm {inf}}$, defined as the radius where the minimum of $v_{\mathrm {rad}}$ occurs, is between $2R_{200}$ and $3R_{200}$, independently of the cluster mass and redshift.

We estimate the mean radial velocity profile at discrete values of the radius $r$. For the infall velocity $v_i$ of the shell (Eq.~\ref{thickness}), we take the radial velocity associated with the bin $[2-2.5]R_{200}$.
The choice of this velocity is consistent with the shell adopted for the estimate of the MAR: the internal radius of the shell is $R_i = 2R_{200}$, comparable with the splashback radius \citep{More_2015,2014JCAP...11..019A}, and the thickness typically returned by Eq.~(\ref{thickness}) is $\sim 0.5R_{200}$.  
For both bins of cluster mass, this radial bin corresponds to the radial range where the radial velocity profile has its minimum thus capturing the largest contribution to the MAR. 

Our prescription for the MAR estimate clearly depends on the choice of both $R_i$ and $v_i$. We will discuss the effect 
of the value of $v_i$ on the MAR in Sect. \ref{Results} below. 
Here, we show that using the mean infall velocity rather than the infall velocity of each cluster has very little impact on the estimate of the MAR.
We compare the spreads of the velocity profiles shown in Fig. \ref{fig:radvel} with the spreads of the MAR estimates obtained with the mean $v_i$ and either the three-dimensional mass profiles or the caustic mass profiles that we compute in Sect. \ref{MarSimulated} below (and shown in Fig. \ref{marsimulated}). 

We use the $1\sigma$ relative standard deviation of the mean $\epsilon_x=\sigma_x/(\sqrt{N}\bar{x})$, where $\bar{x}$ is the mean of a sample of $N$ measurements of the quantity $X$. For the low-mass bin, the infall velocity standard deviation $\epsilon_{v_i}\approx 1.2\%$ propagates into the standard deviation of the three-dimensional MAR, $\epsilon_{\rm {MAR_{3D}}}\approx 0.95\%$; this spread is well within the standard deviation $\epsilon_{\rm {MAR_{caustic}}}\approx 1.9\%$ of the MAR distribution estimated from the caustic mass profiles. For the high-mass bin, where the number of clusters decreases from 2000 to 50, the velocity standard deviation becomes $\epsilon_{v_i}\approx 4.0\%$, which implies an $\epsilon_{\rm {MAR_{3D}}}\approx 6.6\%$ standard deviation of the real three-dimensional MAR; this $\epsilon_{\rm {MAR_{3D}}}$ is smaller than the $\epsilon_{\rm {MAR_{caustic}}}\approx 10.9\%$ standard deviation of the MAR estimated from the caustic mass profiles. We thus conclude that assuming the same mean radial velocity profile for every cluster in a given mass and redshift bin does not introduce any systematic bias in the MAR estimate. The uncertainty in the MAR is actually dominated by the uncertainty in the mass profile.

To estimate the MAR of galaxy clusters, we also need the accretion time $t_{\rm inf}$ (see Eq.~\ref{MAR_recipe}).
Following \citet{2016ApJ...818..188D}, we adopt $t_{\rm inf}=10^9$~yrs, which is comparable with the dynamical time $t_{\rm dyn}=R/\sigma$ for the clusters we consider here, where $R$ and $\sigma$ are their size and one-dimensional velocity dispersion, respectively. 
To see that this value of $t_{\rm inf}$ is a reasonable choice, consider a homogeneous spherical system which is $\alpha$ time denser than  the  critical  density $3H_0^2/8\pi G$ and  thus  has  mass $GM=\alpha H_0^2 R^3/2$. When the system is in virial equilibrium, its potential energy $W=-3GM^2/5R$ and kinetic energy $K=3M\sigma^2/2$ can  be  combined  in  the  virial  relation $K=\vert W\vert/2,$ which  yields $GM=5R\sigma^2/4$.   
Therefore, $t_{\rm dyn}=R/\sigma = (5/2\alpha)^{1/2}H_0^{-1} \sim 10^9$~yrs when $\alpha\sim 200$. The size and the velocity dispersion of the cluster progenitors at earlier times decrease by comparable factors; therefore, $t_{\rm dyn}$ remains roughly constant thus justifying the choice $t_{\rm inf}=10^9$~yrs for any cluster at any redshift.

In principle, we could adopt a different value for $t_{\rm inf}$.
However, the thickness of the shell derived to estimate the MAR is proportional to $t_{\rm inf}$, and adopting  $t_{\rm inf}$ larger than $10^9$~yrs would generally return a shell thicker than $~\sim 0.5 R_{200}$, which is the typical value we estimate in our analysis below. Thicker shells would thus include regions beyond $\sim 2.5-3 R_{200}$. In these regions, real clusters suffer from spectroscopic incompleteness (see Sect. \ref{sec:selection}) and $t_{\rm inf}> 10^9$~yrs would thus return a biased estimate of the MAR. Adopting $t_{\rm inf}$ smaller than $10^9$~yrs would instead return thinner shells; in this case, the shell would contain a handful number of galaxies in the redshift diagram and the MAR estimate would be affected by relevant shot noise.

\subsection{Estimates of the MAR of simulated clusters}
\label{MarSimulated}

We now apply the caustic technique to each of our mock catalogues to derive the caustic mass profile and to estimate the MAR with our recipe (Eq.~\ref{MAR_recipe}, Sect. \ref{model}). 
We bin the mock catalogues according to the cluster redshift and mass. 

As we show with real clusters (Sect. \ref{mar}), some of the individual ${R-v_{\rm los}}$ diagrams do not support an estimate of the MAR. 
Our recipe requires that we estimate the mass of the shell of radii $R_i=2R_{200}$ and $(1+\delta_s)R_i\sim 2.5R_{200}$. The caustic method estimates this mass from 
\begin{equation}
        GM_{\rm shell} = {\cal F}_\beta \int_{R_i}^{(1+\delta_s)R_i} {\cal A}^2(R) {\rm d}R,
\end{equation}
where ${\cal F}_\beta$ is the filling factor \citep{1999MNRAS.309..610D,2011MNRAS.412..800S} and ${\cal A}(R) $ is the amplitude of the caustics, namely, the vertical separation of the upper and lower caustics at radius $R$.  At such large radii, a system can return a caustic amplitude ${\cal A}(R)=0 $, either because of poor sampling (especially in real systems) or because of galaxy-rich background or foreground structures that inhibit the caustic technique from properly identifying the caustic location. In these cases, the mass of the shell, and thus the MAR, cannot be estimated.

\begin{figure}[htbp]
\centering
\includegraphics[scale=0.54]{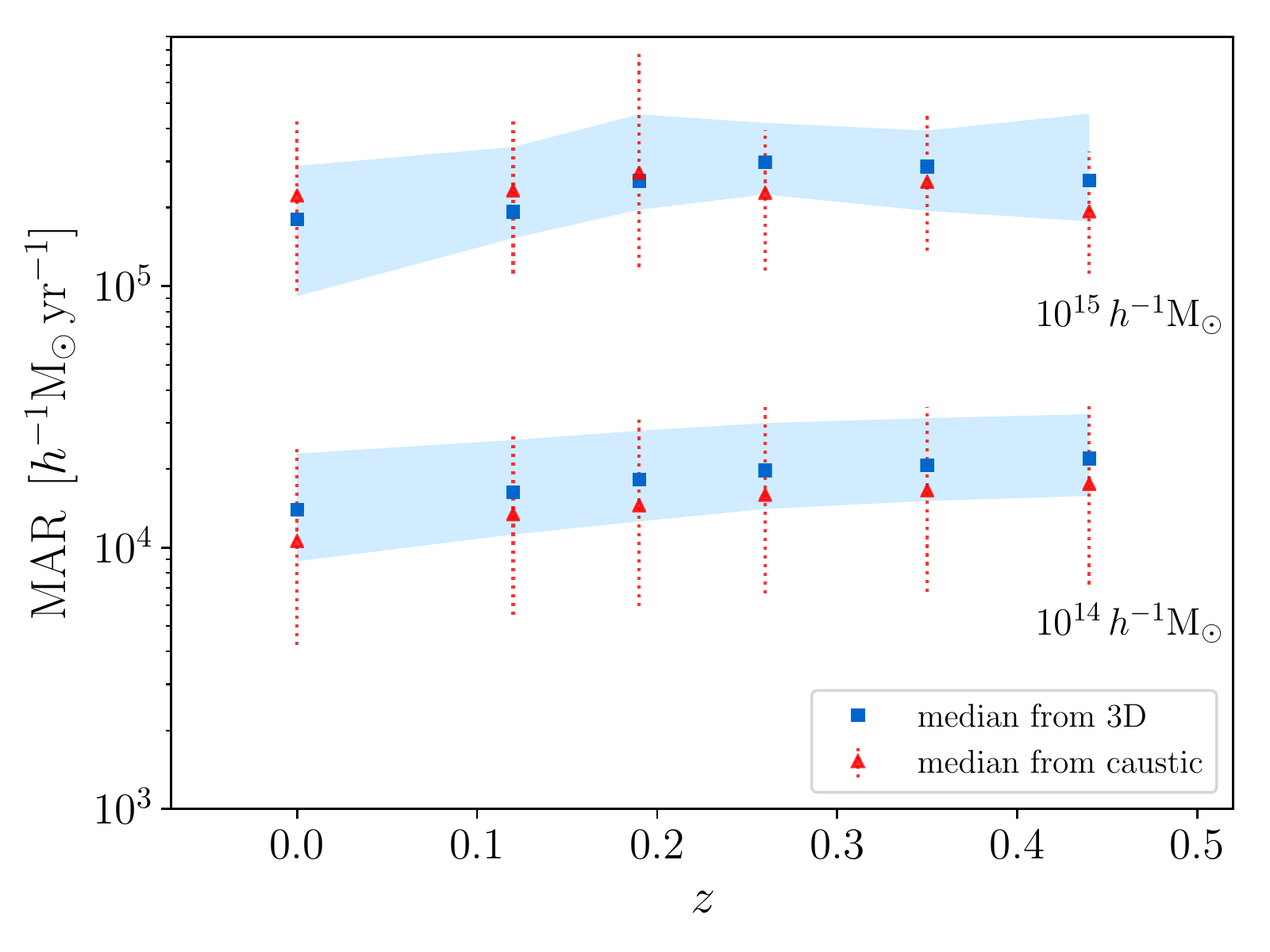}
\caption{ MAR of simulated clusters in the low- (lower set of points) and high- (upper set of points) mass bins.
        The blue squares and the red triangles show the median MAR based on the three-dimensional and the caustic mass profiles, respectively.
        The blue shaded areas show the $68$th percentile range of the distribution of the MAR derived from the three-dimensional mass profiles; the red error bars show the $68$th percentile ranges of the estimates obtained with the caustic mass profiles.}\label{marsimulated}
\end{figure}

In the samples of real clusters, we visually inspect the ${R-v_{\rm los}}$ diagrams to identify systems where the caustic technique fails. With the mock catalogues, we adopt an automatic procedure: to be conservative we remove the ${R-v_{\rm los}}$ diagrams where, in the range $[R_i;(1+\delta_s)R_i]$, the caustic amplitude ${\cal A}(R)<200$~km~s$^{-1}$ and ${\cal A}(R)<400$~km~s$^{-1}$, for the low- and high-mass bin, respectively. The caustic technique algorithm also prohibits unphysical increases of ${\cal A}(R) $ with increasing $R$. This feature of the algorithm is more effective in real clusters than in mock clusters; in fact, the simulated clusters tend to have less sharp separation between the cluster members and the foreground and background galaxies \citep{1999MNRAS.309..610D} and the caustic amplitude ${\cal A}(R) $ can remain artificially large at large $R$. Therefore, we also remove the ${R-v_{\rm los}}$ diagrams where the caustic amplitude ${\cal A}(R)>2000$~km~s$^{-1}$ in the range $[R_i;(1+\delta_s)R_i]$. This procedure removes 31\% and 16\% of the ${R-v_{\rm los}}$ diagrams for the low- and high-mass bin, respectively.
In Fig. \ref{marsimulated}, the red triangles show the median of the caustic MARs of the clusters at each redshift bin. The upper and lower sets of points are relative to the two cluster mass bins, as indicated in the figure. The error bars also show the $68$th percentile range of the distribution of the estimated caustic MARs.

To quantify the systematic errors introduced by the projection effects in realistic observations, the blue squares in Fig. \ref{marsimulated} show the MAR estimated with the recipe of Sect. \ref{model} but with the correct three-dimensional mass profile.
The procedure applied to obtain these estimates coincides with the procedure described in \citet{2016ApJ...818..188D}. The two estimates of the average MAR agree at all redshifts and for both mass bins: the average relative difference is $\lesssim 19\%$.
 
The difference between the three-dimensional MAR and the caustic MAR appears in their relative spreads. The relative spreads $\sigma_{\mathrm {MAR}}/{\mathrm {MAR}}$ of the MAR obtained with the three-dimensional mass profiles are  $\sim 42$\% and $\sim 46$\% for the low- and the high-mass bin, respectively. In contrast, using the mass profiles estimated with the caustic method, the relative spreads are $\sim 86\%$ and $\sim 75\%$, respectively.
These spreads are consistent with the spread of the caustic mass profile around the true mass profile estimated in $N$-body simulations \citep[][Fig. 12]{2011MNRAS.412..800S}. 
According to \citet{2011MNRAS.412..800S}, this spread mainly originates from projection effects. Consequently, projection effects are also mainly responsible for the spreads of the MAR estimated with the caustic method shown in Fig. \ref{marsimulated}.

In addition, the larger spread in the low-mass bin originates from the fact that in $N$-body simulations, less massive systems have less well defined structure in redshift space compared to more massive systems \citep{1999MNRAS.309..610D}. Consequently, the identification of the caustics is prone to larger random errors. The algorithm takes this effect into account by associating a larger uncertainty with the caustic location and with the mass profile. Redshift space structures are sharper in the real universe \citep{2000MNRAS.312..638S,10.1111/j.1365-2966.2006.11010.x} and applications of the caustic technique to numerous real clusters have indeed shown that the errors in the caustic mass profiles estimated with $N$-body simulations are probably upper limits \citep{Geller_1999,diaferio2005caustic,Geller_2013}.

Figure \ref{marsimulated} demonstrates that the caustic technique should return an unbiased estimate of the median three-dimensional MAR of real clusters at any redshift. In addition, the uncertainties overestimate the spread in the MAR based on three-dimensional mass profiles of a sample of clusters of comparable mass. We conclude that applying our recipe to real clusters should return a robust estimate of their MAR.

\section{Catalogues of real clusters}
\label{data}

Estimating the MAR of real clusters requires a dense redshift survey of the cluster outer regions. The largest catalogues currently available that satisfy this condition are the Cluster Infall Regions in the Sloan Digital Sky Survey (CIRS) \citep{2006AJ....132.1275R} and the Hectospec Cluster Survey (HeCS) \citep{2013ApJ...767...15R}. The former catalogue contains clusters at $z<0.1$ and the latter contains clusters in the redshift range $0.1<z<0.3$. These redshift ranges enable us
to measure the MAR as a function of redshift.
In Sects. \ref{cirs} and \ref{hecs}, we review the main features of the two catalogues; Sect. \ref{sec:selection} discusses some systematic effects in the selection of the galaxy samples and Sect. \ref{mar} describes the estimated MARs. 

\subsection{CIRS}\label{cirs}

The CIRS project extended the analysis of the CAIRNS survey \citep{2003AJ....126.2152R}, which pioneered the study of the infall region of clusters; CAIRNS used nine nearby galaxy clusters observed by the 2MASS survey \citep{jarrett_2004}, exploiting extensive spectroscopy and near-infrared photometry. 

CIRS is based on the Fourth Data Release (DR4) of SDSS, a photometric and spectroscopic wide-area survey at high galactic latitudes and low redshifts \citep{2002AJ....123..485S}. The DR4 includes 6670 deg$^2$ of imaging data and 4783 deg$^2$ of spectroscopic data \citep{Adelman_McCarthy_2006}. It was thus possible to extend the study of infall patterns around clusters initiated by CAIRNS to a larger number of clusters. By matching four X-ray cluster catalogues derived from the ROSAT All Sky Survey (RASS; \citealt{1999A&A...349..389V}) to the spectroscopic area covered by the DR4, \citet{2006AJ....132.1275R} obtained the CIRS catalogue, a sample of 74 clusters at $z<0.1$, perfectly suited to the study of infall regions with the caustic technique. 

\begin{figure}
\centering
\includegraphics[scale=0.57]{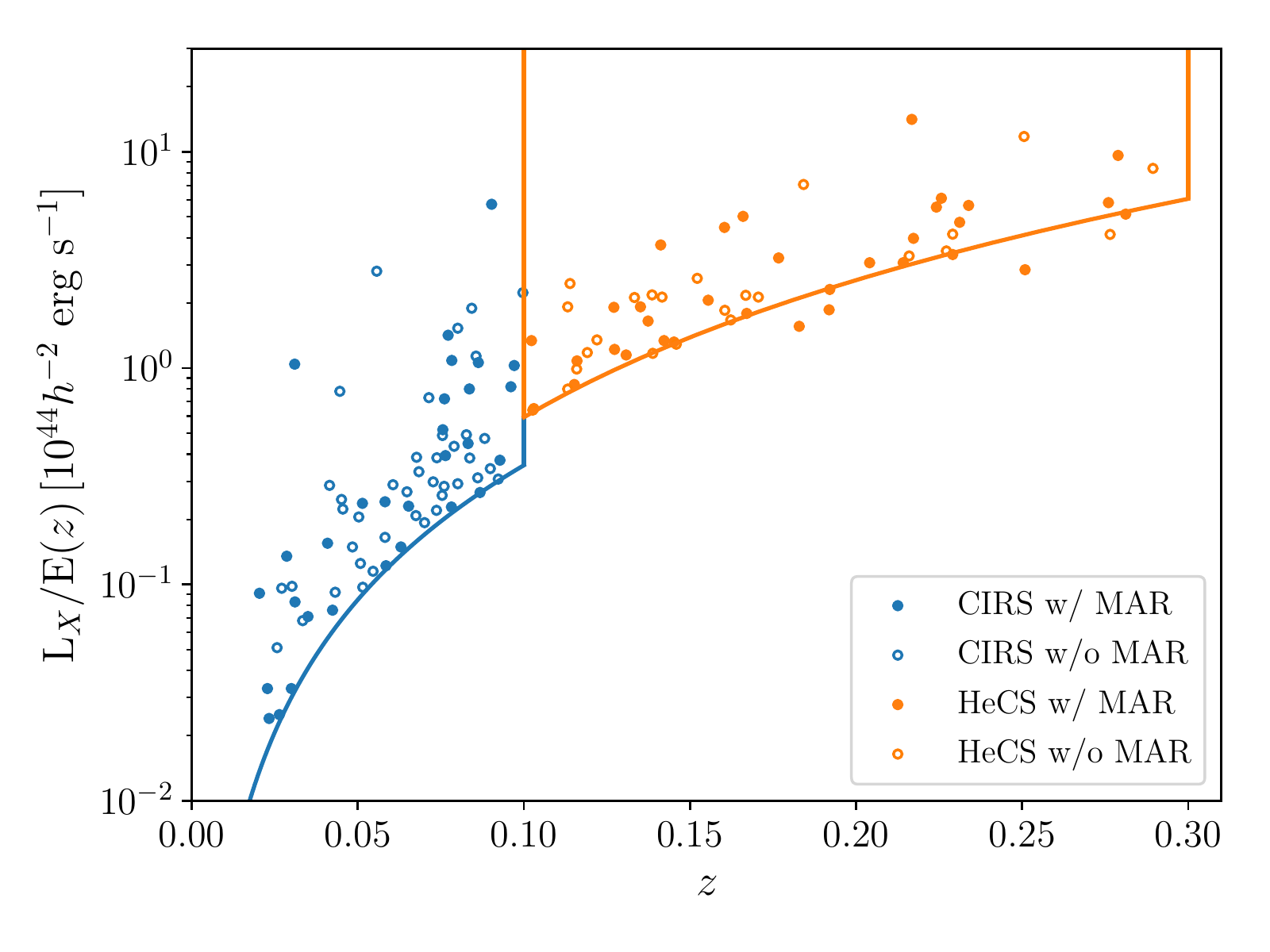} 
\caption{Luminosities of the CIRS (blue points) and HeCS (orange points) clusters as a function of redshift. The superimposed curves show the flux and redshift selection functions of the two catalogues. Filled points refer to the clusters where we can compute an individual MAR.}\label{selection}
\end{figure}

In our analysis, we used the updated catalogues of the CIRS dataset obtained by compiling the SDSS DR14 spectroscopic sample; these new catalogues are now part of the HeCS-omnibus survey \citep{sohn2019velocity}.
Here, three clusters of the original sample are removed: NGC4636, NGC5846 and Virgo. These clusters have redshifts $< 0.01$ and are poorly sampled. The remaining 71 clusters constitute our CIRS sample.
Figure \ref{selection} shows the selection function of the CIRS clusters: their X-ray flux limit in the 0.1-2.4 keV band is $3\times 10^{-12}$~erg~s$^{-1}$~cm$^{-2}$ and their redshift range is $[0,0.1]$.

\begin{figure}[htbp]
\centering
\includegraphics[scale=0.57]{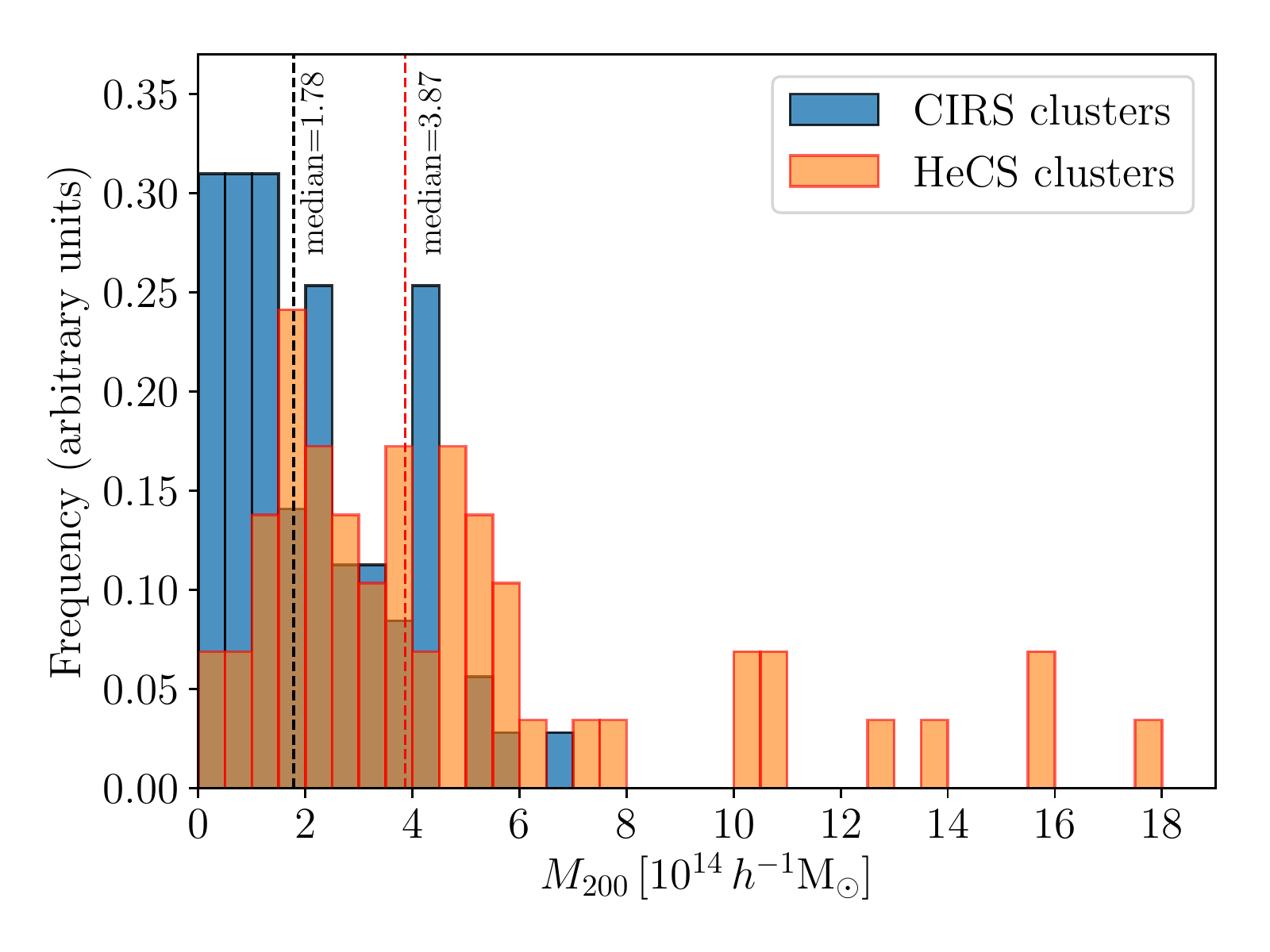}
\caption{Mass distribution of the CIRS (blue histogram) and HeCS (orange histogram) clusters. The black and red dashed lines show the median masses for the CIRS and HeCS catalogues, respectively. The area under each histogram is normalised to unity.}
\label{ch-mass}
\end{figure}

Table {\ref{CIRSsample}} lists the celestial coordinates, redshift, size $R_{200}$, and the corresponding mass $M_{200}$. It also lists the individual MARs estimated with the caustic method and discussed in Sect. \ref{Individuals}.\footnote{The properties listed here are derived from the analyses of the updated catalogue of \citet{sohn2019velocity}; these updated quantities are consistent with the values reported in \citet{2006AJ....132.1275R}.}   
The blue histogram in Fig. \ref{ch-mass} shows the mass distribution of the CIRS clusters.
Table \ref{cat info} lists the medians and percentile ranges of the redshift and mass distributions for both the complete catalogue and the subset of the CIRS clusters for which we estimate the individual MAR.

\begin{table*}[htbp] \tiny
\caption{\label{CIRSsample} \footnotesize CIRS clusters. }
\centering
\begin{tabular}{ccccccc}
\hline
\hline
        cluster & RA  & DEC & $z$ & $R_{200}$ & $M_{200}$ & ${\mathrm {MAR}}$ \\
  \quad & \scriptsize{[deg]} & \scriptsize{[deg]} &  & \scriptsize{[$h^{-1}$~Mpc]} & $\scriptstyle{\left[10^{14} h^{-1}~\text{M}_{\odot }\right]}$ & $\scriptstyle{\left[10^3 h^{-1}~\text{M}_\odot\text{yr}^{-1} \right]}$ \\
 \hline
\\
A0085 & 10.44 & -9.46 & 0.055 & $1.216 \pm 0.026$ & $4.37 \pm 0.28$ & $\cdots$ \\
A0119 & 14.06 & -1.28 & 0.044 & $1.3008 \pm 0.0084$ & $5.31 \pm 0.10$ & $\cdots$ \\
A0160 & 18.30 & 15.51 & 0.043 & $0.724 \pm 0.010$ & $0.914 \pm 0.037$ & $\cdots$ \\
A0168 & 18.75 & 0.28 & 0.045 & $0.703 \pm 0.026$ & $0.840 \pm 0.093$ & $\cdots$ \\
A0295 & 30.54 & -1.01 & 0.042 & $0.5253 \pm 0.0049$ & $0.349 \pm 0.010$ & $10.69 \pm 0.32$ \\
A0602 & 118.35 & 29.37 & 0.060 & $1.1903 \pm 0.0014$ & $4.124 \pm 0.014$ & $\cdots$ \\
A0671 & 127.11 & 30.44 & 0.050 & $1.005 \pm 0.016$ & $2.46 \pm 0.12$ & $\cdots$ \\
A0757 & 138.40 & 47.76 & 0.051 & $0.505 \pm 0.012$ & $0.313 \pm 0.021$ & $8.22 \pm 0.86$ \\
A0779 & 139.93 & 33.71 & 0.023 & $0.5065 \pm 0.0014$ & $0.3079 \pm 0.0026$ & $7.705 \pm 0.093$ \\
A0954 & 151.97 & 0.58 & 0.096 & $0.660 \pm 0.068$ & $0.73 \pm 0.22$ & $4.5 \pm 1.6$ \\
A0957 & 153.43 & -0.91 & 0.046 & $0.9558 \pm 0.0016$ & $2.108 \pm 0.011$ & $\cdots$ \\
A0971 & 154.98 & 40.98 & 0.092 & $1.111 \pm 0.011$ & $3.45 \pm 0.10$ & $\cdots$ \\
A1035A & 158.10 & 40.15 & 0.068 & $0.664 \pm 0.041$ & $0.72 \pm 0.13$ & $\cdots$ \\
A1035B & 158.05 & 40.28 & 0.078 & $0.7069 \pm 0.0001$ & $0.8772 \pm 0.0003$ & $\cdots$ \\
A1066 & 159.78 & 5.21 & 0.069 & $0.958 \pm 0.046$ & $2.16 \pm 0.31$ & $\cdots$ \\
A1142 & 165.23 & 10.51 & 0.036 & $0.641 \pm 0.015$ & $0.631 \pm 0.046$ & $9.76 \pm 0.97$ \\
A1173 & 167.38 & 41.57 & 0.076 & $0.6442 \pm 0.0066$ & $0.663 \pm 0.020$ & $\cdots$ \\
A1190 & 167.96 & 40.85 & 0.076 & $0.895 \pm 0.013$ & $1.779 \pm 0.076$ & $\cdots$ \\
A1205 & 168.49 & 2.46 & 0.076 & $0.689 \pm 0.016$ & $0.811 \pm 0.056$ & $56.0 \pm 5.1$ \\
A1291A & 173.09 & 55.96 & 0.051 & $0.9397 \pm 0.0037$ & $2.013 \pm 0.024$ & $\cdots$ \\
A1291B & 173.07 & 56.00 & 0.059 & $1.041 \pm 0.014$ & $2.76 \pm 0.11$ & $\cdots$ \\
A1314 & 173.67 & 49.08 & 0.033 & $0.803 \pm 0.028$ & $1.24 \pm 0.13$ & $\cdots$ \\
A1377 & 176.85 & 55.75 & 0.052 & $0.9151 \pm 0.0077$ & $1.860 \pm 0.047$ & $\cdots$ \\
A1424 & 179.42 & 5.08 & 0.075 & $0.9960 \pm 0.0029$ & $2.449 \pm 0.022$ & $\cdots$ \\
A1436 & 180.05 & 56.24 & 0.064 & $0.777 \pm 0.055$ & $1.15 \pm 0.24$ & $\cdots$ \\
A1552 & 187.47 & 11.78 & 0.086 & $1.062 \pm 0.018$ & $3.00 \pm 0.15$ & $\cdots$ \\
A1650 & 194.65 & -1.75 & 0.084 & $0.779 \pm 0.024$ & $1.18 \pm 0.11$ & $\cdots$ \\
A1663 & 194.67 & -1.73 & 0.084 & $1.201 \pm 0.010$ & $4.32 \pm 0.11$ & $\cdots$ \\
A1728 & 200.58 & 11.23 & 0.089 & $1.0295 \pm 0.0081$ & $2.737 \pm 0.065$ & $\cdots$ \\
A1750 & 202.69 & -1.85 & 0.085 & $1.0691 \pm 0.0086$ & $3.055 \pm 0.073$ & $\cdots$ \\
A1767 & 204.04 & 59.19 & 0.071 & $1.195 \pm 0.039$ & $4.21 \pm 0.41$ & $\cdots$ \\
A1773 & 205.54 & 2.25 & 0.078 & $1.179 \pm 0.025$ & $4.07 \pm 0.26$ & $26.6 \pm 4.5$ \\
A1809 & 208.29 & 5.15 & 0.079 & $0.991 \pm 0.038$ & $2.42 \pm 0.28$ & $\cdots$ \\
A1885 & 213.42 & 43.66 & 0.089 & $0.793 \pm 0.010$ & $1.249 \pm 0.049$ & $\cdots$ \\
A2061 & 230.30 & 30.58 & 0.077 & $1.154 \pm 0.010$ & $3.82 \pm 0.10$ & $86.9 \pm 3.4$ \\
A2064 & 230.24 & 48.66 & 0.074 & $0.929 \pm 0.014$ & $1.985 \pm 0.086$ & $\cdots$ \\
A2067 & 230.28 & 30.58 & 0.077 & $1.098 \pm 0.014$ & $3.28 \pm 0.12$ & $\cdots$ \\
A2110 & 234.90 & 30.71 & 0.097 & $0.707 \pm 0.039$ & $0.89 \pm 0.15$ & $22.9 \pm 2.7$ \\
A2124 & 236.25 & 36.11 & 0.066 & $0.84 \pm 0.15$ & $1.44 \pm 0.80$ & $\cdots$ \\
A2142 & 239.61 & 27.21 & 0.089 & $1.148 \pm 0.051$ & $3.80 \pm 0.51$ & $89 \pm 11$ \\
A2149 & 240.36 & 54.00 & 0.065 & $0.483 \pm 0.011$ & $0.277 \pm 0.020$ & $13.82 \pm 0.59$ \\
A2169 & 243.43 & 49.07 & 0.058 & $0.655 \pm 0.023$ & $0.685 \pm 0.073$ & $12.3 \pm 1.7$ \\
A2175 & 245.10 & 29.89 & 0.096 & $1.1765 \pm 0.0090$ & $4.110 \pm 0.095$ & $21.9 \pm 2.6$ \\
A2197 & 247.46 & 40.66 & 0.030 & $0.787 \pm 0.014$ & $1.163 \pm 0.061$ & $26 \pm 12$ \\
A2199 & 247.12 & 39.51 & 0.031 & $1.236 \pm 0.024$ & $4.50 \pm 0.26$ & $75.3 \pm 6.3$ \\
A2244 & 255.76 & 33.91 & 0.099 & $1.208 \pm 0.039$ & $4.46 \pm 0.43$ & $\cdots$ \\
A2245 & 255.68 & 33.52 & 0.088 & $1.159 \pm 0.022$ & $3.90 \pm 0.22$ & $9.4 \pm 5.3$ \\
A2249 & 257.44 & 34.45 & 0.085 & $1.095 \pm 0.052$ & $3.28 \pm 0.47$ & $53 \pm 22$ \\
A2255 & 258.10 & 64.02 & 0.080 & $1.337 \pm 0.019$ & $5.94 \pm 0.26$ & $\cdots$ \\
A2399 & 329.34 & -7.82 & 0.058 & $0.968 \pm 0.026$ & $2.21 \pm 0.18$ & $6.5 \pm 1.4$ \\
A2428 & 334.09 & -9.34 & 0.084 & $0.7523 \pm 0.0074$ & $1.063 \pm 0.031$ & $27.7 \pm 1.2$ \\
A2593 & 351.11 & 14.65 & 0.042 & $0.9809 \pm 0.0084$ & $2.271 \pm 0.059$ & $\cdots$ \\
A2670 & 358.57 & -10.44 & 0.076 & $1.1782 \pm 0.0083$ & $4.055 \pm 0.085$ & $28.7 \pm 1.1$ \\
MKW04 & 181.12 & 1.87 & 0.020 & $0.787 \pm 0.020$ & $1.151 \pm 0.086$ & $5.74 \pm 0.93$ \\
MKW08 & 220.16 & 3.47 & 0.027 & $0.584 \pm 0.018$ & $0.473 \pm 0.043$ & $25.1 \pm 5.3$ \\
MKW11 & 202.36 & 11.71 & 0.023 & $0.5737 \pm 0.0004$ & $0.4474 \pm 0.0010$ & $\cdots$ \\
MS1306 & 198.05 & -0.98 & 0.083 & $0.8218 \pm 0.0091$ & $1.385 \pm 0.046$ & $35.15 \pm 0.84$ \\
NGC4325 & 185.75 & 10.57 & 0.025 & $0.42 \pm 0.14$ & $0.18 \pm 0.18$ & $\cdots$ \\
NGC6107 & 244.40 & 35.02 & 0.032 & $0.794 \pm 0.034$ & $1.20 \pm 0.15$ & $31.3 \pm 7.3$ \\
NGC6338 & 258.85 & 57.43 & 0.029 & $0.713 \pm 0.018$ & $0.864 \pm 0.066$ & $28.6 \pm 2.5$ \\
RXCJ1022p3830 & 155.60 & 38.55 & 0.054 & $1.000 \pm 0.013$ & $2.429 \pm 0.093$ & $\cdots$ \\
RXCJ1053p5450 & 163.53 & 54.85 & 0.073 & $1.0469 \pm 0.0024$ & $2.838 \pm 0.020$ & $\cdots$ \\
RXCJ1115p5426 & 172.27 & 54.13 & 0.069 & $0.8903 \pm 0.0042$ & $1.738 \pm 0.025$ & $\cdots$ \\
RXCJ1210p0523 & 184.46 & 3.67 & 0.077 & $1.296 \pm 0.036$ & $5.41 \pm 0.45$ & $20.0 \pm 2.0$ \\
RXCJ1326p0013 & 199.83 & -0.91 & 0.084 & $0.784 \pm 0.012$ & $1.203 \pm 0.056$ & $\cdots$ \\
RXCJ1351p4622 & 207.90 & 46.36 & 0.063 & $0.32 \pm 0.39$ & $0.084 \pm 0.303$ & $13 \pm 11$ \\
RXCJ2214p1350 & 333.67 & 13.85 & 0.026 & $0.486 \pm 0.025$ & $0.273 \pm 0.042$ & $1.95 \pm 0.46$ \\
RXJ0137 & 24.34 & -9.24 & 0.041 & $0.4542 \pm 0.0064$ & $0.2253 \pm 0.0096$ & $6.21 \pm 0.67$ \\
SHK352 & 170.41 & 2.89 & 0.049 & $0.8555 \pm 0.0059$ & $1.517 \pm 0.032$ & $\cdots$ \\
Zw1215p0400 & 184.46 & 3.68 & 0.077 & $1.404 \pm 0.051$ & $6.88 \pm 0.75$ & $76.1 \pm 6.4$ \\
Zw1665 & 125.86 & 4.37 & 0.030 & $0.5636 \pm 0.0074$ & $0.427 \pm 0.017$ & $\cdots$ \\

\hline

\end{tabular}
\end{table*}

\begin{table*}[htbp]
\begin{center}
\caption{ \label{cat info} CIRS and HeCS samples. }
\begin{tabular}{lccrrr}
\hline
\hline
sample & median $z$ & $68$th percentile range & median $M_{200}$ & $68$th percentile range \\
        \quad & & [redshift $z$] & $\left[10^{14}h^{-1}~\text{M}_{\odot }\right]$  & $\left[10^{14}h^{-1}~\text{M}_{\odot }\right]$ \\
 \hline
 & & & & \\
all CIRS & 0.068 & 0.033-0.085 & 1.8 & 0.64 - 4.0 \\
CIRS with individual MAR & 0.064 & 0.030-0.086 & 1.1 & 0.31 - 4.1 \\
\hline
 & & & & \\
all HeCS & 0.16 & 0.12-0.23 & 3.9 & 1.5-7.6 \\
all low-mass HeCS & 0.14 & 0.12-0.18 & 1.9 & 1.0-2.9 \\ 
all high-mass HeCS & 0.20 & 0.14-0.25 & 5.6 & 4.4-12 \\
low-mass HeCS with individual MAR & 0.14 & 0.12-0.22 & 2.2 & 1.2-3.1 \\ 
high-mass HeCS with individual MAR & 0.21 & 0.16-0.24 & 5.2 & 4.3-12 \\
\hline
 \end{tabular}
 \end{center}
 \end{table*}

\subsection{HeCS}\label{hecs}

HeCS is the first systematic and extensive spectroscopic survey of the infall regions of clusters at $z\geqslant 0.1$ \citep{2013ApJ...767...15R}.
HeCS takes advantage of the SDSS and RASS surveys. In particular, existing X-ray cluster catalogues based on RASS were used to define flux-limited cluster samples that were then matched to the imaging footprint of the SDSS DR6 \citep{2008ApJS..175..297A}. Figure \ref{selection} shows the selection function of the HeCS clusters: their X-ray flux limit in the 0.1-2.4 keV band is $5\times 10^{-12}$~erg~s$^{-1}$~cm$^{-2}$ and their redshift range is $[0.1,0.3]$. We include four additional clusters below the flux limit but with fluxes $ > 3\times 10^{-12}$~erg~s$^{-1}$~cm$^{-2}$. 

The imaging footprint of the SDSS DR6 includes 8417 deg$^2$ of imaging data. The multicolour photometry enabled the selection of candidate cluster members using the red-sequence technique. At $z\geqslant 0.1$, the SDSS spectroscopic survey is not dense enough for accurate measurements of the cluster masses with the caustic technique; therefore, the MMT/Hectospec instrument \citep{2005PASP..117.1411F} was used to obtain spectroscopic data for the candidate members. Recently, \citet{sohn2019velocity} updated the HeCS dataset by using the spectroscopic data of SDSS DR14 and incorporated the new catalogues in the HeCS-omnibus survey.

The HeCS survey contains 58 clusters in the redshift range $ 0.1 < z <0.3 $, for a total amount of 22,680 observed galaxy redshifts, of which 10,145 are cluster members. Each cluster survey typically includes $\sim 400-550$ redshifts; in general, roughly half of these galaxies are cluster members and the remaining galaxies are foreground or background objects. 

The orange histogram of Fig \ref{ch-mass} shows the mass distribution of the HeCS cluster sample.
This sample includes fewer low-mass clusters than CIRS because it covers a deeper redshift range $[0.1,0.3]$.
The HeCS sample also contains more high-mass clusters as a result of the larger survey volume.
Due to the extended mass range of the HeCS cluster sample (see Fig. \ref{ch-mass}), we separate the 58 clusters sorted by mass into two subsamples of 29 clusters each. 
The low-mass and high-mass samples have a median mass of  $M_{200} = 1.86 \times 10^{14} h^{-1}~\rm{M_\odot}$ and  $M_{200} = 5.61 \times 10^{14} h^{-1}~\rm{M_\odot}$, respectively. 

Table \ref{HeCS sample} lists the HeCS clusters separated into the two subsamples.\footnote{Similarly to CIRS, the properties listed here are derived from the analyses of the updated catalogue of \citet{sohn2019velocity}; these updated quantities are consistent with the values reported in \citet{2013ApJ...767...15R}.}
Table \ref{cat info} lists the medians and the percentile ranges of the redshift and mass distributions of the entire HeCS catalogue and of its subsamples, including the subsets of clusters for which we estimate the individual MARs in Sect. \ref{Individuals}.

\begin{table*}[htbp] \footnotesize
        \caption{\label{HeCS sample} HeCS clusters.}
\begin{center}
\begin{tabular}{ccccccc}
\hline
\hline
 cluster & RA  & DEC & $z$ & $R_{200}$ & $M_{200}$ & $MAR$\\
  \quad & \scriptsize{[deg]} & \scriptsize{[deg]} &  & \scriptsize{[$h^{-1}$~Mpc]} & $\scriptstyle{\left[10^{14} h^{-1}~\text{M}_{\odot }\right]}$ & $\scriptstyle{\left[10^3 h^{-1}~\text{M}_\odot\text{yr}^{-1} \right]}$ \\
\hline 
\\

A0646 & 126.34 & 47.17 & 0.127 & $1.140 \pm 0.014$ & $3.84 \pm 0.14$ & $43.8 \pm 1.6$ \\
A0655 & 126.38 & 47.14 & 0.127 & $1.049 \pm 0.048$ & $3.00 \pm 0.41$ & $40.8 \pm 7.4$ \\
A0667 & 127.02 & 44.80 & 0.145 & $0.9173 \pm 0.0005$ & $2.0384 \pm 0.0032$ & $22.519 \pm 0.027$ \\
A0689 & 129.36 & 14.97 & 0.279 & $0.740 \pm 0.017$ & $1.220 \pm 0.084$ & $82.0 \pm 4.4$ \\
A0750 & 137.23 & 11.02 & 0.164 & $0.967 \pm 0.051$ & $2.43 \pm 0.39$ & $53 \pm 14$ \\
A0990 & 155.90 & 49.16 & 0.141 & $0.965 \pm 0.070$ & $2.36 \pm 0.51$ & $\cdots$ \\
A1033 & 157.91 & 35.06 & 0.122 & $0.9330 \pm 0.0061$ & $2.100 \pm 0.041$ & $\cdots$ \\
A1068 & 160.19 & 39.93 & 0.138 & $0.517 \pm 0.044$ & $0.362 \pm 0.092$ & $\cdots$ \\
A1132 & 164.60 & 56.79 & 0.135 & $1.0028 \pm 0.0030$ & $2.639 \pm 0.024$ & $102.6 \pm 1.1$ \\
A1201 & 168.22 & 13.42 & 0.167 & $1.01 \pm 0.14$ & $2.79 \pm 1.16$ & $53 \pm 21$ \\
A1204 & 168.33 & 17.59 & 0.170 & $0.641 \pm 0.078$ & $0.71 \pm 0.26$ & $\cdots$ \\
A1235 & 170.84 & 19.59 & 0.103 & $0.789 \pm 0.019$ & $1.246 \pm 0.091$ & $31.7 \pm 2.5$ \\
A1302 & 176.20 & 67.43 & 0.116 & $1.033 \pm 0.021$ & $2.83 \pm 0.17$ & $77 \pm 10$ \\
A1361 & 175.91 & 46.35 & 0.116 & $0.8431 \pm 0.0070$ & $1.540 \pm 0.038$ & $\cdots$ \\
A1366 & 176.22 & 67.40 & 0.116 & $0.855 \pm 0.031$ & $1.61 \pm 0.18$ & $118 \pm 25$ \\
A1423 & 179.28 & 33.62 & 0.214 & $1.085 \pm 0.020$ & $3.61 \pm 0.20$ & $80.9 \pm 4.8$ \\
A1902 & 215.44 & 37.29 & 0.163 & $1.088 \pm 0.036$ & $3.46 \pm 0.35$ & $\cdots$ \\
A1918 & 216.26 & 63.16 & 0.140 & $0.481 \pm 0.027$ & $0.293 \pm 0.049$ & $\cdots$ \\
A1930 & 218.18 & 31.58 & 0.131 & $0.8373 \pm 0.0094$ & $1.529 \pm 0.051$ & $22.0 \pm 1.4$ \\
A1978 & 222.79 & 14.63 & 0.146 & $0.828 \pm 0.014$ & $1.502 \pm 0.075$ & $\cdots$ \\
A2055 & 229.70 & 6.24 & 0.103 & $0.852 \pm 0.034$ & $1.57 \pm 0.19$ & $29.4 \pm 5.6$ \\
A2187 & 246.05 & 41.23 & 0.183 & $0.724 \pm 0.030$ & $1.04 \pm 0.13$ & $67.5 \pm 9.4$ \\
A2259 & 260.04 & 26.63 & 0.160 & $0.88 \pm 0.13$ & $1.84 \pm 0.83$ & $\cdots$ \\
A2261 & 260.60 & 32.03 & 0.225 & $0.987 \pm 0.049$ & $2.74 \pm 0.41$ & $162 \pm 28$ \\
RXJ1720 & 260.04 & 26.62 & 0.160 & $0.8853 \pm 0.0049$ & $1.858 \pm 0.031$ & $35.5 \pm 1.1$ \\
RXJ2129 & 322.42 & 0.08 & 0.234 & $1.062 \pm 0.013$ & $3.45 \pm 0.13$ & $33.2 \pm 1.5$ \\
Zw1478 & 119.98 & 54.00 & 0.103 & $0.674 \pm 0.029$ & $0.78 \pm 0.10$ & $\cdots$ \\
Zw3179 & 156.48 & 12.70 & 0.142 & $0.73 \pm 0.13$ & $1.04 \pm 0.55$ & $26.1 \pm 5.2$ \\
Zw8197 & 259.34 & 56.66 & 0.113 & $0.984 \pm 0.053$ & $2.45 \pm 0.39$ & $\cdots$ \\

\\
\hline
\\

A0267 & 28.18 & 0.98 & 0.229 & $1.2290 \pm 0.0028$ & $5.320 \pm 0.036$ & $\cdots$ \\
A0697 & 130.76 & 36.37 & 0.282 & $1.62 \pm 0.10$ & $12.7 \pm 2.4$ & $201 \pm 69$ \\
A0773 & 139.50 & 51.75 & 0.218 & $1.6849 \pm 0.0056$ & $13.54 \pm 0.14$ & $250.1 \pm 4.4$ \\
A0795 & 141.01 & 14.15 & 0.137 & $1.140 \pm 0.013$ & $3.89 \pm 0.13$ & $41.8 \pm 1.7$ \\
A0963 & 154.31 & 39.03 & 0.204 & $1.183 \pm 0.070$ & $4.62 \pm 0.82$ & $230 \pm 47$ \\
A0980 & 155.61 & 50.12 & 0.156 & $1.195 \pm 0.091$ & $4.6 \pm 1.0$ & $36.5 \pm 8.1$ \\
A1246 & 171.06 & 21.47 & 0.192 & $1.162 \pm 0.062$ & $4.34 \pm 0.70$ & $127 \pm 49$ \\
A1413 & 178.83 & 23.41 & 0.141 & $1.201 \pm 0.095$ & $4.6 \pm 1.1$ & $71 \pm 44$ \\
A1437 & 178.83 & 3.33 & 0.133 & $1.607 \pm 0.065$ & $10.8 \pm 1.3$ & $\cdots$ \\
A1553 & 187.67 & 10.56 & 0.167 & $1.2094 \pm 0.0020 $ & $4.768 \pm 0.024$ & $\cdots$ \\
A1682 & 196.72 & 46.53 & 0.227 & $1.3886 \pm 0.0006$ & $7.654 \pm 0.009$ & $\cdots$ \\
A1689 & 197.90 & -1.32 & 0.184 & $1.871 \pm 0.013$ & $17.95 \pm 0.36$ & $\cdots$ \\
A1758 & 203.15 & 50.56 & 0.276 & $1.512 \pm 0.036$ & $10.38 \pm 0.75$ & $109.8 \pm 6.9$ \\
A1763 & 203.83 & 41.00 & 0.231 & $1.55 \pm 0.14$ & $10.7 \pm 2.9$ & $87 \pm 30$ \\
A1835 & 210.27 & 2.87 & 0.252 & $1.758 \pm 0.012$ & $15.94 \pm 0.33$ & $ \cdots$ \\
A1914 & 216.51 & 37.84 & 0.166 & $1.257 \pm 0.023$ & $5.35 \pm 0.30$ & $265 \pm 46$ \\
A2009 & 225.09 & 21.37 & 0.152 & $1.272 \pm 0.040$ & $5.47 \pm 0.52$ & $\cdots$ \\
A2034 & 227.52 & 33.47 & 0.113 & $1.3628 \pm 0.0037$ & $6.487 \pm 0.052$ & $\cdots$ \\
A2050 & 229.07 & 0.07 & 0.119 & $1.3010 \pm 0.0033$ & $5.673 \pm 0.043$ & $\cdots$ \\
A2069 & 231.04 & 29.87 & 0.114 & $1.32 \pm 0.16$ & $5.9 \pm 2.1$ & $\cdots$ \\
A2111 & 234.93 & 34.41 & 0.230 & $1.3597 \pm 0.0050$ & $7.207 \pm 0.079$ & $171.6 \pm 3.7$ \\
A2219 & 250.06 & 46.71 & 0.226 & $1.776 \pm 0.056$ & $16.0 \pm 1.5$ & $244 \pm 67$ \\
A2396 & 328.91 & 12.50 & 0.192 & $1.267 \pm 0.018$ & $5.61 \pm 0.24$ & $82.9 \pm 5.5$ \\
A2631 & 354.40 & 0.25 & 0.277 & $1.494 \pm 0.011$ & $10.02 \pm 0.22$ & $\cdots$ \\
A2645 & 335.32 & -9.03 & 0.251 & $1.101 \pm 0.039$ & $3.91 \pm 0.41$ & $66 \pm 21$ \\
MS0906 & 137.28 & 10.94 & 0.177 & $1.2247 \pm 0.0034$ & $4.998 \pm 0.041$ & $48.59 \pm 0.50$ \\
RXJ1504 & 226.02 & -2.81 & 0.216 & $1.15 \pm 0.12$ & $4.3 \pm 1.4$ & $63 \pm 20$ \\
Zw2701 & 148.25 & 51.85 & 0.215 & $1.116 \pm 0.034$ & $3.92 \pm 0.36$ & $\cdots$ \\
Zw3146 & 155.91 & 4.19 & 0.290 & $1.200 \pm 0.076$ & $5.26 \pm 0.99$ & $\cdots$ \\

\hline

\end{tabular}
\end{center}
\footnotesize{{\bf Notes.} The upper (lower) part of the table lists the low-mass (high-mass) subsample of the HeCS clusters.}
\end{table*}

\begin{figure*}
\centering
\includegraphics[scale=0.73]{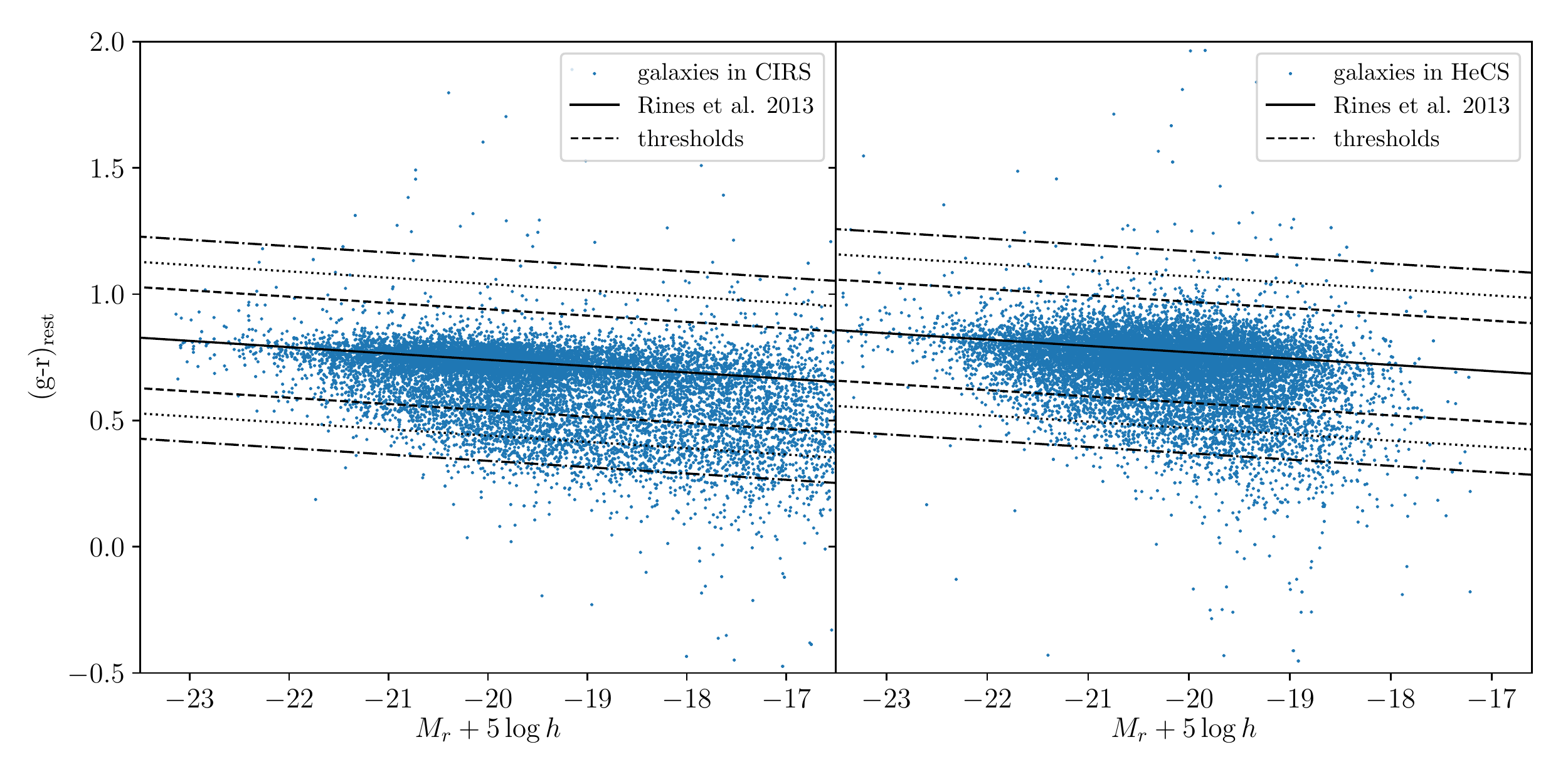}
\caption{$(g-r)$ colour-magnitude diagrams of the CIRS ({\it left}) and HeCS ({\it right}) galaxies, including $k$-corrections. The galaxies have a cluster-centric distance smaller than 3$R_{200}$ and line-of-sight velocity $|v_\mathrm{los}-v_\mathrm{cl}|<3000$~km~s$^{-1}$. The black solid curve in the right panel shows the fit by \citet{2013ApJ...767...15R} derived from the HeCS galaxies. The black dashed, dotted, and dash-dotted lines show the $\pm 0.30,\pm 0.40,\pm 0.50$ shifts of the same fit. In the left panel, the black solid line shows the Rines et al.'s fit with an offset of -0.03 mag. The black dashed, dotted, and dash-dotted lines show the $\pm 0.30,\pm 0.40,\pm 0.50$ mag shifts of this line.}\label{cmd}
\end{figure*}

\subsection{Effects of the selection of the galaxy samples}
\label{sec:selection}

\subsubsection{Photometric completeness}

The galaxies in the HeCS clusters are selected according to their red colours, whereas the galaxies in CIRS are from a magnitude-limited survey. Therefore, in principle, unlike CIRS,  a substantial number of blue galaxies could be missing in HeCS.

To quantify the impact of these different selections, Fig. \ref{cmd} shows the colour-magnitude diagram of the two catalogues. We only show the galaxies at projected distance smaller than 3$R_{200}$ from the cluster centre and with a line-of-sight velocity $|v_\mathrm{los}-v_\mathrm{cl}|<3000$~km~s$^{-1}$, where $v_\mathrm{cl}$ is the line-of-sight velocity of the cluster. In the right panel, the black solid line shows the fit of \citet{2013ApJ...767...15R} to the red sequence for all the member galaxies in HeCS; in the left panel, the solid line shows the red sequence fit of \citet{2013ApJ...767...15R} shifted by -0.03 mag. 
The black dashed, dotted, and dash-dotted lines show the boundaries of the stripes used for sample selection obtained by changing the intercept of the fit by $\pm 0.30, \pm 0.40, \pm 0.50$ mag. 

Some of the galaxies outside the selection stripes are extremely red. However, they are a tiny minority and they are almost never cluster members \citep{2013ApJ...767...15R}. The impact of red objects can thus be ignored here. The spatial distribution of the red galaxies peaks within $R_{200}$; the blue galaxy distribution peaks at significantly larger radius.
The distributions of the line-of-sight velocities of both red and blue galaxies are centred on zero and  are approximately Gaussian; the width of these distributions is smaller for the red galaxies than for the blue galaxies, which is in agreement with previous observations of galaxies in clusters \citep[e.g.][]{Dickens1976A1367,Carlberg1997,Geller2014A383,Barsanti2016} and with the expectations from simulations of galaxy formation \citep[e.g.][]{Diaferio2001}.

\begin{table}[htbp] \tiny
\begin{center}
\caption{\label{bluegal} \footnotesize Relative difference of blue galaxies in CIRS and HeCS.}
\centering
\begin{tabular}{cccc}
\hline
\hline
stripe & $\Xi_\mathrm{CIRS}$ & $\Xi_\mathrm{HeCS}$ & $\delta_\Xi$  \\
 \hline
& & & \\
$\pm 0.30$ mag & 0.1072 & 0.0764 & 0.287 \\
$\pm 0.40$ mag & 0.0387 & 0.0319 & 0.176 \\
$\pm 0.50$ mag & 0.0150 & 0.0147 & 0.020 \\
\hline
\end{tabular}
\end{center}
\end{table}

Table \ref{bluegal} lists the ratio $\Xi$ between the number of the blue galaxies outside each of the stripes shown in Fig. \ref{cmd} and the total number of galaxies in CIRS and HeCS.
The fraction of blue galaxies is $\lesssim 10\%$ in both catalogues, for any stripe. The relative difference in blue galaxies between CIRS and HeCS, $\delta_\Xi=1-\Xi_\mathrm{HeCS}/\Xi_\mathrm{CIRS}$, decreases from 29\% to 2\% from the $\pm 0.30$ mag-stripe to the $\pm 0.50$ mag-stripe. 

These numbers suggest that HeCS might roughly miss, at most, one third of blue galaxies compared to CIRS. The effect on the caustic location should thus be mild. In addition, the caustic technique locates the caustics by adopting a threshold of the number density distribution of the galaxies in the ${R-v_{\rm los}}$ diagram that is set by the galaxies within $\sim R_{200}$. The spatial distribution of the red galaxies peaks within this radius, whereas the blue galaxy distribution peaks further out, as mentioned above. 

We prove that missing this fraction of blue galaxies in HeCS does not affect our estimates of the MAR by taking ten CIRS clusters and randomly removing 35\% of the galaxies outside the $\pm 0.30$ mag-stripe, a slightly larger fraction of the upper limit of $\delta_\Xi$ listed in Table \ref{bluegal}. The differences between these caustic mass profiles and the original profiles are within the uncertainties of the caustic technique and they are thus statistically indistinguishable. 

Our conclusion is further supported by \citet{2013ApJ...767...15R}. They present a case study of three HeCS clusters: with additional spectroscopic observations including blue galaxies, they can  quantify how these galaxies can affect the estimate of the velocity dispersion and  the dynamical mass. They find that a small fraction of blue galaxies are actual cluster members and their inclusion only increases the velocity dispersion by 0.3\%.  
\citet{2013ApJ...767...15R} conclude that targeting the galaxies on the red-sequence alone does not produce any bias on the velocity dispersion and the mass estimates.

\begin{figure*}
\centering
\includegraphics[scale=0.6]{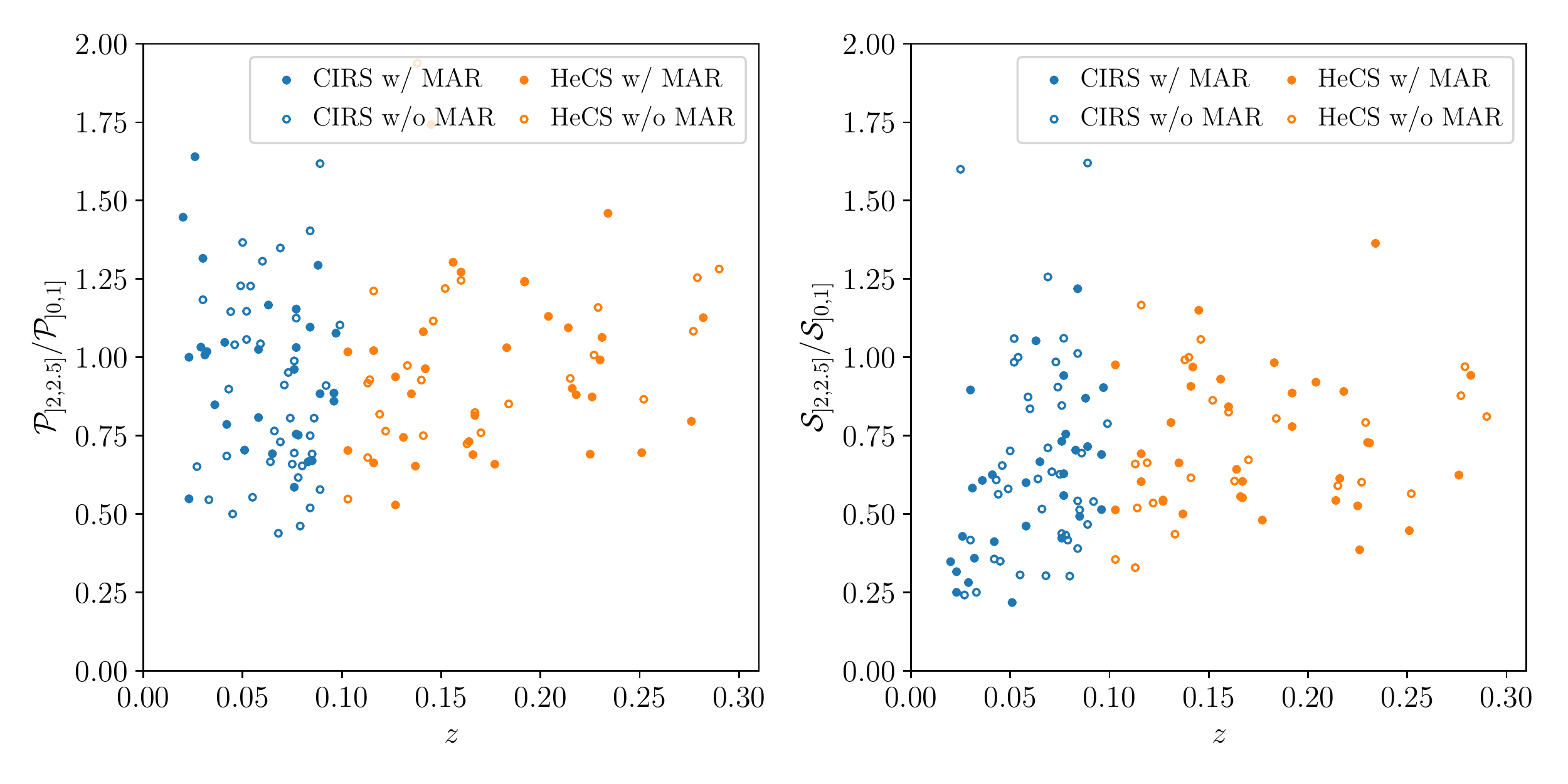} 
\caption{{\it Left panel}: Ratio of the numbers of galaxies with photometric data in the range $r/R_{200}\in ]2,2.5]$ and within $r/R_{200}=1$ within each cluster, against the cluster redshift. {\it Right panel}: Ratio between the numbers of galaxies with spectroscopic redshifts within the same regions. Blue and orange points show the CIRS and HeCS clusters, respectively. Solid points show the clusters for which we estimate the individual MAR.}\label{photoSpectro}
\end{figure*}

\begin{figure}
\centering
\includegraphics[scale=0.6]{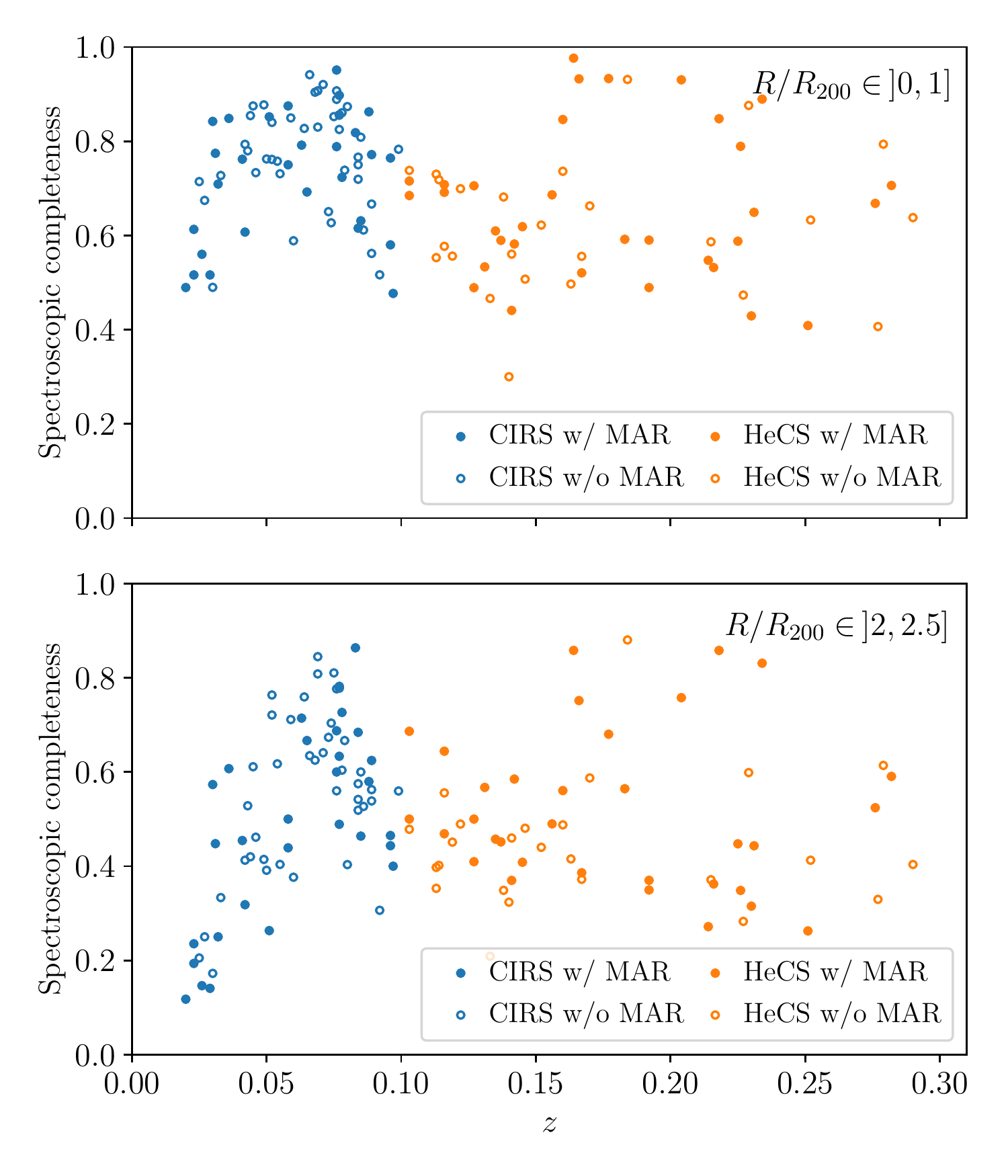} 
\caption{Spectroscopic completeness of the CIRS (blue points) and HeCS (orange points) clusters, against their redshifts, within $r/R_{200}=1$ from the cluster centre ({\it upper panel}) and within the range $r/R_{200}\in ]2,2.5]$ ({\it lower panel}). Solid points show the clusters for which we estimate the individual MAR.}\label{spect_compl}
\end{figure}

\subsubsection{Spectroscopic completeness}

In general, the spectroscopic completeness, namely, the ratio between the number of galaxies with spectroscopic measurements and the number of galaxies with photometric measurements, can decrease with increasing distance from the cluster centre.
In CIRS, the spectroscopic completeness could be affected by the edges of the footprint of SDSS, whereas in HeCS, spectroscopic redshifts are only available within 30 arcmin from the cluster centre. Hence, the spectroscopic measurements can be incomplete at large radii. For the caustic method, this incompleteness can cause an underestimate of the caustic amplitude and thus an underestimation of the cluster mass and of the MAR. However, when the sample is too sparse, the caustic method refrains from locating the caustics and returns no mass estimates. In contrast, if the incompleteness is not too severe to prevent the location of the caustics, the underestimate of the mass might be within the caustic mass uncertainty. Towards the end of this section, we confirm that this is indeed the case for the CIRS and HeCS clusters. 

We are first concerned with the fact that the spectroscopic incompleteness could  affect the two catalogues differently and, thus,   bias our MAR estimates in different ways. We show that this case does not occur and that the spectroscopic incompleteness is comparable in the two CIRS and HeCS samples.

The left panel of Fig. \ref{photoSpectro} shows the ratio $\mathcal{P}=\mathcal{P}_{]2,2.5]}/\mathcal{P}_{]0,1]}$ between the numbers of galaxies with photometric measurements in the range of $r/R_{200}\in\, ]2,2.5]$ and in the range $r/R_{200}\in\, ]0,1]$. The right panel shows the ratio $\mathcal{S}=\mathcal{S}_{]2,2.5]}/\mathcal{S}_{]0,1]}$ of the numbers of galaxies with spectroscopic redshift in the same two regions. Blue and orange points show the CIRS and HeCS clusters, respectively. The ratios, $\mathcal{P}$ and $\mathcal{S,}$ of each cluster are computed with the galaxies brighter than $M_r=-20$. This limit is $\sim M^\ast_r+1$, where $M^\ast_r\approx -21$ is the characteristic red-band magnitude of the Schechter luminosity function, as indicated by photometric studies of A2029 and Coma \citep{sohn2017velocity}. The left panel shows that, on average, the photometric samplings of the CIRS and HeCS surveys in the central and outer regions of the clusters are comparable: $\Braket{\mathcal{P}}=0.99 \pm 0.61$ and $\Braket{\mathcal{P}}=0.97 \pm 0.27$ for CIRS and HeCS, respectively. A similar result holds for the spectroscopic samplings: $\Braket{\mathcal{S}}= 0.65 \pm 0.30$ and $\Braket{\mathcal{S}}= 0.73 \pm 0.22$ for CIRS and  HeCS, respectively. 
 
The values of the ratio $\mathcal{S}<1$ suggest that the spectroscopic completeness might decrease with distance from the cluster centre. In fact, Fig. \ref{spect_compl} shows the spectroscopic completenesses $\mathcal{S}_{]0,1]}/\mathcal{P}_{]0,1]}$ and $\mathcal{S}_{]2,2.5]}/\mathcal{P}_{]2,2.5]}$ in the central and outer regions of the clusters. It also confirms a spectroscopic incompleteness at large radii.  
Nevertheless, the spectroscopic completenesses in the central and outer regions are comparable in the two catalogues. In the centre, CIRS and HeCS have mean spectroscopic completeness $0.75\pm 0.12$ and  $0.65\pm 0.15$, respectively. In the outer region, the mean completenesses are $0.53\pm 0.19$ and $0.49\pm 0.15$. We obtain similar results by considering the subsample of clusters for which, as we illustrate in the next section, we estimate the MAR individually. The similarity of the ratios $\mathcal{P}$ and $\mathcal{S}$ of CIRS and HeCS in Fig. \ref{photoSpectro} shows that the comparable completenesses are not a fluke originating from different photometric samplings in the central and outer regions of the two catalogues.

Based on these results, we conclude that the spectroscopic incompleteness as a function of radius is present in our samples, but it might similarly bias our MAR estimates of the CIRS and HeCS clusters. This conclusion is further supported by the fact that the fraction of clusters with null amplitude beyond $2R_{200}$ is $\sim 50$\% in both catalogues, as we show in the next section. In turn, these fractions are larger than the fractions $\sim 13-18$\% found for the simulated clusters (Sect. \ref{MarSimulated}). 

We now quantify the possible systematic error resulting from the spectroscopic incompleteness as a function of radius. 
We create mock catalogues by stacking the simulated clusters at $z=0.12$, the intermediate redshift between the average redshifts of the CIRS and HeCS clusters. We stack all the 141 clusters of the high-mass bin, whereas we stack a random sample of only 35 clusters of the low-mass bins, to limit the computational time. The ${R-v_{\rm los}}$ diagrams of these two stacked clusters contain $\sim$50,000 particles, and simulate catalogues that are spectroscopically complete. 
We also create mock catalogues by randomly removing some particles within $r/R_{200}=1$ and within the range  of $r/R_{200}\in\, ]2,2.5]$ to simulate the spectroscopic incompleteness as a function of radius. We consider two undersampled catalogues: in the first catalogue, we remove $25\%$ and $47\%$ of the particles in the central and outer regions, respectively; in the second catalogue, we remove $35\%$ and $51\%$ of the particles in the two regions. These fractions are chosen accordingly to the mean spectroscopic incompleteness estimated above for the CIRS and HeCS clusters.

The undersampled mock catalogues return mass and MAR estimates consistent with the estimates obtained from the complete catalogues: with the undersampled catalogues, the mass $M_{200}$ is underestimated by $\sim 13\%$, whereas the MARs is overestimated by $\sim 22\%$, on average. Both values are within the uncertainties of the estimates obtained from the complete catalogues and are thus statistically indistinguishable.
The MAR is, on average, overestimated rather than underestimated, as one might have naively expected; this overestimation confirms that the spectroscopic incompleteness as a function of radius in the CIRS and HeCS clusters can generate statistical fluctuations on the MAR estimates, rather than a systematic error.    

\section{Measure of the mass accretion rate of real clusters}
\label{mar}

Here, we discuss our estimates of the MAR of individual clusters (Sect. \ref{Individuals}) and the average MAR of the cluster samples 
(Sect. \ref{Results}) as a function of mass and redshift.  

\subsection{Individual MARs} \label{Individuals}

As anticipated in Sects. \ref{cirs} and \ref{hecs}, we only estimate the individual MAR of a subset of clusters, selected by visually inspecting their ${R-v_{\rm los}}$ diagrams. In fact, we remove (i) the clusters whose caustic amplitude shrinks to zero within the infalling mass shell and (ii) the clusters whose caustics have unphysical spikes. These cases usually occur in the presence of galaxy-rich background or foreground structures that prevent the caustic technique algorithm from properly identifying the caustic location. This procedure is similar to the automatic procedure adopted for the mock catalogues in Sect. \ref{MarSimulated}. 
Visual inspection selects 30 clusters (out of 71) for the CIRS sample. We selected 18 out of 29 and 16 out of 29 clusters, respectively, for the low- and high-mass HeCS samples.

The initial infall velocity $v_i$ entering Eq.~(\ref{thickness}) depends on the cluster mass and redshift. Rather than consider the mass and
redshift of each cluster, we consider the same $v_i$ for all the clusters within each sample. We thus estimate the value of $v_i$ appropriate for the
median redshift and the median mass of each cluster sample. 

The median mass and redshift of the CIRS subsample are $M_{200}=1.1\times 10^{14}h^{-1}$~M$_\odot$ and $z=0.064$ (Table \ref{cat info}). The closest redshifts of the simulated snapshot are $z=0$ and $z=0.12$. We thus estimate $v_i$ appropriate for the CIRS median mass with three linear interpolations on the simulation information. For each of the two simulated samples of mass $10^{14}$ and $10^{15} h^{-1}$~M$_\odot$, we first interpolate between the two median masses at redshifts $z=0$ and $z=0.12$ listed in Table \ref{binStat} to estimate the appropriate median masses at $z=0.064$, $M_{0.064}^{\rm low}$ and $M_{0.064}^{\rm high}$.  
The second interpolation returns the  velocities appropriate for $z=0.064$, $v_{0.064}^{\rm low}$ and $v_{0.064}^{\rm high}$, for each of the two simulated samples: for each sample, we consider the radial velocity profiles at the two redshifts $z=0$ and $z=0.12$ and consider the value of the velocity at the single radius lying in the range $[2-2.5]R_{200}$; $v_{0.064}^{\rm low}$ and $v_{0.064}^{\rm high}$ are derived from the interpolation between the two values at redshifts $z=0$ and $z=0.12$ for
each sample. Finally, to obtain $v_i$ appropriate for the median mass $M_{200}=1.1\times 10^{14}h^{-1}$~M$_\odot$ at $z=0.064$, we interpolate between the two median masses, $M_{0.064}^{\rm low}$ and $M_{0.064}^{\rm high}$, and the two velocities, $v_{0.064}^{\rm low}$ and $v_{0.064}^{\rm high}$. We find  $v_i=-170\pm 3$~km~s$^{-1}$. 

For the sake of completeness, the value of $v_i$ above also includes  the uncertainty derived with an analogous interpolation based on the profile of the standard deviation of the mean radial velocity profile; the standard deviation is comparable to the $68$th percentile range shown by the shaded bands in Fig. \ref{fig:radvel}.
As discussed in Sect. \ref{radialvel}, the spread on $v_i$ is roughly a factor three smaller than the uncertainty on the MAR derived from the caustic technique and does not generate any bias on the MAR itself. Therefore, we ignore the uncertainty on $v_i$ 
when computing the shell thickness (Eq.~\ref{thickness}).

For the two HeCS subsamples, we adopt the same procedure. The median mass and redshift of the low-mass HeCS subsample are $M_{200}=2.2\times 10^{14}h^{-1}$~M$_\odot$ and $z=0.14$ (Table \ref{cat info}), and we use the snapshots of the simulation at redshifts $z=0.12$ and $z=0.19$. Finally, we find that $v_i=-288\pm 8$~km~s$^{-1}$. 

For the high-mass HeCS subsample, the median mass is  $M_{200}=5.2\times 10^{14}h^{-1}$~M$_\odot$ and the median redshift is $z=0.21$ (Table \ref{cat info}). The closest redshifts of the simulated snapshots are $z=0.19$ and $z=0.26$. We thus find $v_i=-566\pm 19$~km~s$^{-1}$.

The uncertainty in the MAR only depends on the uncertainty, $ \sigma_{M_{\mathrm{shell}}}$,
on the mass $M_{\mathrm {shell}}$ of the infalling shell because we adopt a value for $t_{\mathrm {inf}}$ without an uncertainty. The caustic method estimates the mass profile from the  caustic amplitudes,  $\mathcal{A}_i$, measured at a set of radii, $r_i$. 
The mass, $M_i$, of the $i$-th shell is proportional to $\mathcal{A}_i^2$. According to \citet{1999MNRAS.309..610D} and \citet{2011MNRAS.412..800S},
we thus estimate the uncertainty  on the mass of the infalling shell as:
\begin{equation}\label{mass-uncert}
\sigma_{M_{\mathrm{shell}}} = \sum 2\, M_i \frac{\sigma_{\mathcal{A}_i}}{\mathcal{A}_i},
\end{equation} 
where the sum extends over the shells of the caustic mass profile within the infalling shell, and $\sigma_{\mathcal{A}_i}$ is the
uncertainty on the caustic amplitude of the $i$-th shell; this uncertainty increases with decreasing ratio between the number
of galaxies within the caustics and the number of galaxies within the ${R-v_{\rm los}}$ diagram at each radius $r_i$.
Below, we find that the relative uncertainties of the individual MARs are 17\% on average; for only 7 out of 64 clusters we measure a MAR with a relative uncertainty larger than 35\%. The uncertainties mostly come from 
(1) the assumption of spherical symmetry and (2) the presence of very populated ${R-v_{\rm los}}$ diagrams that make it challenging to locate the caustics. 

\begin{figure}[htbp]
  \centering 
  \includegraphics[scale=0.57]{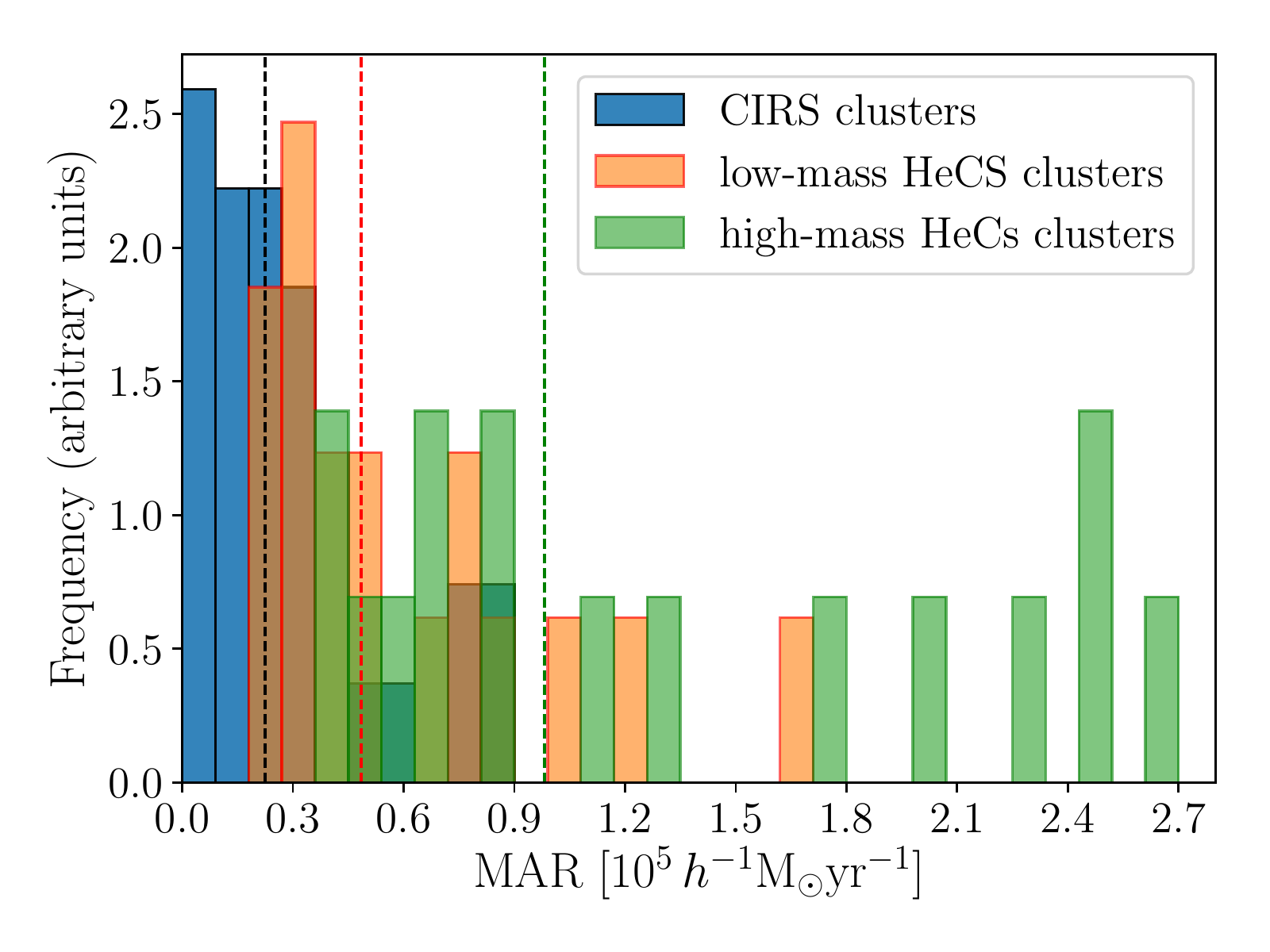}
        \caption{Distributions of the individual MARs. The blue, orange and green histograms refer to the CIRS, low-mass HeCS, and high-mass HeCS clusters, respectively. The dashed lines with the same colours show the median MAR of each sample. The area under each histogram is normalised to unity.}
\label{histograms}
\end{figure}

\begin{figure*}[htbp]
  \centering
  \includegraphics[scale=0.55]{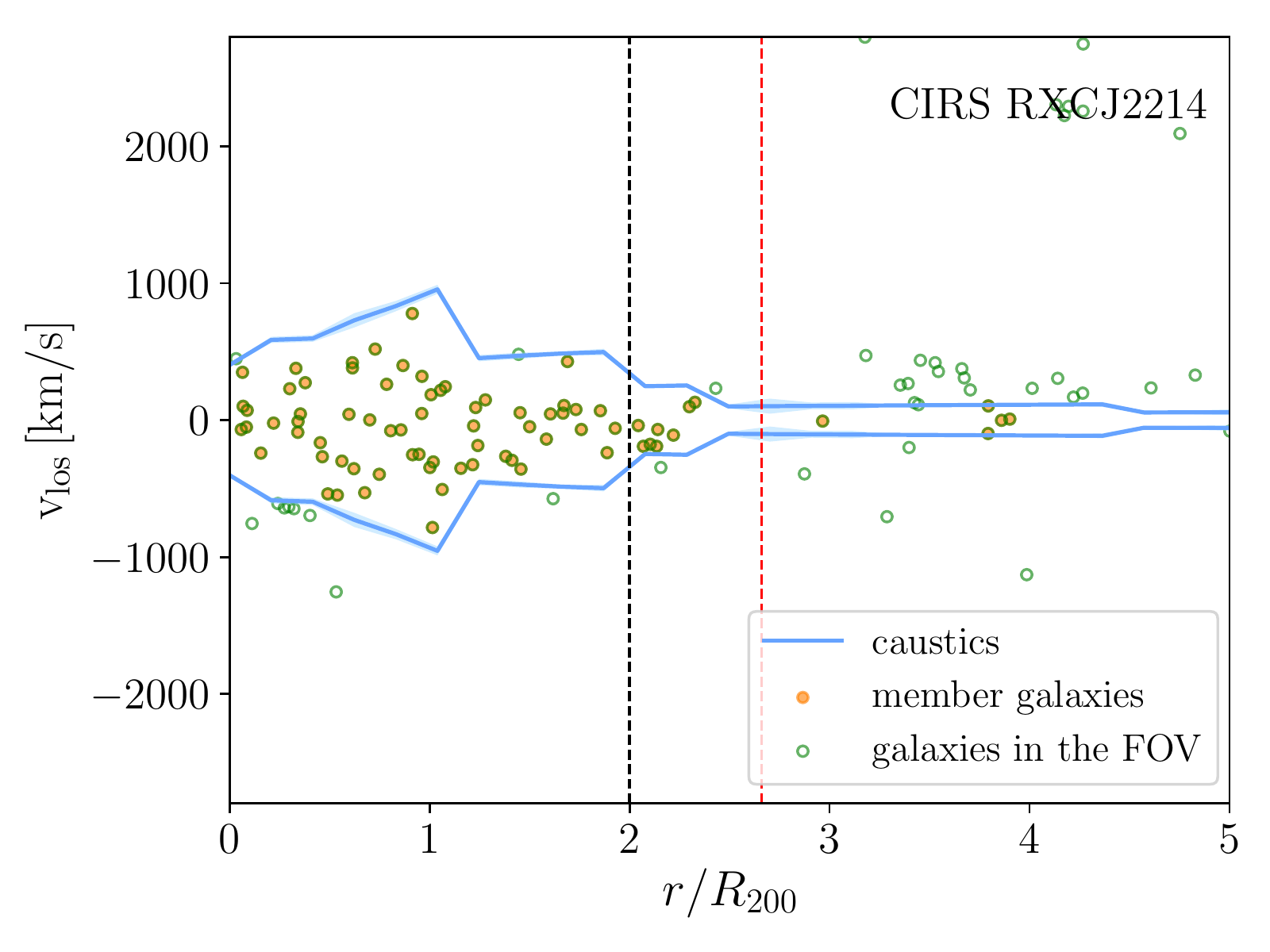}  %
  \includegraphics[scale=0.55]{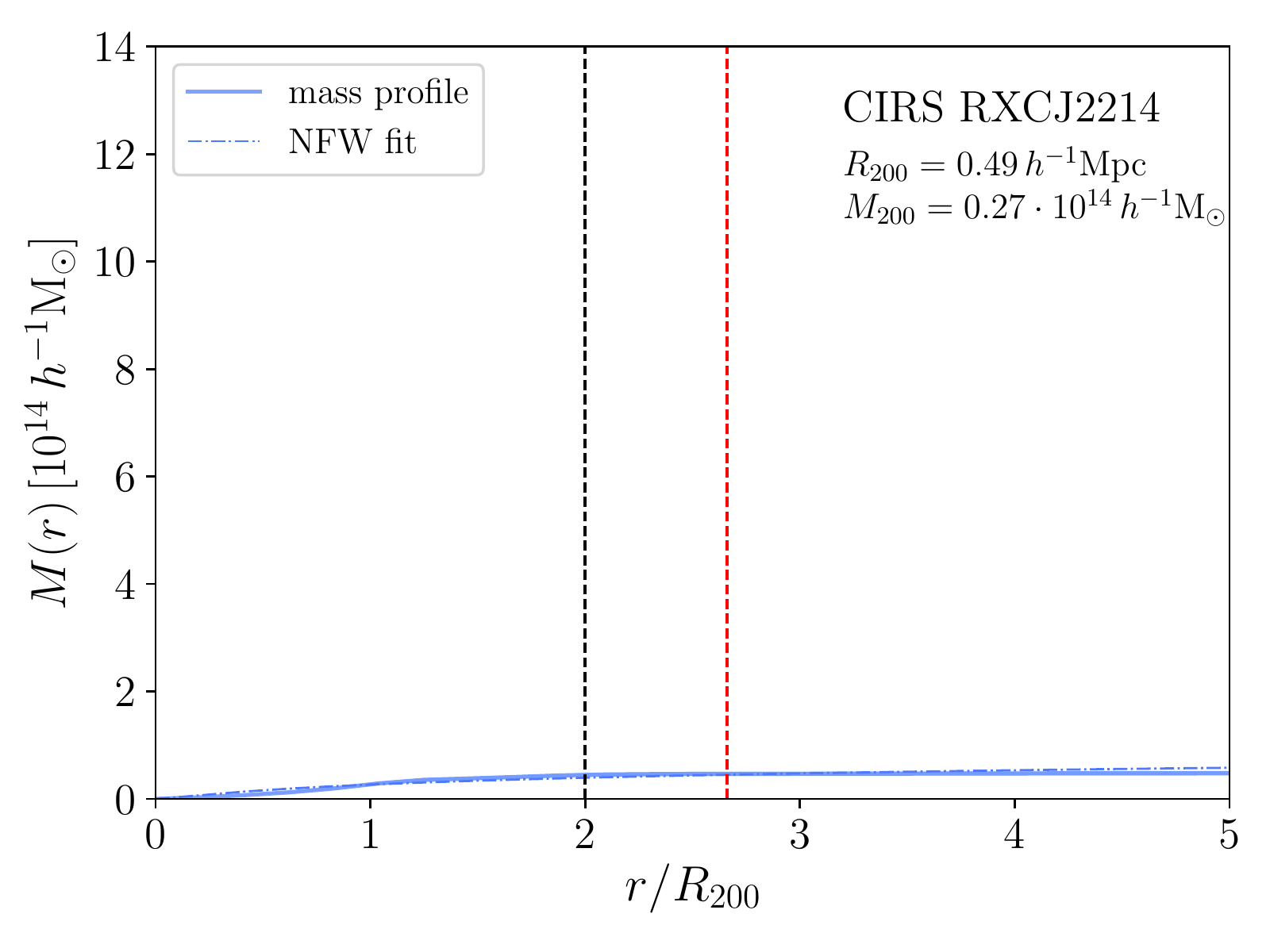} \\
  \includegraphics[scale=0.55]{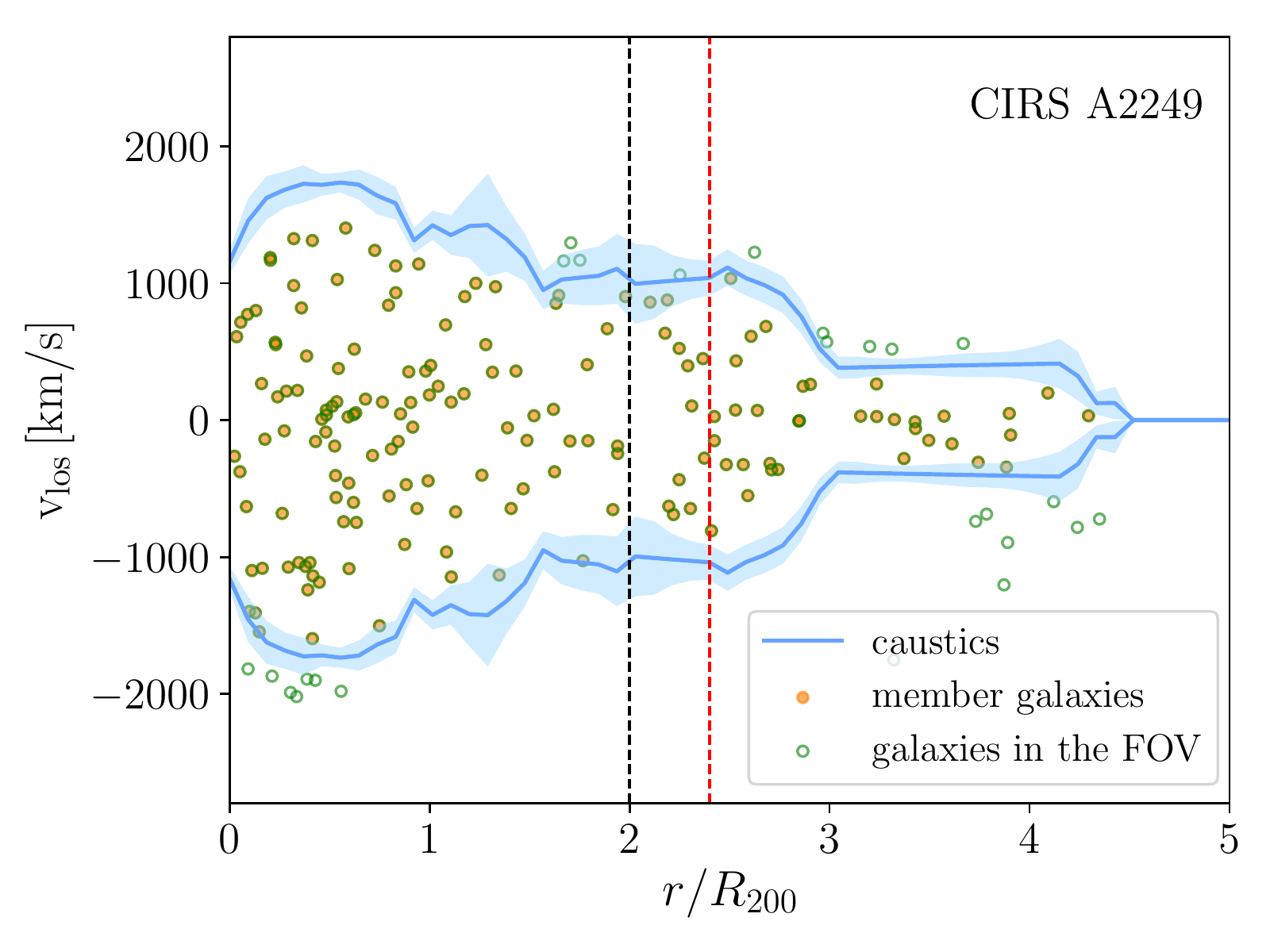}  %
  \includegraphics[scale=0.55]{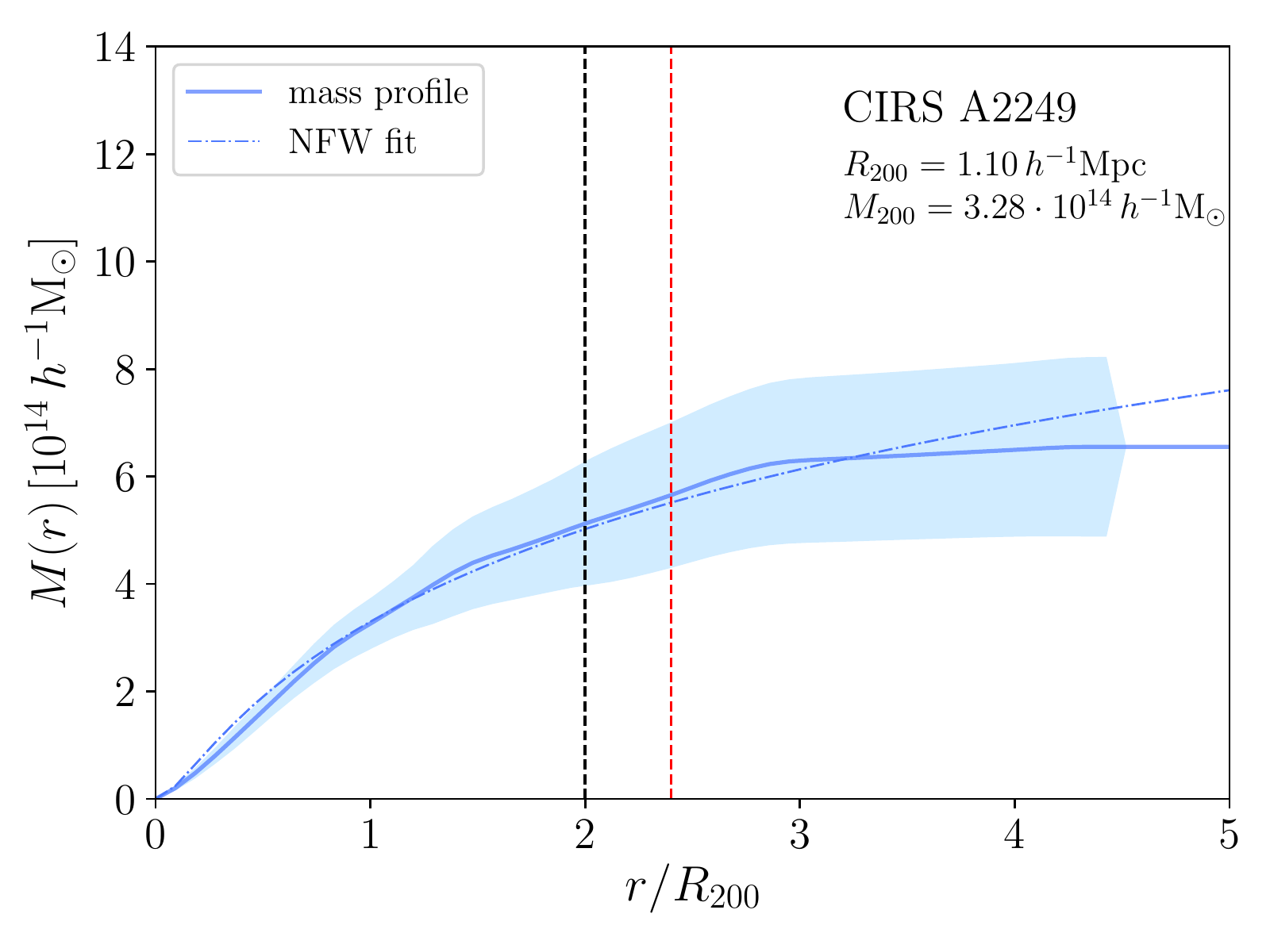} \\
  \includegraphics[scale=0.55]{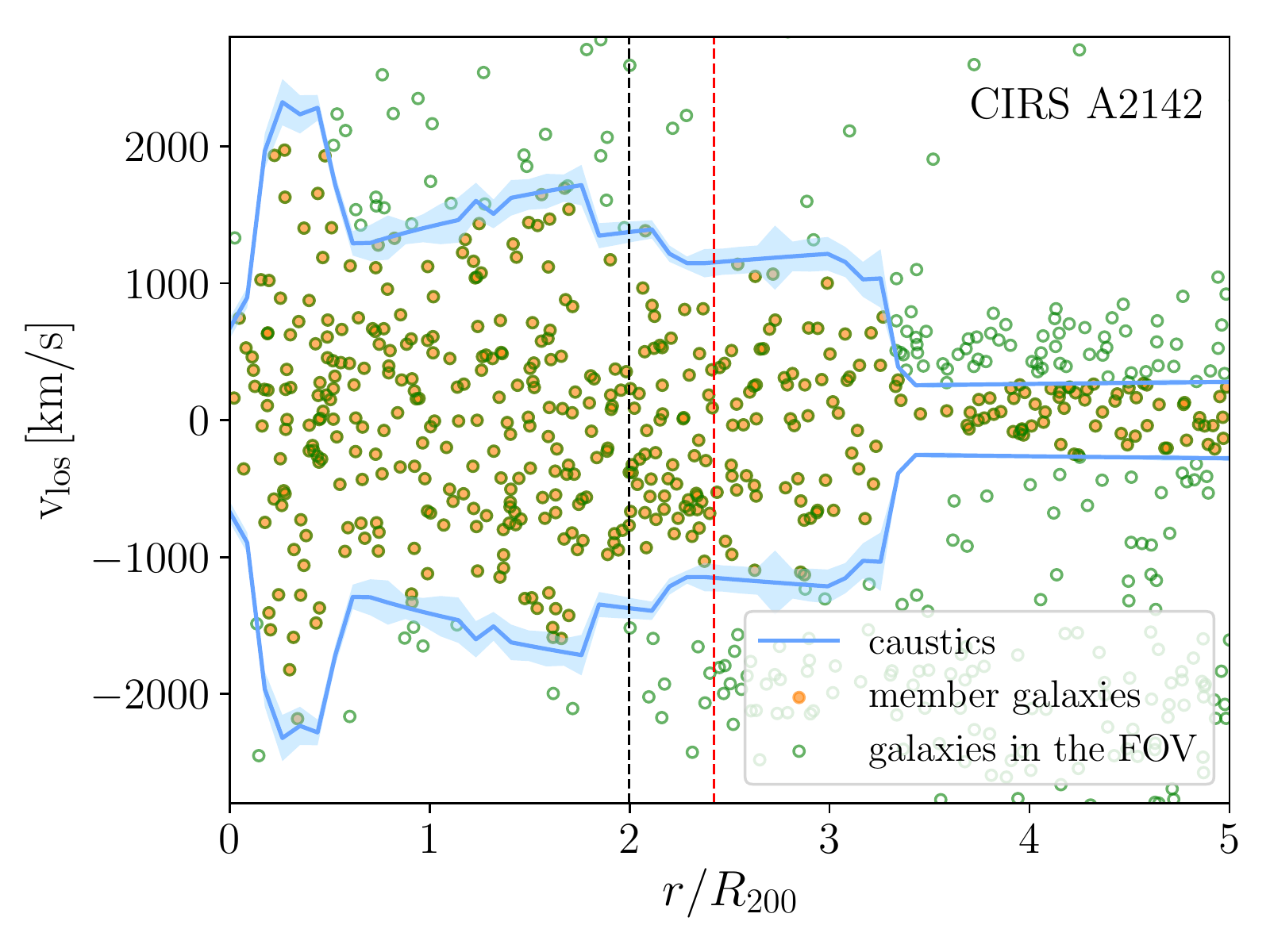}  %
  \includegraphics[scale=0.55]{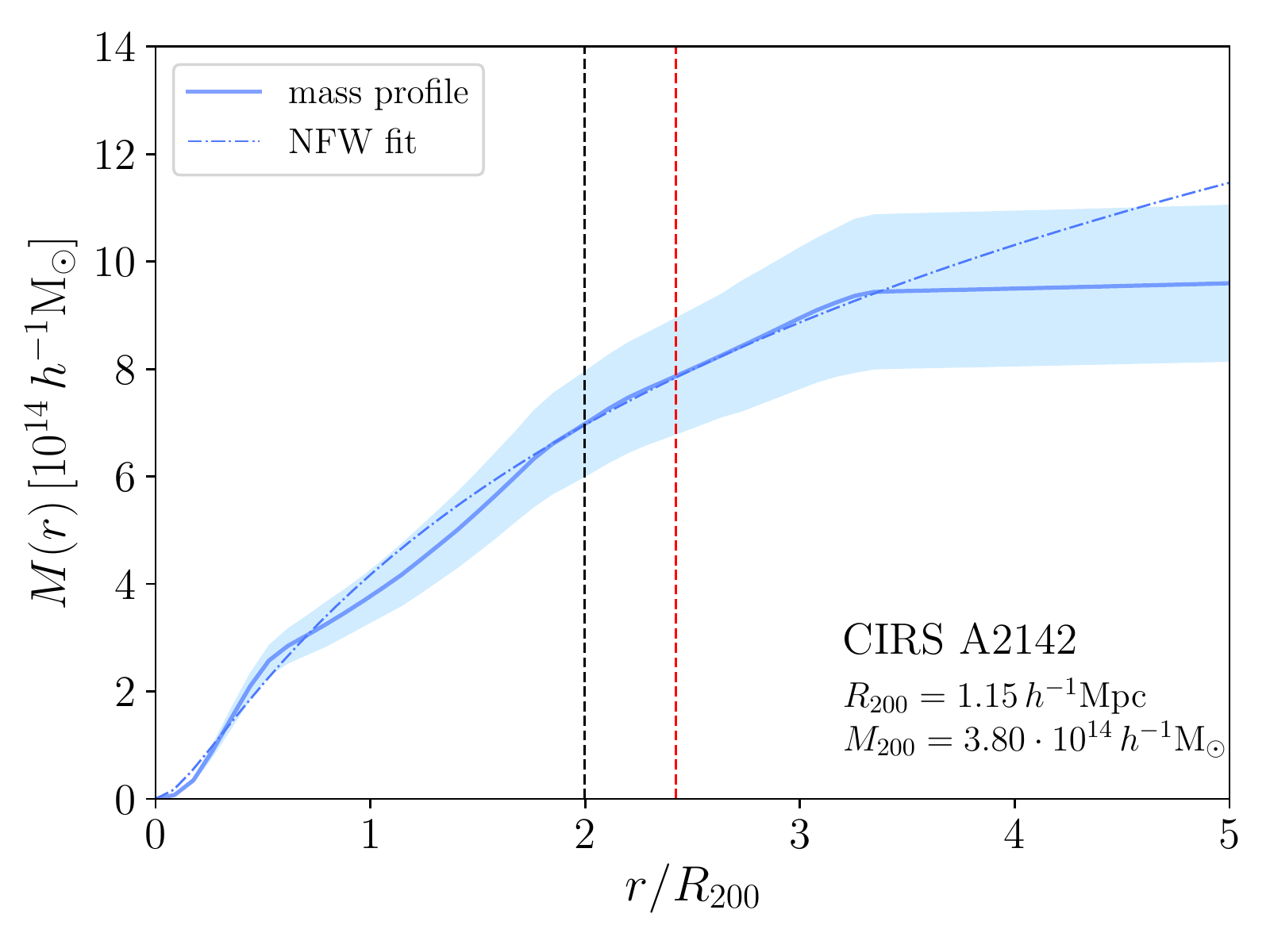} 
\caption{Three examples of the estimation of the MAR for individual clusters in CIRS. The {\it left column} shows the ${R-v_{\rm los}}$ diagram for each cluster. The {\it right column} shows the corresponding caustic mass profile (solid line) and the NFW fit (dot-dashed line). The black and red dashed lines show the inner and outer radius of the infalling spherical shell. The $M_{200}$ of the clusters increases from top to bottom.}
\label{cuts}
\end{figure*}

Tables \ref{CIRSsample} and \ref{HeCS sample} list the individual MARs and 
Fig. \ref{histograms} shows the distributions of the individual MARs of our three cluster subsamples. The density distributions 
are normalised to unity for an easier comparison. The blue, orange, and green histograms refer to the CIRS, low-mass HeCS, and high-mass HeCS clusters, respectively; the same colours are used for the vertical dashed lines showing their median MAR. 
The distributions generally show a peak at small MARs and an extended tail of large MARs. In general, the small and large MARs are associated with low- and high-mass clusters, respectively.

\begin{figure*}[htbp]
  \centering
  \includegraphics[scale=0.55]{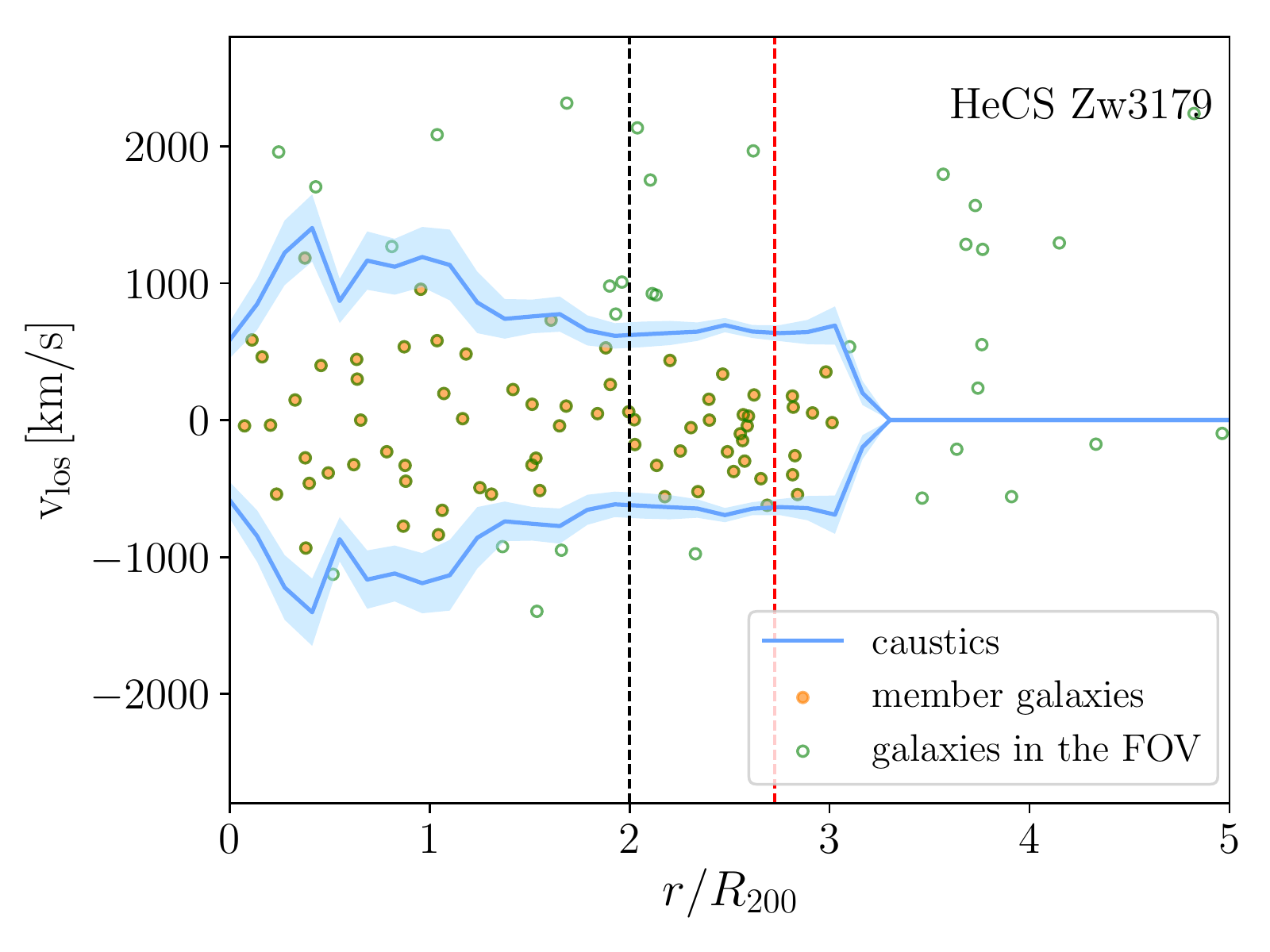}  %
  \includegraphics[scale=0.55]{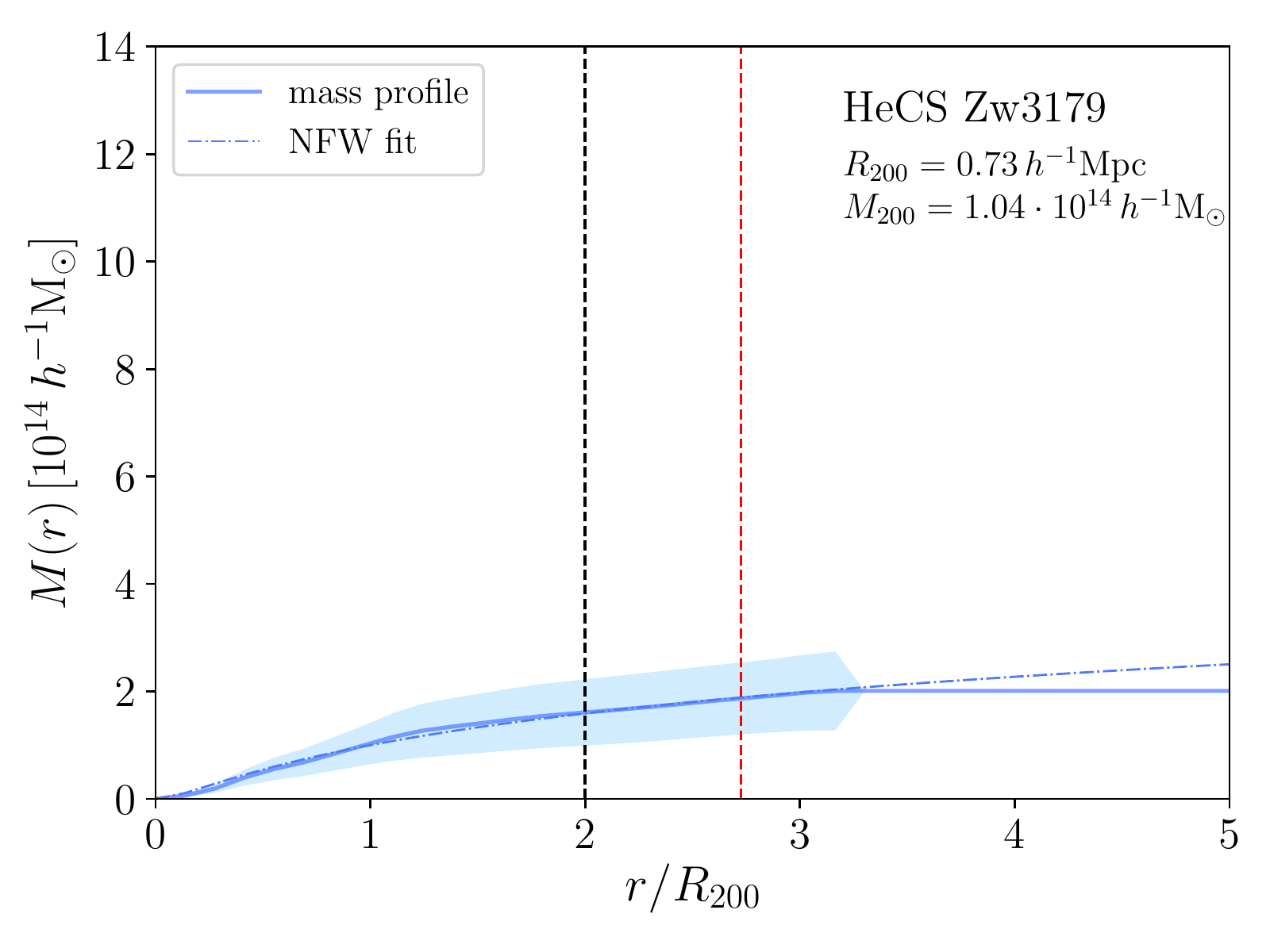} \\
  \includegraphics[scale=0.55]{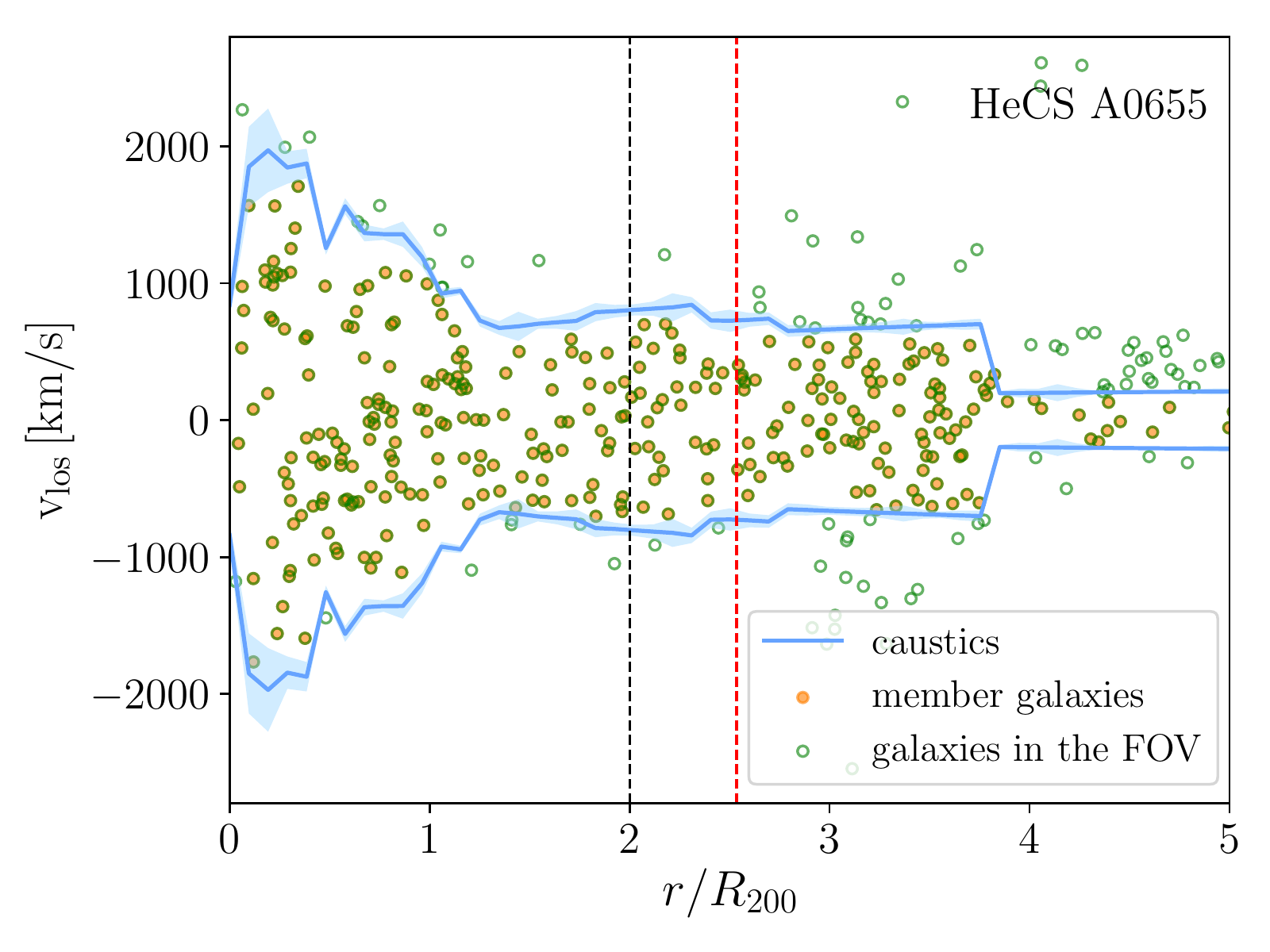}  %
  \includegraphics[scale=0.55]{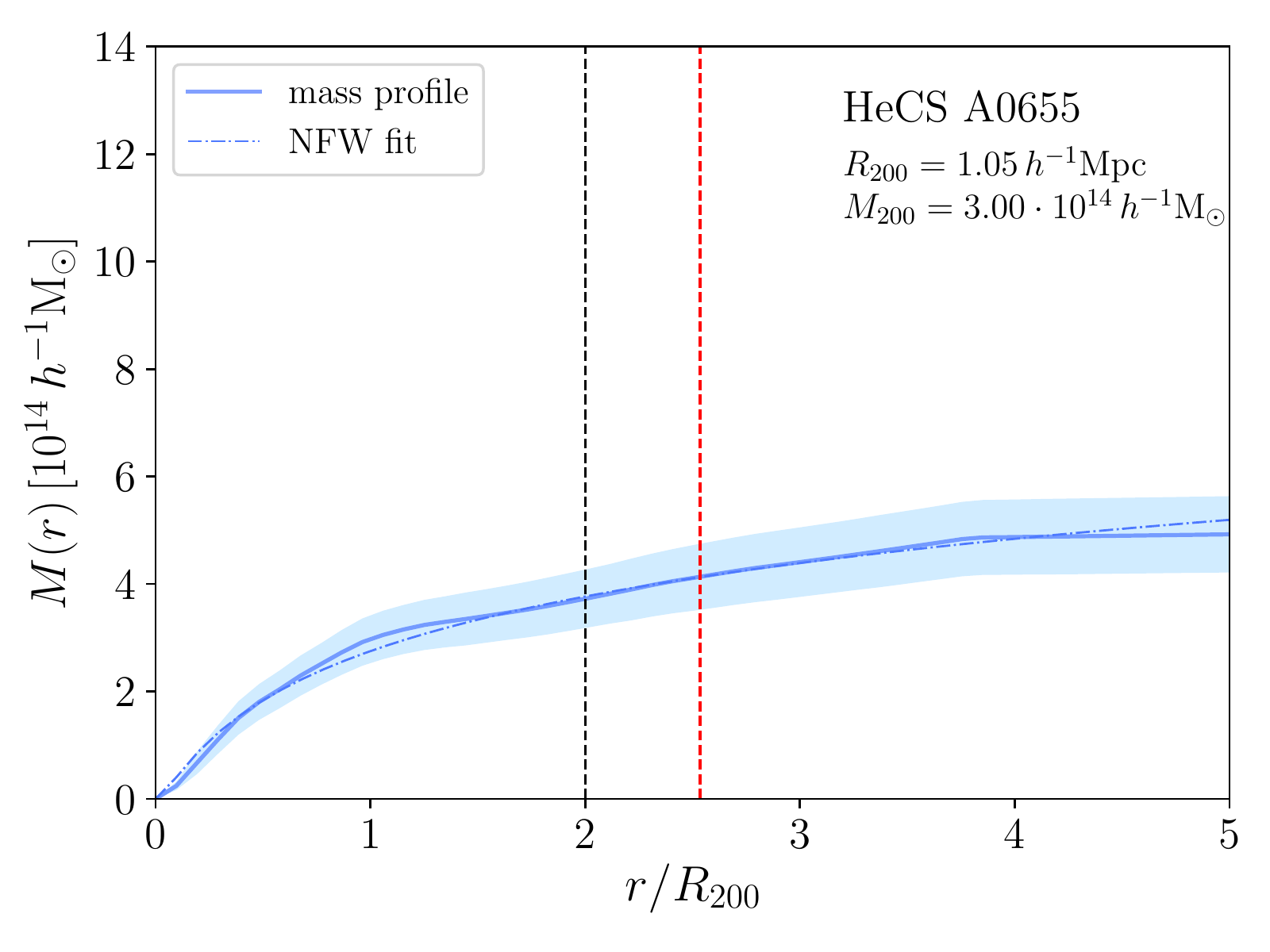} \\
  \includegraphics[scale=0.55]{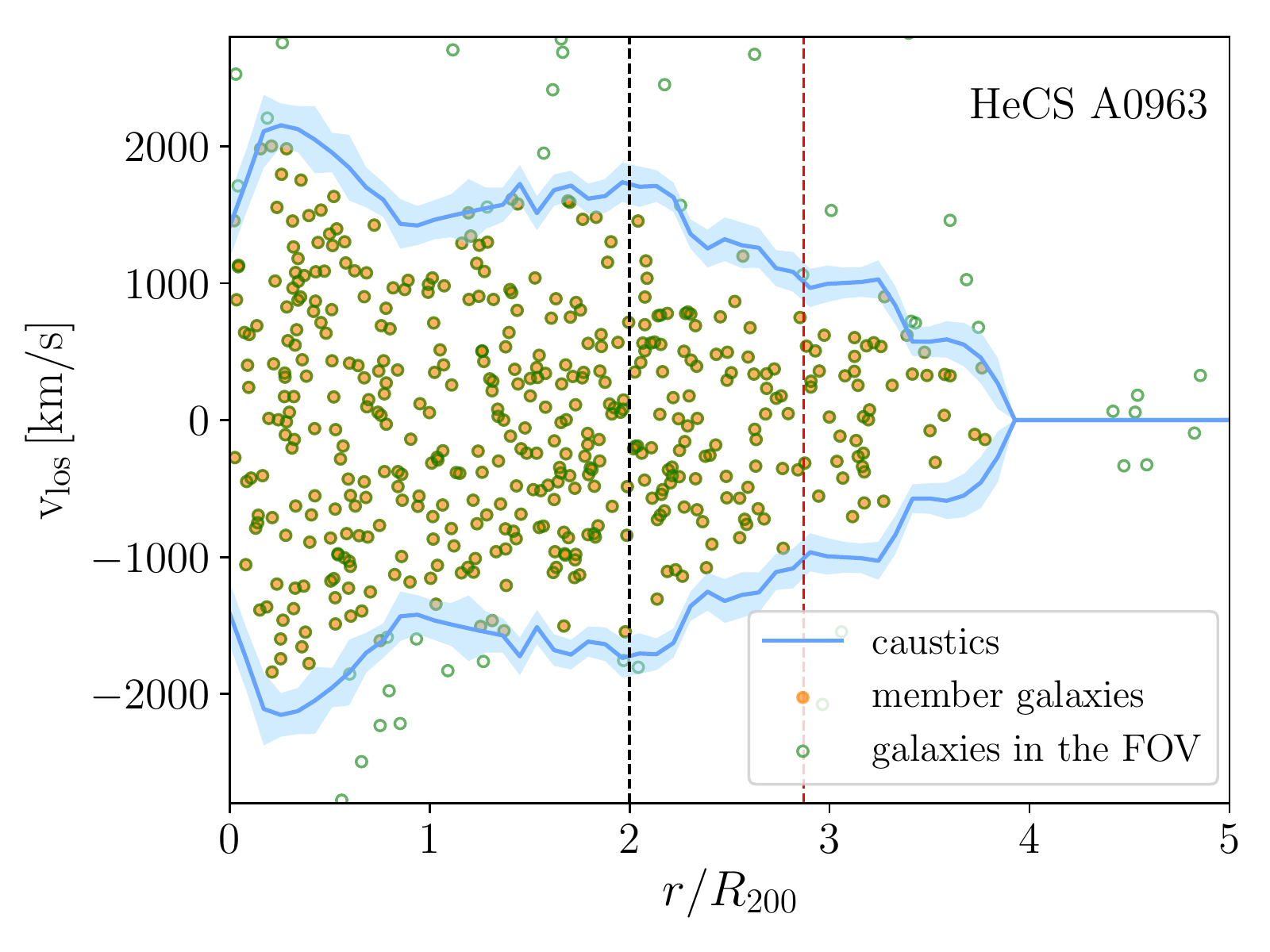}  %
  \includegraphics[scale=0.55]{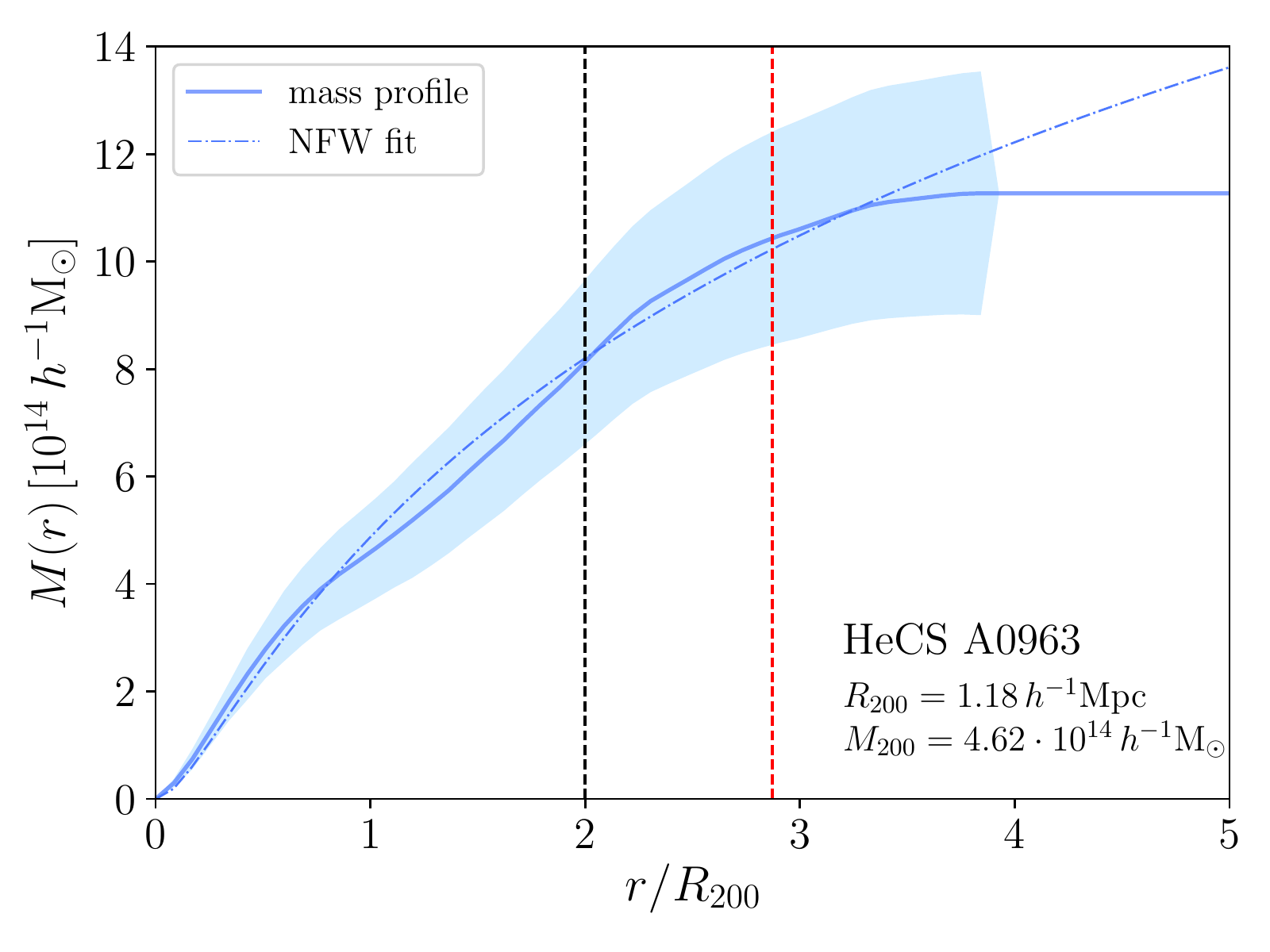} 
        \caption{Same as Fig. \ref{cuts} for three HeCS clusters. Zw3179 and A655 belong to the low-mass sample, whereas A963 is in the
        high-mass sample.}
\label{cutshecs}
\end{figure*}

Figures \ref{cuts} and \ref{cutshecs} show ${R-v_{\rm los}}$ diagrams and cumulative mass profiles for six sample clusters.
In all the panels, the black dashed line shows the inner radius of the infalling shell; the red dashed line shows the outer radius of the shell (Eq.~\ref{thickness}).
In the two figures, the mass of the cluster, $M_{200}$, increases from top to bottom. More massive systems tend to be surrounded by larger amounts of mass and the caustic amplitude decreases more slowly with radius. Consequently, we expect that the mass of the infalling shell, and thus the MAR, is correlated with $M_{200}$. 

\begin{figure}[htbp]
\centering
\includegraphics[scale=0.57]{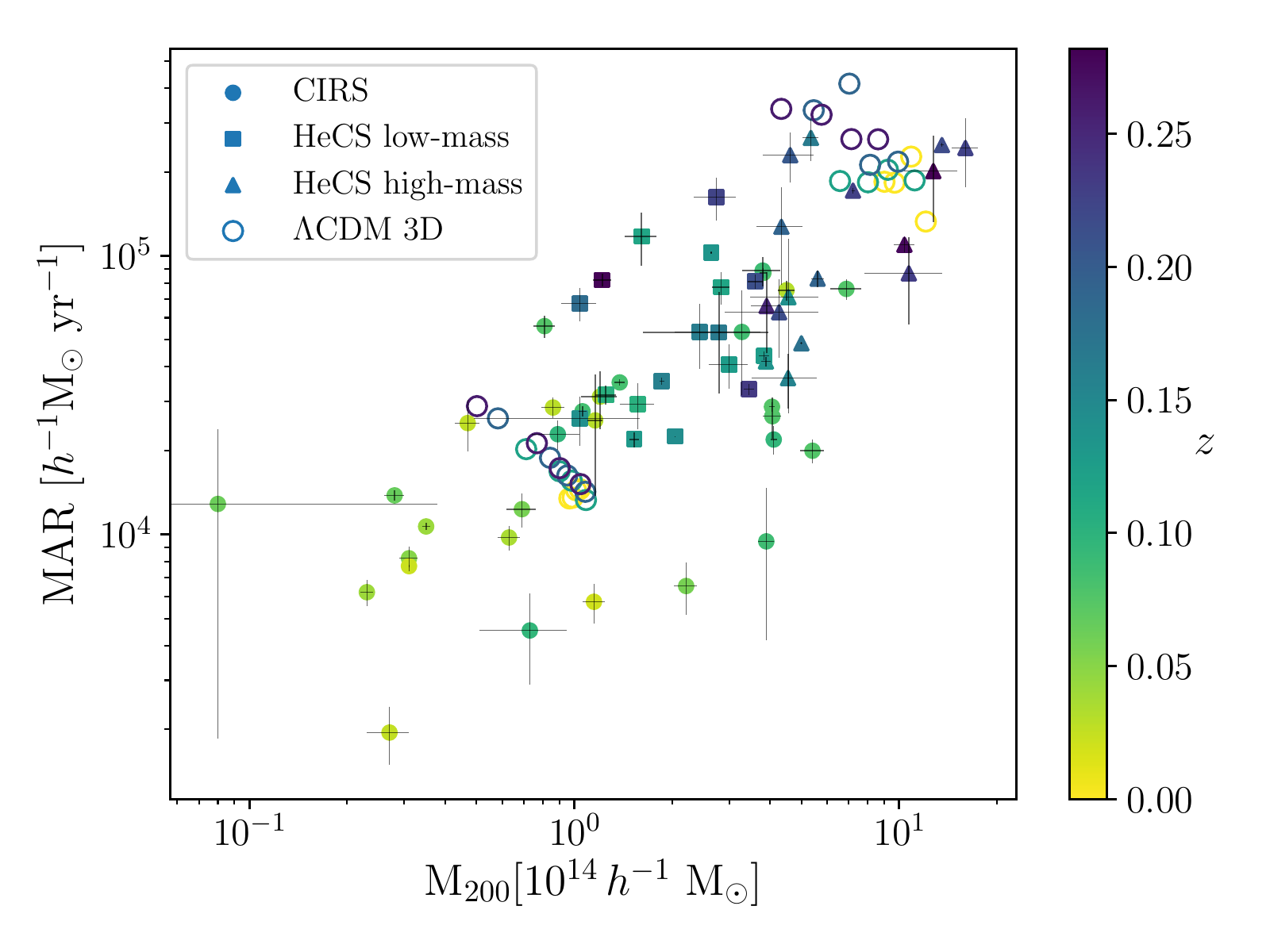}  
\caption{MAR of individual clusters as a function of their mass $M_{200}$; the colour code shows the dependence on redshift. Filled circles, squares and triangles refer to the CIRS, low- and high-mass HeCS sample, respectively. The open circles show the median MARs of the $N$-body clusters of our two simulated samples; for these clusters, we estimate the MAR from their three-dimensional mass profiles.}
        \label{mar-m200wsim}
\end{figure}

Figure \ref{mar-m200wsim} shows the correlation between $M_{200}$ and the MAR. There is also, as expected, a correlation with redshift.
The positive correlation with redshift and $M_{200}$ is clearly expected in the hierarchical clustering scenario where more massive halos lie in higher density regions and they are surrounded by larger amounts of mass \citep{bardeen1985statistics,laceyCole93,2009MNRAS.398.1858M,2010MNRAS.406.2267F,2014MNRAS.445.1713V}. Similarly, halos with comparable masses are expected to have larger MARs at larger redshifts. The correlations are statistically significant according to Kendall's test: the coefficients, $\tau$, are $0.516$ and $0.528$ for the MAR versus $M_{200}$ and versus $z$, respectively, with corresponding significance levels of $p_{M_{200}}=1.7\cdot 10^{-9}$ and $p_z=7.6\cdot 10^{-10}$.

To compare these measurements with our simulated clusters, we consider each simulated
cluster sample at the four redshifts $z=0.0$, $0.12$, $0.19$, and $0.26$, and separate each sample into four mass bins; these splittings yield 16 subsamples for each of the $10^{14}$ and $10^{15}h^{-1}$~M$_\odot$ sample of simulated clusters. The open circles in Fig. \ref{mar-m200wsim} show the median MARs of these 32 subsamples. The $\Lambda$CDM expectations appear fully consistent with our measurements. 

We close this section with a brief comment about two specific clusters, namely: A750 and A1758. In agreement with \cite{2013ApJ...767...15R}, we find that the galaxies in the redshift catalogue of A750 belong to two different clusters at different redshifts: A750 at $z= 0.164$, and MS0906 at $z = 0.177$. The caustic technique successfully identifies the two clusters and estimates the two MARs separately. Their individual MARs are listed
in Table \ref{HeCS sample}. A750 belongs to the low-mass HeCS sample, whereas MS0906 is in the high-mass sample.

Similarly, a weak lensing analysis, based on $B$ and $V$ passbands images of A1758 at redshift $z = 0.28$, by \cite{ragozzine2011weak}, suggests that this cluster is a system of four gravitationally bound substructures currently undergoing two separate mergers.
The mass estimate derived by the caustic technique is not affected by the presence of substructures \citep{1999MNRAS.309..610D} and we can thus estimate the MAR of A1758 without any particular precaution (Table \ref{HeCS sample}). 

\subsection{Average MAR} \label{Results}

\begin{figure}[htbp]
\centering
\includegraphics[scale=0.56]{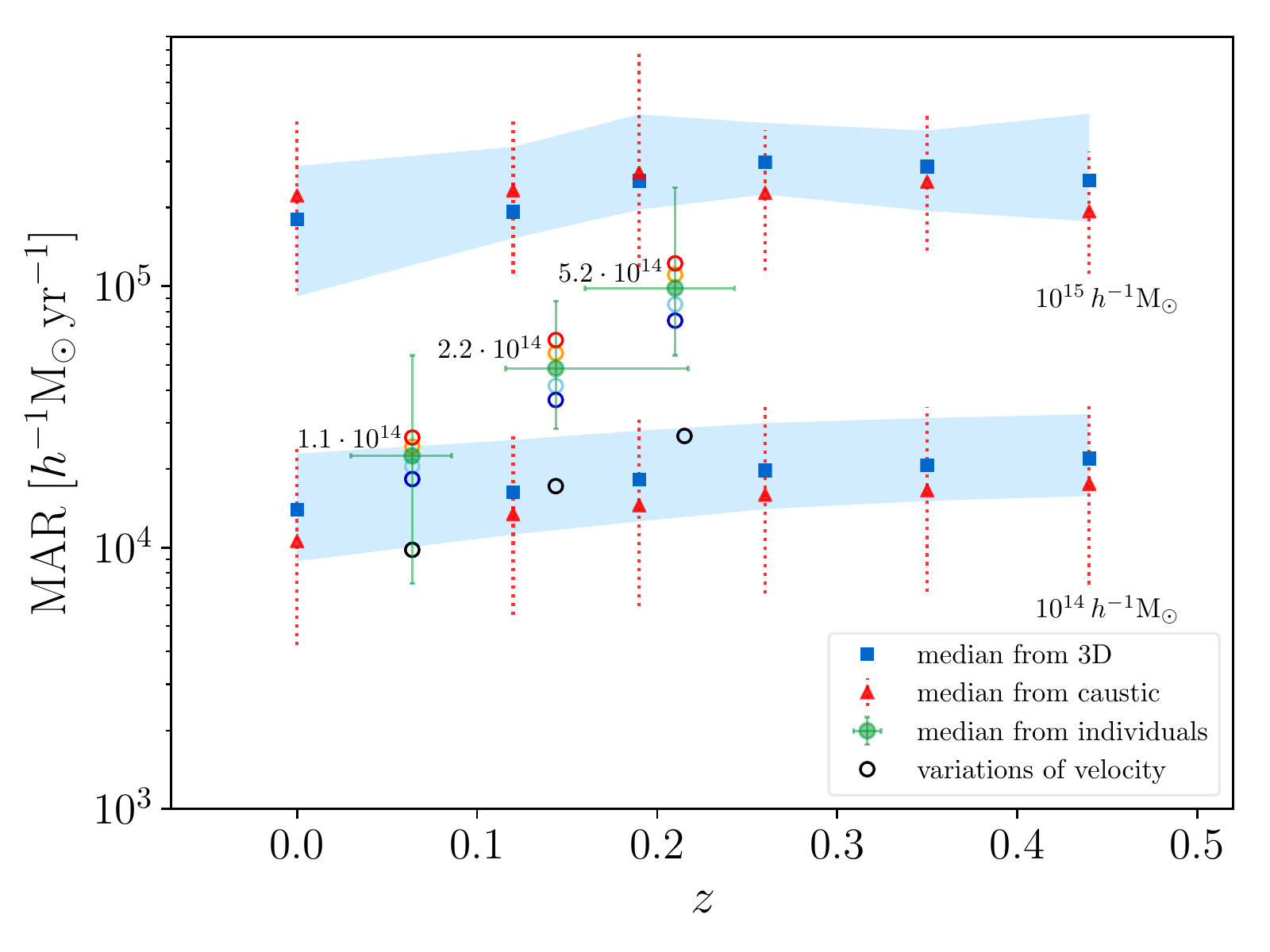}
        \caption{Median MARs of real (green) and simulated (blue and red) clusters. The green error bars show the $68$th percentile ranges of the MAR and redshift distributions of the real clusters. The value close to each green filled circle is the median mass $M_{200}$ of the real clusters in units of $h^{-1}$~M$_\odot$. The blue squares and red triangles are for the MAR based on the three-dimensional and the caustic mass profiles of simulated clusters, respectively (from Fig. \ref{marsimulated}).
        The light-blue shaded areas and the red error bars are the $68$th percentile range of the distribution of the individual MAR of the simulated clusters. The open circles show the median MARs of real clusters for different values of the infall velocity. We show the results where we changed the initial infall velocity that we adopted by: -40\% (blue) , -20\% (cyan), +20\% (orange), and then +40\% (red). The black circles show the MAR estimated with an initial infall velocity of $v_i=0$.} 
\label{marscomplete}
\end{figure}

Figure \ref{marscomplete} shows the median of the individual MARs as a function of redshift. The green filled circle at redshift $z=0.064$ shows the median MAR of the CIRS clusters; the green filled circles at redshift $z=0.14$ and $z=0.21$ show the median MARs of the two HeCS subsamples.
We plot each green circle at the median redshift of the subsample. The number close to each of them shows the median mass of each sample in units of $h^{-1}$~M$_\odot$. 
The error bars show the $68$th percentile ranges of the distributions  of the MAR and of the redshift of the clusters.

To compare these measured MARs with the expectations of the $\Lambda$CDM model, Fig. \ref{marscomplete} shows the results of Fig. \ref{marsimulated} for the simulated clusters extracted from the L-CoDECS simulation (Sect. \ref{MarCodecs}). The blue squares show the medians of the MARs derived from the three-dimensional mass profiles. The red triangles show the medians of the MARs derived from the caustic mass profiles estimated from the mock redshift surveys. 

This figure confirms the result of Fig. \ref{mar-m200wsim}: the medians of the MARs of real clusters fall within the range of the MAR of simulated clusters. The three median masses of the real cluster samples, $1.1$, $2.2$, and $5.2\times 10^{14}h^{-1}$~M$_\odot$, are in between the two median masses,  $10^{14}$ and $10^{15}h^{-1}$~M$_\odot$,  of the two samples of the simulated clusters. 

Figure \ref{marscomplete} also shows that the spreads in the median MARs of the real clusters are comparable with or even smaller than the spreads of the mock catalogues. This result supports the conclusion that the caustic technique returns robust estimates of the average MAR of clusters and that mock catalogues tend to overestimate the expected uncertainties because the caustics are usually less well-defined in the simulations than in the data (Sect. \ref{MarSimulated}).

We now quantify the effect of the value of the initial velocity $v_i$ on the estimate
of the MAR. The open circles in Fig. \ref{marscomplete} show the median MARs of each cluster sample when we decrease $v_i$ by 20 or 40\% (cyan 
and blue circles) or increase  $v_i$ by 20 or 40\% (orange and red circles) with respect to the adopted $v_i$ (green dots 
with error bars).
The estimated MAR does indeed depend on $v_i$, but the resulting MARs remain within the $68$th percentile range of the MAR distribution.
The extreme and unrealistic choice of $v_i=0$ makes the estimated MAR (black circles) decrease substantially: the MAR then disagrees significantly with the $\Lambda$CDM model expectations. Nevertheless, the correlations between the MAR, redshift, and cluster mass persist regardless of the choice of $v_i$. 

\begin{figure}[htbp]
\centering
\includegraphics[scale=0.57]{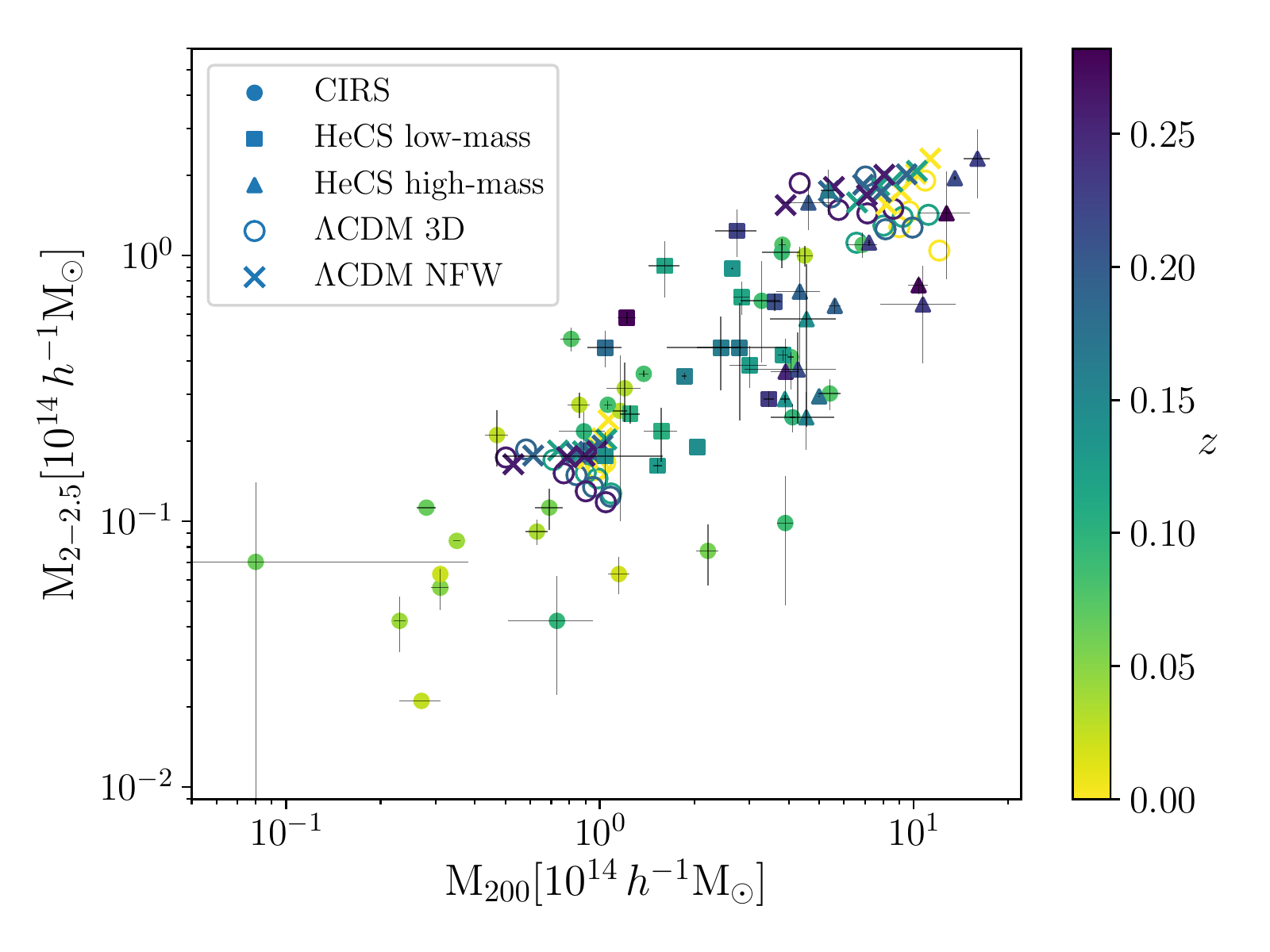}
\caption{Mass $M_{2-2.5}$ of the shell with radii $2R_{200}$ and $2.5R_{200}$ as a function of $M_{200}$. The colour code shows the dependence on redshift. Circles, squares and triangles refer to the CIRS, low- and high-mass HeCS sample, respectively. The open circles show the median $M_{2-2.5}$ of the $N$-body clusters of our two simulated samples, estimated from their three-dimensional mass profiles. The crosses show the median $M_{2-2.5}$ obtained from the NFW fits to the three-dimensional mass profiles: on average, they overestimate the individual true mass of the shell by $\sim 22-28\%$, depending on the mass bin. } \label{m2p5}
\end{figure}

\begin{table*}[h]
\begin{center} 
        \caption{\label{cat_stk_info} CIRS and HeCS subsamples for the stacked clusters.}
\begin{tabular}{lccrrc}
\hline
\hline
cluster sample & median $z$ & $68$th percentile range & median $M_{200}$ & $68$th percentile range & $M_{200}^{\rm stk}$ \\
        \quad & & [redshift $z$] & $\left[10^{14}h^{-1}~\text{M}_{\odot }\right]$  & $\left[10^{14}h^{-1}~\text{M}_{\odot }\right]$ & $\left[10^{14}h^{-1}~\text{M}_{\odot }\right]$ \\
 \hline
 & & & & \\
CIRS-STK1 & 0.055 & 0.025-0.083 & 0.88 & 0.31-1.2 & 1.07 $\pm$ 0.43\\
CIRS-STK2 & 0.076 & 0.052-0.087 & 3.3 & 2.2-4.4 & 3.3 $\pm$ 1.1\\
HeCS-STK1 & 0.14 & 0.12-0.18 & 1.9 & 1.0-2.9 & 2.8 $\pm$ 0.57\\ 
HeCS-STK2 & 0.20 & 0.14-0.25 & 5.6 & 4.4-11.8 & 7.3 $\pm$ 1.8 \\
\hline
 \end{tabular}
 \end{center}
 \end{table*}
 
A simple analysis explains qualitatively why the correlations between MAR, redshift, and cluster mass are so robust. 
Around each cluster, we consider the mass $M_{2-2.5}$ of the shell with inner and outer radius 
$2R_{200}$ and $2.5R_{200}$, respectively. This shell is comparable to the shell usually identified by the solution of Eq.~(\ref{thickness}). Figure \ref{m2p5} shows $M_{2-2.5}$ as a function of the cluster mass $M_{200}$. The colour and symbols are the same as in Fig. \ref{mar-m200wsim}. The two figures are qualitatively similar and show, as expected, that the correlations we see for the MAR derives directly from the correlations between $M_{2-2.5}$, $M_{200}$, and $z$.
We also show the shell mass $M_{2-2.5}$ derived from the NFW fits to the three-dimensional mass profiles of our simulated samples. On average, for the clusters in the low-mass bin, the shell mass derived from the NFW fits is $\sim 22\%$ larger than the actual mass. This average overestimate increases to $\sim 28\%$ for the clusters in the high-mass bin. Because this shell mass strongly correlates with the MAR, we expect a comparable overestimation of the MAR if we adopt the NFW fits rather than the mass profiles estimated with the caustic method. In Sect. \ref{MarCodecs}, we show that, over the entire radial range $[0-4]R_{200}$, the NFW mass profile overestimates the true mass profile by $\sim 5\%$, on average. The mass of the shell estimated with the NFW mass profile overestimates the true shell mass by a larger factor, $\sim 22-28\%$, because the shell mass derives from the difference of the masses estimated at two different radii, $2R_{200}$ and $\sim 2.5R_{200}$. 

A different strategy for estimating the mean MAR of real clusters is to apply our recipe to an average cluster obtained by stacking the real clusters \citep{2006AJ....132.1275R,2011MNRAS.412..800S}. This approach has the advantage of considering all the clusters of each sample, including those clusters where the individual MAR cannot be estimated. In addition, stacking the clusters averages out the deviations from spherical symmetry of the individual clusters. 

Similarly to HeCS, we separated the CIRS sample into a low-mass and a high-mass subsample, according to the cluster median mass $M_{200}=1.78\times 10^{14} h^{-1}$~M$_\odot$. Table \ref{cat_stk_info} lists the median masses and the $68$th percentile ranges of the mass and redshift distributions of the four subsamples.
For each cluster subsample, we built a ${R-v_{\rm los}}$ diagram containing all the galaxies in the field of view with line-of-sight velocity within $4000$~km~s$^{-1}$ from the cluster velocity and within $10h^{-1}$~Mpc from the cluster centre. The `Total' column of Table \ref{cat_stk_nrs} shows the number of galaxies within these four stacked clusters.

\begin{table}[h]
\begin{center} 
        \caption{ Galaxies within the stacked clusters.}\label{cat_stk_nrs}
\begin{tabular}{lcccc}
\hline
\hline
cluster sample & Total & Members-only & Equally-weighted \\
 \hline
 & & & \\
CIRS-STK1 & 17361 & 5243 & 7151\\
CIRS-STK2 & 26868 & 12161 & 15615\\
HeCS-STK1 & 11686 & 5300 & 6589\\ 
HeCS-STK2 & 12745 & 7982 & 6456\\
\hline
 \end{tabular}
 \end{center}
 \end{table}

\begin{figure*}[htbp]
\centering
\includegraphics[scale=1]{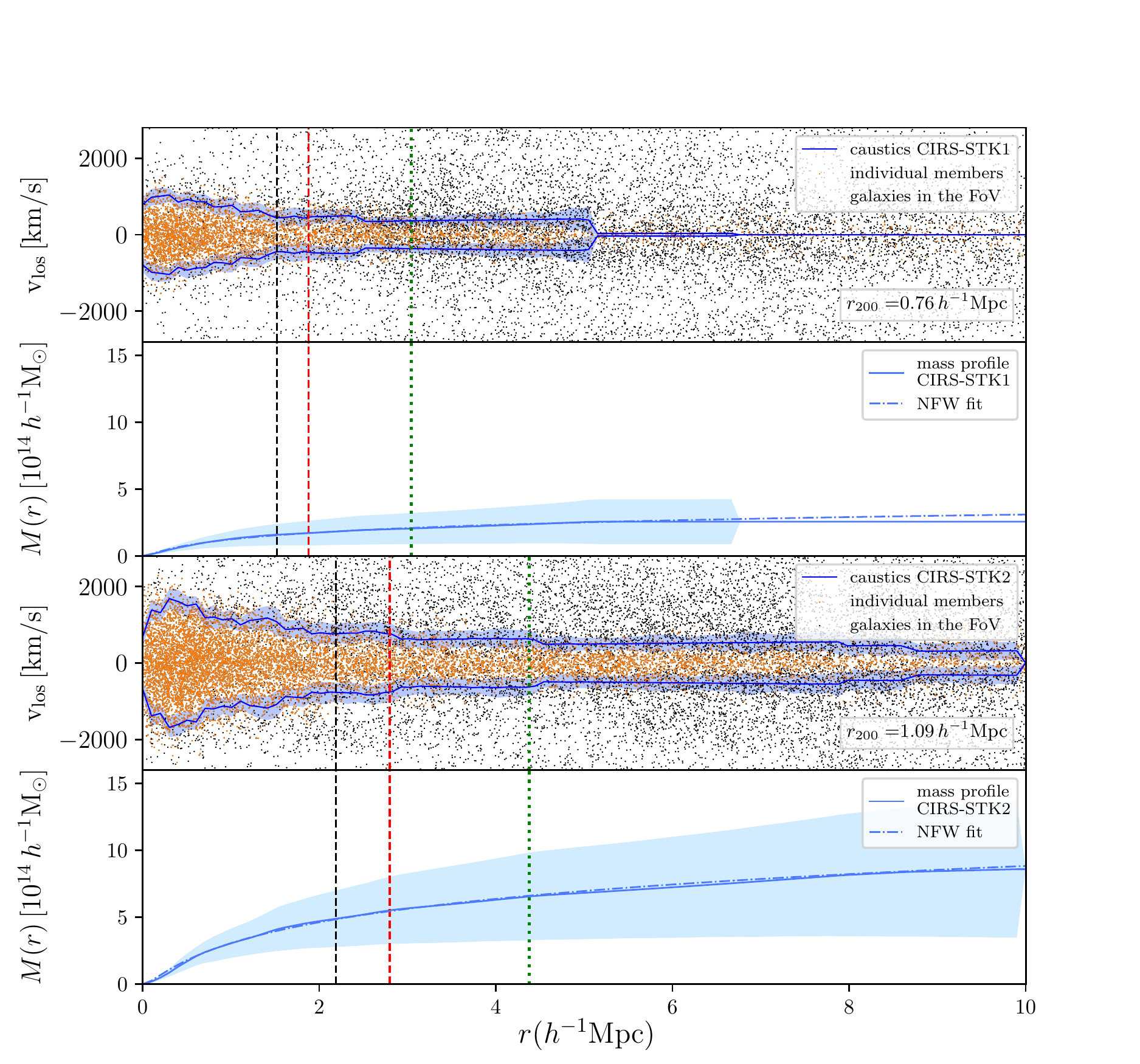}   

        \caption{${R-v_{\rm los}}$ diagrams and corresponding mass profiles of the two stacked clusters from the CIRS catalogue. The {\it upper (lower) panels} refer to the low-mass (high-mass) subsample. The black and red dashed vertical lines show the inner and outer radius of the spherical shell used to estimate the MAR. The shaded areas show the 50\% confidence level of the caustic location and of the caustic mass profile according to the caustic technique recipe. The dot-dashed lines show the NFW fits to the caustic mass profiles; the green vertical dotted lines show the upper limit of the radial range over which the NFW fits have been performed. } \label{stkC} 
\end{figure*}

\begin{figure*}[htbp]
\centering  
\includegraphics[scale=1]{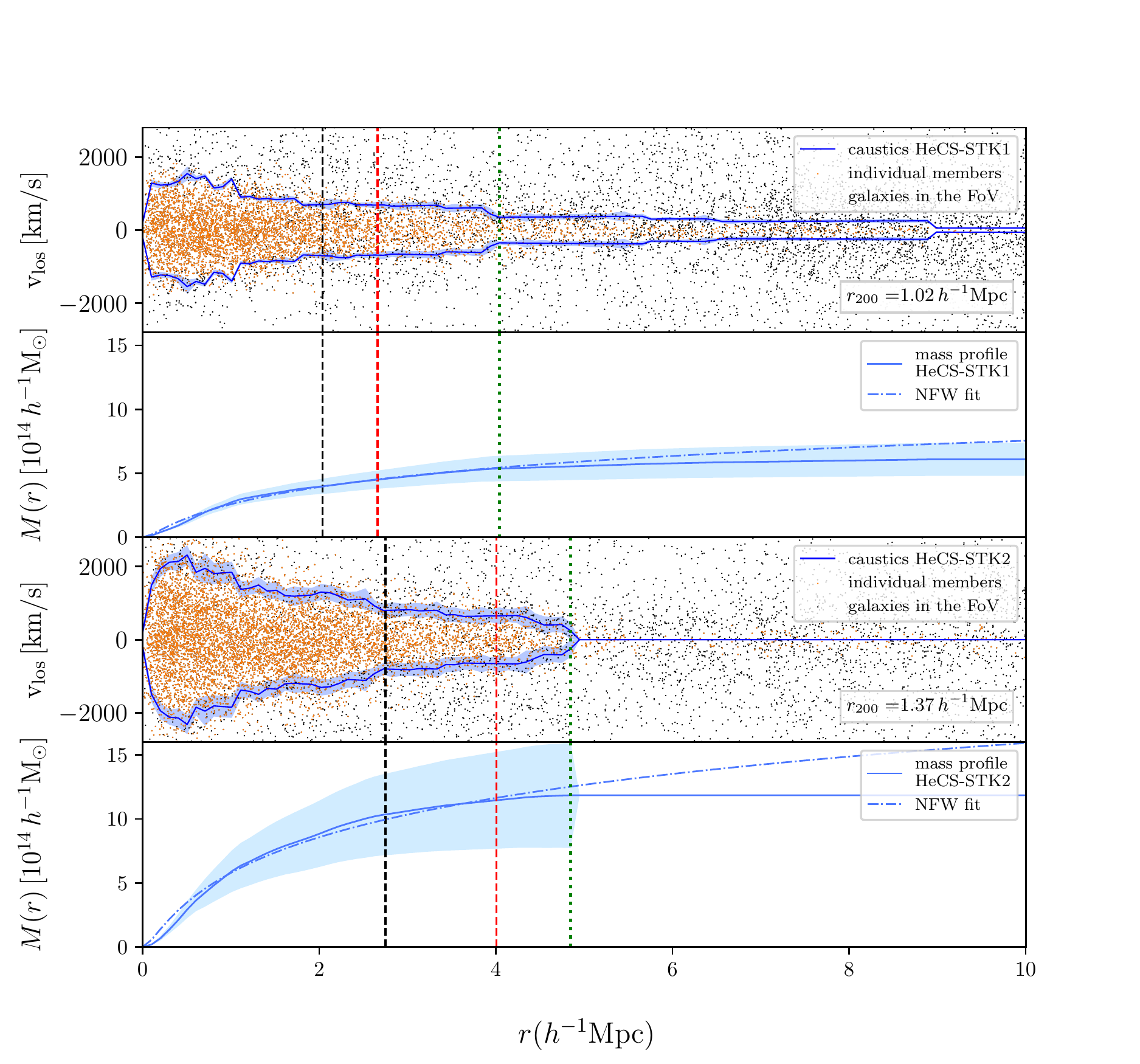} 

\caption{Same as Fig. \ref{stkC} for the HeCS catalogue. } \label{stkH} 
\end{figure*} 

In the ${R-v_{\rm los}}$ diagram of the stacked cluster, the velocities and the radial distances are expressed in the reference frame of each cluster without any additional normalisation.
This simple approach is justified by the homogeneity of mass and size for the clusters within each subsample. This stacking procedure avoids introducing errors from the estimate of the size and velocity dispersion of the clusters. If we stacked clusters of widely different mass we would require an additional normalisation. The ${R-v_{\rm los}}$ diagrams of the stacked clusters are shown in the first and third panels of Figs. \ref{stkC} and \ref{stkH}.
The mass profiles estimated with the caustic technique applied to these ${R-v_{\rm los}}$ diagrams are shown in the second and fourth panels of Figs. \ref{stkC} and \ref{stkH}. The dot-dashed lines in these panels show the NFW fits to these profiles performed over the radial range within $4R_{200}$ or within the radius where the caustic amplitude shrinks to zero, if this radius is smaller than $4R_{200}$. The masses $M_{200}$ of the stacked clusters are listed in Table \ref{cat_stk_info}. 

To estimate the MAR, we need to set the value of the initial infall velocity $v_i$. With the same procedure described in Sect. \ref{Individuals}, we use the L-CoDECS simulations to find $v_i=-164\pm 3$~km~s$^{-1}$, and $v_i=-303\pm 12$~km~s$^{-1}$ for the low- and high-mass CIRS subsample, and $v_i=-324\pm 10$~km~s$^{-1}$ and $v_i=-735\pm 27$~km~s$^{-1}$, for the low- and high-mass HeCS subsample, respectively.

Figures \ref{stkC} and \ref{stkH} show the inner (black dashed line) and outer radius (red dashed line) of the infalling shell, according to Eq.~(\ref{thickness}).
Averaging out the deviations from spherical symmetry and increasing the number of member galaxies 
make the caustic location and the mass profiles appear smoother than those of individual clusters (e.g. Figs. \ref{cuts} and \ref{cutshecs}).

\begin{figure}[htbp]
\centering
\includegraphics[scale=0.56]{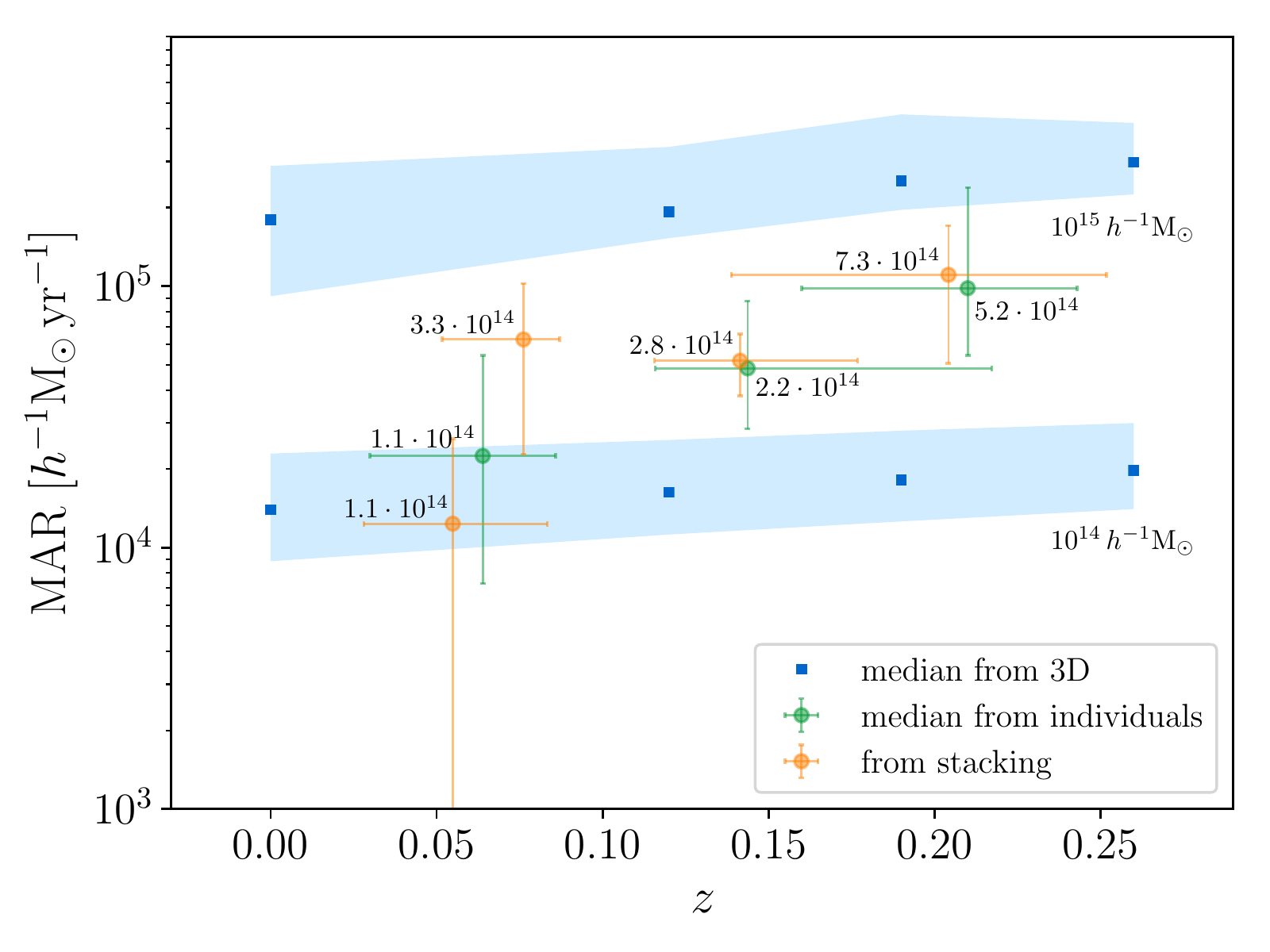} 
        \caption{MARs of the stacked clusters of CIRS (the two orange circles at low redshift) and HeCS (the two orange circles at high redshift). The median MARs of the individual clusters from Fig. \ref{marscomplete} are shown in green. The value close to each circle is the mass $M_{200}$ in units of $h^{-1}$~M$_\odot$. The blue squares with the shaded bands are from Fig. \ref{marsimulated} and show the median MARs and their spreads of the simulated clusters estimated with the three-dimensional mass profiles.} \label{stackedmar}
\end{figure}

Figure \ref{stackedmar} shows the MAR of the stacked clusters (orange circles), superimposed on the results obtained from the individual clusters (Fig. \ref{marscomplete}, green circles). The MARs of the stacked clusters are consistent with the median MARs of the individual clusters; they confirm the dependence of the MAR on cluster mass and redshift and agree with the predictions of the $\Lambda$CDM model. 

The uncertainty in the MARs of the stacked clusters are also consistent with the spread of the distributions of the individual MARs. 
The relative uncertainty in the MARs of the stacked clusters are $\sim 64$\%, an uncertainty that is larger than the relative uncertainties of $\sim 17\%$ of the individual MARs. Although the stacked clusters more closely satisfy the assumption of spherical symmetry, the increase of the number of background and foreground galaxies in the ${R-v_{\rm los}}$ diagram makes the location 
of the caustics more uncertain and, consequently, increases the uncertainty on the mass profile.\footnote{The number density of background and foreground galaxies in the ${R-v_{\rm los}}$ diagram affects the estimate of the uncertainties in the mass profile. The relative uncertainty in the caustic amplitude is $\varepsilon = {\sigma_{\mathcal{A}}/\mathcal{A}} = \kappa / \max\{f(r,v_\mathrm{los})\}$, where $f(r,v_\mathrm{los})$ is the galaxy number density in the ${R-v_{\rm los}}$ diagram and $\kappa$ is the threshold identified by the algorithm to locate the caustics. Larger number densities of background and foreground galaxies tend to decrease $\max\{f(r,v_\mathrm{los})\}$; $\varepsilon$ thus increases and, in turn, the uncertainty on the mass increases (Eq.~\ref{mass-uncert}). For example CIRS-STK2 has $\varepsilon \sim 0.35$, and a relative error in the MAR of $\sim 70$\%.} 
Nevertheless, as expected, most members of the stacked clusters are members of the individual clusters: 75\%, 81\%, 80\%, and 94\% of the members of CIRS-STK1, CIRS-STK2, HeCS-STK1, and HeCS-STK2, respectively, are members of the individual clusters. Hence, the location of the caustic curves appears to be robust.

We can reduce the uncertainty on the mass profile by adopting a different strategy to build the stacked clusters. For each cluster, we then include only the member galaxies identified by the caustic technique, namely, the galaxies within the caustics in each ${R-v_{\rm los}}$ diagram. Again, we include all the clusters, irrespective of the fact that their caustic amplitude can be zero beyond $2R_{200}$. The `Members-only' column of Table \ref{cat_stk_nrs} shows the number of galaxies within these four stacked clusters.  
Clearly, the number of galaxies in the ${R-v_{\rm los}}$ diagrams substantially drops compared to the previous stacked clusters: for the two mass bins of CIRS, the number of galaxies is reduced by 70\% and 55\%, respectively; for the HeCS subsamples, these numbers decrease by 55\% and 43\%. 
The ${R-v_{\rm los}}$ diagrams of these new stacked clusters have now well-defined trumpets and the caustics identify substantially fewer interlopers. 
As expected, the mean relative uncertainty in the estimated MAR drops to $\sim36\%$, approximately half of the relative uncertainty of the previous stacked clusters. Although the new MAR estimates are $\sim 12$\% smaller, on average, than the previous MARs, because the ${R-v_{\rm los}}$ diagrams are now slightly less sampled than the original stacked clusters, the difference between the two MAR estimates is within the uncertainties and the estimates are thus statistically consistent with each other.

In the stacking procedure, for each cluster, we include all the galaxies in the field of view within $4000$~km~s$^{-1}$ along the line of sight and within $10 h^{-1}$~Mpc from the cluster centre without any further normalisation of the line-of-sight velocity and projected distance of the cluster galaxies. 
The properties of the stacked cluster might thus mirror the properties of the clusters that contribute most galaxies and are presumably more massive. To quantify this effect, we then create new stacked clusters. For each cluster, we include the same number $N$ of galaxies randomly sampled from the field of view: for each mass bin, we set $N$ close to the number of galaxies of the cluster with the smallest number of galaxies. 
For the two mass bins of CIRS, we have $N=263$ and $N=571$; for HeCS, we have $N=402$ and $N=408$. These values of $N$ preserve the optimal number of galaxies $\gtrsim 200$ within $3R_{200}$ \citep{2011MNRAS.412..800S}. Only 5 out of 36 clusters in the low-mass subsample of CIRS have fewer galaxies than $N=263$; the clusters with fewer galaxies than $N$ are 2 out of 35, 1 out of 26, and 2 out of 26 for the high-mass subsample of CIRS, and for the low- and high-mass subsamples of HeCS, respectively. The `Equally-weighted' column of Table \ref{cat_stk_nrs} shows the number of galaxies within these four stacked clusters. 
These equally weighted stacked clusters have a ${R-v_{\rm los}}$ diagram that is slightly less sampled than the original stacked clusters. As a consequence, their MARs are, on average, 9\% lower than the previous MARs and their relative uncertainties are, on average, $\sim 25$\% larger. Nevertheless, their MAR estimates are still consistent with the results of Fig. \ref{stackedmar}.

\section{Discussion} \label{discussion} 

Our result is the first attempt to measure the MAR of clusters of galaxies by estimating the amount of mass in the outer regions. Our approach is crucially different from the approach usually adopted in the theoretical investigations of the accretion of dark matter halos. 
Most attention is focused on the connection between the accretion history of halos and their global properties at a particular epoch \citep[e.g.][]{2019MNRAS.485.1906R}. These properties include the concentration parameter of the NFW profile \citep{1997ApJ...490..493N} and the amount of substructure \citep{2000ApJ...535...30J}. More recently, \citet{2013MNRAS.432.1103L} show that the ubiquity of the NFW density profile in CDM models results from the mass independence of the accretion histories of dark matter halos.

The accretion histories of dark matter halos can, in principle, be a very sensitive probe of the nature of the dark matter particles and of the theory of gravity. For example, both self-interacting dark matter particles \citep{2018MNRAS.474..746B} and warm dark matter particles \citep{10.1093/mnras/stw1046} predict smaller concentrations for similar halo masses, which, in turn, affect the determination of the formation redshift of the dark matter halos. On the other hand, \citet{2019MNRAS.489.4658O} show that in the \citet{PhysRevD.76.064004} form of $f(R)$ gravity, the linear relation between concentration and formation time that is seen in CDM models becomes more complicated due to gravitational screening. 

Unfortunately, when moving from dark matter halos to real cluster of galaxies, the differences detected in the theoretical relations might appear out of reach because 
the formation redshift is an unobservable quantity and the concentration parameter is prone to various systematic errors and sample biases \citep{Mandelbaum_2008,bartelmann2010gravitational,10.1111/j.1365-2966.2011.20248.x}. 
Moreover, the properties of the central region of clusters, say at $\lesssim 0.1R_{200}$, are complicated by the physics of baryons, as they can play a relevant role in the dynamics of the cluster centre \citep{1986ApJ...301...27B,kravtsov2005effects}.

The $N$-body simulations suggest that the mass that is falling onto the cluster for the first time is located beyond the splash-back radius $\sim 2R_{200}$ \citep{2014ApJ...789....1D,2015ApJ...810...36M}. 
However, to date most investigations of the accretion rate of real clusters focus on regions inside this radius \citep[e.g.][]{Lemze_2013,2018MNRAS.477.4931H}. Therefore, comparing their results with the MAR derived from $N$-body simulations is not straightforward. 

On  the  contrary,  the  estimation  of  the  mass  at  distances $\gtrsim 2R_{200}$ is a more straightforward approach because most of this mass is falling onto the cluster for the first time and is thus a direct estimate of the accretion rate. In addition, in this region, baryons are expected to have a limited effect because gravity is the main driver of the properties of the density and velocity fields. Estimating the mass in these outer regions is inaccessible to strong lensing measurements and to current X-ray observations, but this can be tackled with the caustic technique or weak gravitational lensing. Thanks to the caustic technique, estimating the mass at these large radii, where dynamical equilibrium does not hold, becomes feasible when appropriately dense spectroscopic surveys are available  \citep{2011AJ....142..133G,2013ApJ...768..116S,2018ApJ...856..172S,2018ApJ...862..172R}. 

According to $N$-body simulations, at radii larger than $\sim 2R_{200}$, not all the mass is falling onto the cluster for the first time \citep[e.g.][]{Aung2020}, but $\sim 20-25\%$ of this mass has already been within $R_{200}$  \citep{2009PhDT........13L,2017ApJ...843..140D, 2020arXiv200805475B,2020MNRAS.499.3534X}. When estimating the MAR of real clusters from the amount of mass in this outer region, this evidence can, in principle, prove problematic.  
In our approach, we could avoid this problem by adopting an inner radius of the shell greater than $2R_{200}$; this choice would  decrease the fraction of the mass that has already been within $R_{200}$ but it would yield a shell extending to radii larger than $\sim 2.5 - 3R_{200}$; here, real clusters currently suffer from spectroscopic incompleteness and the estimate of the MAR would be biased. Adopting $2R_{200}$ as the inner radius of the shell is thus a trade-off between the maximisation the fraction of infalling mass and the current observational limitations.

With our approach, the comparison between the observed MAR and the expectations derived from the $N$-body simulations is immediate -- once exactly the same procedure is applied to the real and mock clusters. Moreover, the identification of the systematic errors is straightforward. On the contrary, making a  clean comparison of our estimated MAR with the expectations available in the literature is not possible: the theoretical modelling is based on the mass accretion history of dark matter halos derived from their merger trees \citep[e.g.][]{2009MNRAS.398.1858M,2010MNRAS.406.2267F,10.1046/j.1365-8711.2002.05171.x,2014MNRAS.445.1713V,2013MNRAS.432.1103L,2014MNRAS.441..378L,10.1093/mnras/stw1046}  or from the application of the extended Press-Schechter theory \citep{Bond1991,laceyCole93,2015MNRAS.450.1514C}. All these models relate the MAR of a cluster of mass $M(z)$ at redshift $z$ to the mass $M(z_i)$ of the same cluster at redshift $z_i$. For real clusters, the knowledge of the cluster mass at two different redshifts clearly is unavailable.

The technique we present here has two potential observational limitations: the use of $N$-body simulations for setting the value of the infall velocity and the removal of the real clusters with a sparse spectroscopic sampling when computing the individual MAR. At present, the former issue cannot be solved because  the measurement of the radial velocity of cluster galaxies is not currently feasible. We can, in principle, remove the requirement of including the infall velocity from $N$-body simulations, as well as the assumption of the infall time, by adopting the mass within the shell $[2,2.5]R_{200}$ as an appropriate prior of the MAR, as Figs. \ref{m2p5} and \ref{mar-m200wsim} suggest.  

The issue of removing clusters from the samples of clusters with individual MAR concerns the distribution of the galaxies within the ${R-v_{\rm los}}$ diagram. The caustics of $\sim 65\%$ of the removed clusters have unphysical spikes that lead the caustic technique to refrain from estimating the mass at large radii. Usually, a spike emerges when there is an asymmetric lack of galaxies with respect to the line $v_{\mathrm{los}}=v_{\mathrm{cl}}$ in the ${R-v_{\rm los}}$ diagram. This lack of galaxies can originate either from a poor spectroscopic sampling or from a real underdense region around the cluster: clusters are generally not spherically symmetric, as assumed by the caustic technique, and asymmetric galaxy distributions on the sky can generate inhomogeneous galaxy distributions in the ${R-v_{\rm los}}$ diagrams. For clusters with a spike resulting from poor sampling rather than an actual underdense region, we estimate that the caustic technique could properly locate the caustics if the spectroscopic sampling were increased by $\sim 15\%$ in a stripe, of the ${R-v_{\rm los}}$ diagram, of thickness $\sim 0.5\,h^{-1}$~Mpc centred on the spike.  
In principle, we could thus be able to recover  $\sim 55$\% of the clusters removed in both CIRS and HeCS. The remaining clusters are either: (1)  less massive and thus have a poorly populated ${R-v_{\rm los}}$ diagram or (2) embedded in a particularly dense region and the caustic algorithm is unable, at large radii, to properly distinguish between galaxy members and interlopers and, thus, to locate the caustics.  

The growth of cosmic structures on linear scales can be different for different dark energy models and modified gravity models \citep[e.g.][]{burrage2017beyond,giocoli2018weak,perenon2019optimising}. In principle, the MAR, which measures the growth on non-linear scales, can be a tool for discriminating among different theories, similarly to the cluster mass function \citep{kopp2013spherical,lombriser2013modeling,cataneo2016cluster,barreira2013spherical}. However, the sensitivity of the MAR on the theory of gravity remains to be investigated in detail \citep[e.g.][]{zentner2007excursion}. Similarly, it remains to be seen whether the MAR estimated with the caustic method for clusters at different redshift and with different mass are accurate enough to distinguish among the models. Specifically, we need to assess the optimal balance between the number of observed clusters and the spectroscopy sampling required to reach the necessary sensitivity. However, estimating the mass in the cluster outer regions might not even be necessary for probing the theory of gravity and the velocity field may suffice. For example, in his cosmological $N$-body simulations in the MOND framework, \citet{2016MNRAS.460.2571C} shows that the velocity field in the outskirts of clusters is significantly enhanced compared to standard gravity.
We plan to tackle these issues in future work.

\section{Conclusion}\label{conclusion}

We use the dense redshift surveys of the CIRS \citep{2006AJ....132.1275R} and HeCS \citep{2013ApJ...767...15R} clusters to make the first measurement of the MAR of 129 clusters in the redshift range of $0.01< z< 0.3$ and in the mass range of $10^{14}-10^{15}h^{-1}$~M$_\odot$.

Our measurement is based on the spherical accretion recipe suggested by \citet{2016ApJ...818..188D}, where the MAR is estimated from the mass of a shell, of appropriate thickness, surrounding the cluster. 
We set the inner radius of the shell at $2R_{200}$. The shell thickness is typically $\sim 0.5R_{200}$. To estimate the shell mass at these large radii, we use the caustic technique.
When applied to mock redshift surveys of simulated clusters in the same range of redshift and mass, our procedure returns an unbiased MAR within $\sim 19 \% $ of the true MAR, on average. 

Our estimates of the MARs of the CIRS and HeCS clusters agree with the MARs of the dark matter halos extracted from a $\Lambda$CDM $N$-body model \citep{2012MNRAS.422.1028B} in the same mass and redshift range. As in the simulations, the observed cluster MARs increase with cluster mass and redshift. 

The observed correlations of the MAR with cluster mass and redshift are shared by the mass of the infalling shell. We thus suggest that the mass of the shell can be a proxy of the MAR.

The MAR estimates of the individual clusters have typical uncertainties of $\sim 17\%$; for the stacked clusters, when we only stack the members of each individual cluster identified with the caustic technique, the MAR uncertainty is $\sim 36\%$. Systematic errors resulting from sample selection could impact these results. We show that neither CIRS nor HeCS have significant photometric incompleteness as a function of radius. However, HeCS misses at most $\sim 30\%$ of the blue galaxies compared to CIRS; the spectroscopic completeness of both CIRS and HeCS drops to $\sim 50\%$ at radii larger than $\sim 2 R_{200}$. Nevertheless, neither the photometric nor the spectroscopic incompletenesses lead to a bias in the MAR estimates. The systematics are within the random errors of the caustic technique. 

The measurement of the MAR holds promise for more stringent tests of the standard $\Lambda$CDM paradigm. With much larger cluster samples, it may also be able to distinguish between different models of dark matter and of modified gravity. These tests require  dense, complete spectroscopic surveys that extend to large cluster-centric radius.

The extension of the cluster survey to a significantly higher redshift provides a broader baseline for these tests. For example, N-body simulations predict that the accretion rate at $z=0.7$ is $\sim 2.6$ times larger than at the current epoch \citep{2009MNRAS.398.1858M}.  Both the caustic technique and the weak lensing method can readily be applied to this redshift. Deep photometric surveys, along with such multi-object spectrographs as the Prime Focus Spectrograph on Subaru \citep{takada2014extragalactic} will enable this extension on the observational side. Advances in simulations \citep{giocoli2018weak} will be important for refining the model testing.

\begin{acknowledgements}
We sincerely thank the referee whose sensible comments helped us to substantially clarify some relevant issues of our analysis. An early version of this work was the Master thesis in Physics at the University of Torino of SDG. CDB thanks Benedikt Diemer for useful discussion on the splashback radius. The graduate-student fellowship of MP is supported by the Italian Ministry of Education, University and Research (MIUR) under the {\it Departments of Excellence} grant L.232/2016. We also acknowledge partial support from the INFN grant InDark. This research has made use of NASA’s Astrophysics Data System Bibliographic Services.
\end{acknowledgements}

\bibliographystyle{aa}
\bibliography{michele}

\end{document}